\documentclass{aa}

\usepackage[utf8]{inputenc}
\usepackage[T1]{fontenc}
\usepackage{amsmath}
\usepackage{wasysym}
\usepackage{bm}
\usepackage{multirow}
\usepackage{graphicx}
\usepackage{txfonts}
\usepackage[usenames,dvipsnames,svgnames,table]{xcolor}
\usepackage{float}
\usepackage[
hyperindex=true,
colorlinks=true,
linkcolor=Sepia,
citecolor=Purple]{hyperref}
\usepackage{comment}

\newcommand{\Msun}{~\text{M}_{\astrosun}}
\newcommand{\E}[1]{\times 10^{#1}}
\newcommand{\pd}[2]{\dfrac{\partial #1}{\partial #2}} 
 
\newcommand{\coldens}{\Msun\cdot\mathrm{pc}^{-2}}
\newcommand{\kms}{\mathrm{km}\cdot\mathrm{s}^{-1}}
\newcommand{\ssfr}{\Msun\cdot\mathrm{yr}^{-1}\mathrm{kpc}^{-2}}

\makeatletter
\renewcommand*\aa@pageof{, page \thepage{} of \pageref*{LastPage}}
\makeatother

\defcitealias{brucyLargescaleTurbulentDriving2020}{Paper I}

\graphicspath{{./figures/}{}}

\titlerunning{Influence of magnetic field and turbulent compressive fraction}

\begin{document}

\title{Large-scale turbulent driving regulates star formation in high-redshift gas-rich galaxies II: Influence of the magnetic field and the turbulent compressive fraction}

\date{Received 8 September 2022 / Accepted 28 May 2023}

\author{Noé Brucy\inst{1,2}, Patrick Hennebelle\inst{1}, Tine Colman\inst{1}, Simon Iteanu\inst{1,3}}
\institute{Université Paris-Saclay, Université Paris Cité, CEA, CNRS, AIM, 91191, Gif-sur-Yvette, France 
\and 
Universität Heidelberg, Zentrum für Astronomie, Institut für Theoretische Astrophysik, Albert-Ueberle-Str 2, D-69120 Heidelberg, Germany
\and
ENS Lyon, 15 parvis René Descartes, BP 7000 69342 Lyon Cedex 07, France}

   \abstract
{The observed star formation rate (SFR) in galaxies is well below what it should be if gravitational collapse alone were at play. 
There is still no consensus about the main process that cause the regulation of the SFR.}
{It has recently been shown that one candidate that might regulate star formation, the feedback from massive stars, is suitable only if the mean column density at the kiloparsec scale is  lower than $\approx 20 \Msun\cdot\mathrm{pc}^{-2}$. On the other hand, intense large-scale turbulent driving might slow down star formation in high-density environments to values that are compatible with observations.
In this work, we explore the effect of the nature and strength of the turbulent driving, as well as the effect of the magnetic field. }
{We performed a large series of feedback-regulated numerical simulations of the interstellar medium in which bidimensional large-scale turbulent driving was also applied. We determined the driving intensity needed to reproduce the Schmidt-Kennicutt relation for several gas column densities, magnetization, and 
driving compressibility.}
{We confirm that in the absence of turbulent forcing and even with a substantial magnetic field, the SFR is too high, particularly
at a high column density, compared to the Schmidt-Kennicutt relation.
We find that the SFR outcome strongly depends on the initial magnetic field and on the compressibility of the turbulent driving. As a consequence, a higher magnetic field in high column density environment may lower the energy necessary to sustain
 a turbulence that is sufficiently intense to regulate star formation. 
}
{Stellar feedback does not seem to be sufficient to regulate star formation in gas-rich galaxies where large-scale turbulent driving 
may be needed. The sources of this large-scale turbulence as well as its characteristics, such as its intensity, compressibility, and anisotropy, need to be understood and quantified. }


\maketitle

\section{Introduction}\label{sec:intro}

Stars are complex objects that connect small- and large-scale physics.
Key characteristics of galaxies, such as their structure and shape, 
the amount of gas, their chemistry, and the velocity field, depend on the 
efficiency and on the rate of star formation.
However, understanding and predicting the star formation rate (SFR) remains both an observational and theoretical challenge.
On the theoretical side, the difficulty is the very broad span of scales involved  that cannot be captured in just one simulation \citep{duboisOnsetGalacticWinds2008,hopkinsSelfregulatedStarFormation2011}, and the high degeneracy between the multiple processes at play.

A considerable amount of work has been done to understand 
the interstellar medium (ISM) at the kiloparsec scale, 
which is an obligatory step on the path towards 
determining how the various physical processes 
are combined and influence the structure of the ISM.
Since modelling a full galaxy with parsec or sub-parsec resolution is very challenging, 
many studies have considered
computational boxes with a size of about 1 kpc. The advantage is that this allows describing the disk stratification and 
the large molecular cloud complexes while still resolving the clouds sufficiently well.

The complexity has been progressively increased in these simulations. 
The first series of calculations
\citep[e.g.][]{rosenGlobalModelsInterstellar1995, deavillezGlobalDynamicalEvolution2005,joungTurbulentStructureStratified2006,hillVerticalStructureSupernovadriven2012} 
solved the equations of Magnetohydrodynamics with an interstellar cooling curve in the presence
of an external gravitational potential representing the influence of stars and dark matter. 
In these calculations,
supernova (SN) explosions were placed randomly at a rate aiming to reproduce 
the expected galactic SN rate.
These models produced appealing turbulent multi-phase ISM structures 
with a velocity dispersion close to 
the observations. 
In a second series of calculations
\citep[e.g.][]{kimRegulationStarFormation2011,kimThreedimensionalHydrodynamicSimulations2013,hennebelleSimulationsMagnetizedMultiphase2014,walchSILCCSImulatingLifeCycle2015,iffrigStructureDistributionTurbulence2017,gattoSILCCProjectIII2017}, 
self-gravity was introduced, together with sink particles 
that were to represent stellar clusters. This important step 
allowed addressing the question of the SFR and also the issue of 
the galactic thickness, which requires the correct balance 
between gravity and turbulence. 
In these calculations, the stellar feedback essentially due to SNe
was spatially correlated 
to the sink particles, but no time delay of a few to some tens of million years
was introduced. 
That is to say, the feedback was immediately delivered 
when enough mass was accreted to form a massive star.
Whereas these models could successively reproduce the star 
formation rate typical 
of Milky Way-type galaxies, it was noted that the scheme used for 
the SN explosions had a strong influence on the results. 
In particular, 
if the SN exploded far away from its parent sink particle 
or simply was forced to explode in the diffuse gas, it would 
become very inefficient.

In the last generation
 of models \citep[e.g.][]{padoanSupernovaDrivingOrigin2016,kimThreephaseInterstellarMedium2017,collingImpactGalacticShear2018,kannanEfficacyEarlyStellar2020,ostrikerPressureRegulatedFeedbackModulatedStar2022,rathjenSILCCVIIGas2023}, the 
 massive stars were tracked more self-consistently and the SNe explode after a few millions to tens of million years, depending 
 on the mass of their progenitor. In combination with other processes such as stellar wind, HII regions, or galactic shear, it has been concluded 
 by various teams that the SFR was compatible with Milky Way values. 
 
A fundamental assumption made in these feedback-regulated models
is that the ISM regulation is relatively local.
More precisely, the energy injected in the medium 
at a given point is entirely due to stellar 
feedback, and it typically comes from less than 1 kpc.
However, several recent studies have shown that injection of turbulence from large galactic scales has to be taken into account in order to explain the observed velocity dispersion and SFR, particularly in gas-rich galaxies
\citep{bournaudISMPropertiesHydrodynamic2010, 
renaudStarFormationLaws2012,
goldbaumMassTransportTurbulence2015,
goldbaumMassTransportTurbulence2016,
krumholzUnifiedModelGalactic2018,
meidtModelOnsetSelfgravitation2020,
nusserRegulationStarFormation2022}.
There is also evidence that large-scale driving is needed to reproduce the statistics of the dense gas, in particular, 
the large-scale structures in the Large Magellanic Cloud 
\citep[e.g. ][]{colmanSignatureLargeScaleTurbulence2022}.
Several observational studies have indeed derived turbulence-injection scales of more than 1 kpc \citep{dibStructureCharacteristicScales2021, chepurnovTurbulenceVelocityPower2015} and indicated large-scale driving \citep{szotkowskiMappingSpatialVariations2019, besserglikPowerSpectrumStructure2021}.
Possible sources of turbulence include the orbital energy 
\citep{wadaGravitydrivenTurbulenceGalactic2002,agertzLargescaleGalacticTurbulence2009} and the mass accretion 
onto the galaxies \citep{klessenAccretiondrivenTurbulenceUniversal2010,forbesGasAccretionCan2023}.
The former in particular requires a mechanism such as an instability 
to convert this source of free energy into turbulent energy,
which might  be the gravitational instability, for instance 
\citep{wadaGravitydrivenTurbulenceGalactic2002,agertzLargescaleGalacticTurbulence2009,goldbaumMassTransportTurbulence2015,fenschUniversalGravitydrivenIsothermal2023}, or the magneto-rotational 
instability \citep{piontekModelsVerticallyStratified2007}. 

An important challenge of numerical simulations for the SFR in particular is reproducing 
the Schmidt-Kennicutt relation 
\citep[hereafter SK relation;][]{kennicuttjr.GlobalSchmidtLaw1998,kennicuttStarFormationMilky2012,kennicuttRevisitingIntegratedStar2021} that links the SFR to the column density of gas. 
In a recent paper, \cite{brucyLargescaleTurbulentDriving2020} \citepalias[hereafter][]{brucyLargescaleTurbulentDriving2020} tested the effect of the injection of turbulence by adding a large-scale turbulent driving similar to the one used by \cite{schmidtNumericalSimulationsCompressively2009}.
Importantly, \cite{brucyLargescaleTurbulentDriving2020} have shown that whereas
the SK relation can be reproduced if the added turbulence 
is strong enough, the SFR
appears to be too high, particularly for gas-rich galaxies, 
when stellar feedback alone is accounted for.

The goal of the present paper is to complement the study presented 
in \citetalias{brucyLargescaleTurbulentDriving2020} by studying the 
effects of the magnetic field strength and the compressive fraction 
of the turbulent driving on the SFR. 
Both aspects are known to significantly influence the gas distribution 
\citep[e.g.][]{molinaDensityVarianceMachNumber2012} and likely influence the SFR values. 
We ran simulations of a local region of a galactic disk within a cubic box of 1 kpc. 
We used a numerical setup very similar to the one used by \cite{collingImpactGalacticShear2018} 
and \citetalias{brucyLargescaleTurbulentDriving2020}. 
We simulated regions of galaxies with a wide range of gas column densities that are
representative for Milky Way-like galaxies up to gas-rich galaxies at redshift $z =$~1--3 
\citep{genzelRingsBulgesEvidence2008,genzelStudyGasstarFormation2010,daddiVeryHighGas2010}.

The paper is organised as follows: We start by discussing the orders of magnitude of the turbulent energy that can be dissipated by various processes in the ISM in section \ref{sec:energy_diss}. Next, we present our numerical setup in section 
\ref{sec:setup}, and we explore the effects of the turbulence and the magnetic field in sections 
\ref{sec:turbexplo}  and \ref{sec:mag}, respectively.
We then discuss the main goal of this study, 
that is, the reproduction of the SK relation in section \ref{sec:sk}. 
The caveats of this study are discussed in section \ref{sec:caveats}.
Finally, we discuss the results and conclude in section \ref{sec:conclusion}.

\section{Energy dissipation and turbulent driving in the ISM, orders of magnitude}
\label{sec:energy_diss}
As discussed above, we further investigate  
whether large-scale turbulent driving can help explain the SK relation. 
It is thus of primary importance to discuss the possible sources of turbulent driving 
in galaxies. Below, we discuss three mechanisms, namely gravitational instability \citep{wadaGravitydrivenTurbulenceGalactic2002,agertzLargescaleGalacticTurbulence2009}, accretion-driven 
turbulence \citep{klessenAccretiondrivenTurbulenceUniversal2010, forbesGasAccretionCan2023}, and 
SN driving \citep[e.g.][]{kimThreephaseInterstellarMedium2017,gattoSILCCProjectIII2017}.
We estimate the amount of energy  $P$ that can be extracted per unit
of time through each of the mechanisms, as well as their efficiency $\epsilon$, which is the fraction of that energy that is eventually converted into turbulent kinetic energy.

We start by estimating the amount of turbulent energy that is dissipated per unit of time
in a region of size $L \simeq$ 1 kpc within a galactic disk, with a column density $\Sigma=10-100$ M$_\odot$ pc$^{-2}$ and a velocity dispersion $\sigma \simeq 10-100$ km s$^{-1}$. 
Since the mass contained in a surface $L^2$ is $\Sigma L^2$, whereas the specific kinetic energy 
is $\sigma^2 /2$ and the crossing time is $ L/\sigma$, we obtain
\begin{align}
P_{\mathrm{diss}} &\simeq    \Sigma L^2 \dfrac{\sigma^3 }{2 L}, \\
  &\simeq  3 \times 10^{40} \mathrm{ erg \, s^{-1}}
  \left(  \dfrac{\Sigma}{ 100 \, \mathrm{ M_\odot pc^{-2}}}  \right)  
  \left( \dfrac{ \sigma }{100 \, \mathrm{ km \, s^{-1}}  }\right)^3
   \left( \dfrac{ L }{\mathrm{ 1 \, kpc} } \right). 
\nonumber
\end{align}

The largest amount of energy that can be extracted per unit of time from galactic rotation is typically equal 
to the rotation energy $V_{\mathrm{rot}}^2 / 2 $, divided by a rotation time that is simply given 
$2 \pi R_\mathrm{ gal} / V_{\mathrm{rot}}$ , where $R_\mathrm{ gal} $ is the galactic radius. This leads to 
\begin{align}
P_{\mathrm{rot}} \simeq&\;   
\Sigma L^2 \dfrac{V_{\mathrm{rot}}^3 }{4 \pi R_\mathrm{ gal}},   \\
  \simeq&\;  2 \times 10^{40} \mathrm{ erg \, s^{-1}} \nonumber \\
  &\left( \dfrac{ \Sigma + \Sigma_\star}{100 \, \mathrm{ M_\odot pc^{-2}}  }\right)   \left( \dfrac{ V_{\mathrm{rot}} }{200 \, \mathrm{ km \, s^{-1}}  }\right)^3
    \left( \dfrac{ L }{\mathrm{ 1 \, kpc} } \right)^2 
    \left( \dfrac{ R_{\mathrm{gal}} }{\mathrm{ 2 \, kpc} } \right)^{-1} 
\nonumber
\end{align}
where $\Sigma_\star$ is the stellar column density. The efficiency of turbulent driving $\epsilon_{\mathrm{rot}}$ through gravitational instability clearly depends on the disk stability, which is quantified by the Toomre parameter \citep{toomreGravitationalStabilityDisk1964},
\begin{equation}
    Q = \dfrac{\sigma \kappa}{\pi G \Sigma},
\end{equation} where $\kappa$ is the epicyclic frequency.
Since a galactic disk is composed of gas and stars, it is likely that the efficiency $\epsilon_{\mathrm{rot}}$ depends on the gas column density $\Sigma$ and the gas fraction $\Sigma / (\Sigma + \Sigma_\star)$
\citep{fenschRoleGasFraction2021}. 
Galaxies with a high gas fraction are more prone to gravitational instabilities, and we can therefore expect the efficiency $\epsilon_{\mathrm{rot}}$  in these galaxies to be high, probably close to 1. 
However, Milky Way-type galaxies can make several rotations without shrinking significantly due to angular momentum loss. 
For these galaxies, we expect the efficiency $\epsilon_{\mathrm{rot}}$ to be well below~1.

Accretion at a given mass rate $\dot{M}_\mathrm{accr}$ and a infall velocity $v_{\mathrm{inf}}$ not only brings mass into the galaxy, but also delivers 
energy at a rate $\dot{M}_\mathrm{accr} v_{\mathrm{inf}}^2 / 2$. 
To estimate the mass accretion rate $\dot{M}_\mathrm{accr}$, we assume that the total amount of gas is roughly constant, that is, the accretion of gas balances the SFR,
\begin{equation}
    \dot{M}_\mathrm{accr} \simeq L^2 \Sigma_\mathrm{SFR}.
\end{equation}
We estimate the surface density of the SFR by assuming that the SK relation \citep{kennicuttStarFormationMilky2012} is verified,
\begin{equation}
\label{eq:SK_odg}
    \Sigma_\mathrm{SFR} \simeq 0.1 \mathrm{ M_\odot \, yr^{-1} \, 
 kpc^{-2}} \left(\dfrac{\Sigma}{100 \mathrm{ M_\odot \, pc^{-2}}}\right)^{1.4}.
\end{equation}
On the other-hand, we compute the infall velocity from the gravitational potential,
\begin{equation}
\dfrac{v_{\mathrm{inf}}^2}{2} \simeq G \left(\Sigma + \Sigma_\star + \Sigma_\mathrm{ dm}\right) R_\mathrm{ gal}, \end{equation}
where $\Sigma_\mathrm{dm}$ is the column density of the dark matter. We thus find that 
 \begin{align}
P_{\mathrm{acc}} \simeq&\;  \Sigma_\mathrm{SFR} L^2 G (\Sigma + \Sigma_\star + \Sigma_\mathrm{ dm} ) R_{\mathrm{gal}} \\
\nonumber
\simeq&\;  5 \times 10^{37} \mathrm{ erg \, s^{-1}} \\
 &\left( \dfrac{ \Sigma + \Sigma_\star + \Sigma_\mathrm{ dm}  }{100 \, \mathrm{ M_\odot pc^{-2}}  }\right)  
  \left( \dfrac{ \Sigma }{100 \, \mathrm{ M_\odot pc^{-2}}  }\right)
  ^{1.4}
   \left( \dfrac{ L }{\mathrm{ 1 \, kpc} } \right)^2 
   \left( \dfrac{ R_{\mathrm{gal}} }{\mathrm{ 2 \, kpc} } \right) . 
\nonumber
\end{align}
It is complicated to determine the efficiency of turbulent driving by accretion, $\epsilon_{\mathrm{acc}}$. Based on colliding-flow simulations,
\citet{klessenAccretiondrivenTurbulenceUniversal2010} estimated it to be proportional to the density contrast between 
the accreted material and the medium in which the turbulence is driven. \citet{forbesGasAccretionCan2023} have estimated
it in the context of a galactic disk and found that it varies with galactic radii, with a typical efficiency 
of about $\epsilon_{\mathrm{acc}} \simeq$15-20$\%$. 
The stationary hypothesis is rather conservative, and it is possible that the accretion rate is much higher, in which case the resulting turbulent driving will be stronger as well. 

The amount of energy injected per unit of time by SN remnants is given by the number of SNe produced per unit of time multiplied by the amount of  energy a single SN delivers to the ISM.
This is roughly equal to $(\Sigma_\mathrm{SFR} L^2) / 120$ M$_\odot$, as given by the Salpeter initial mass function, multiplied by $10^{51}$ erg. 

By again assuming that the SFR is in line with the SK relation (Eq. \eqref{eq:SK_odg}), we can estimate
\begin{align}
P_{\mathrm{sn}} &\simeq   
\dfrac{\Sigma_\mathrm{SFR} L^2 }{120 \mathrm{ M_\odot } } 
 10^{51} \, \mathrm{ erg}  \\
  &\simeq   2 \times 10^{40} \mathrm{ erg \, s^{-1}}
   \left( \dfrac{ \Sigma }{100 \, \mathrm{ M_\odot pc^{-2}}  }\right)
   ^{1.4}
   \left( \dfrac{ L }{\mathrm{ 1 \, kpc} } \right)^2 . 
\nonumber
\end{align}
The efficiency of turbulent driving by SN remnants, $\epsilon_{\mathrm{sn}}$, has been estimated from kiloparsec turbulent box
calculations \citep[e.g.][]{iffrigStructureDistributionTurbulence2017} to be about a few percent. 

To summarise, the three envisaged mechanisms could all provide a substantial amount of energy to drive turbulence and 
compensate for turbulent dissipation. They present different scaling with $\Sigma$, however, more precisely,
$P_{\mathrm{rot}} \propto \Sigma$, $P_{\mathrm{acc}} \propto \Sigma^{2.4}$ and $P_{\mathrm{sn}} \propto \Sigma^{1.4}$.
We stress that given the orders of magnitude and the dependence on $\Sigma$, it is likely the case that 
for high values of $\Sigma$, energy rotation  of both gas and stars is likely the dominant mechanism that triggers 
turbulence. This is also the source of energy envisioned by  \citet{krumholzUnifiedModelGalactic2018} and 
\citet{nusserRegulationStarFormation2022}.

\section{Numerical setup}\label{sec:setup}
We used the RAMSES code \citep{teyssierCosmologicalHydrodynamicsAdaptive2002} 
to solve the
equations of Magnetohydrodynamics with a Godunov solver 
\citep{fromangHighOrderGodunov2006} on a uniform cubic grid of $256^3$ cells with 
periodic boundaries in the $x$ and $y$ directions parallel to the disk and 
open boundaries in the vertical direction. The box represents a cubic region 
of the galactic disk of size $L = 1$\,kpc, so that the resolution is about 4\,pc. 
The impact of the resolution on the results is discussed in Appendix \ref{sec:convergence}.

\subsection{Initial conditions}\label{subsec:ic}
We used initial conditions similar to those described in previous 
works \citep{collingImpactGalacticShear2018, brucyLargescaleTurbulentDriving2020}. 
The interstellar gas was initially distributed as a Gaussian along the $z$-axis,
  \begin{equation}
  \label{eq:n0}
n(z) = n_0 \exp \left( - \frac{1}{2} \left( \frac{z}{z_0}  \right)^2 \right),
  \end{equation}
with $n_0$ a free density parameter and $z_0 = 150\ \mathrm{pc}$ the typical scale height.  
The column density of gas (hydrogen and helium), integrated along the z-axis (perpendicular to the disk) is then
\begin{equation}
  \label{eq:Sigma0}
  \Sigma_{\text{gas}, 0} = \sqrt{2\pi} m_p n_0 z_0 
  ,\end{equation}
where $m_p = 1.4 \times 1.66 \cdot 10^{-24}$ g is the mean mass of a gas particle. 
We set the initial temperature at $8000 \ \mathrm{K}$ to match the typical value of the temperature of the warm neutral medium phase of the ISM.
An initial synthetic turbulent velocity field with a 
3D dispersion of $5 \sqrt{n_0/1.5}~\mathrm{km\cdot s}^{-1}$ was also added.
It presents a Kolmogorov power spectrum and was generated using a random phase \citep{kolmogorovLocalStructureTurbulence1941}.
The dependence on $n_0$ ensures that the initial ratio of the kinetic and gravitational energy due to the gas stays the same.
Finally, we added a Gaussian magnetic field, oriented along the $x$-axis,
  \begin{equation} 
B_x(z) = B_0 \exp \left( - \frac{1}{2} \left( \frac{z}{z_0} \right)^2 \right),
\label{eq:B}
  \end{equation}
where $B_0$ is a parameter of our simulation with a value of a few microgauss (see Table \ref{tbl:simu}).

\subsection{Numerical models}
\label{subsec:ismfeed2:num_models}

The gas is subject to an external gravitational potential, corresponding to old stars and dark matter, of the form
\begin{equation}
\label{eq:gext}
g_{\mathrm{ext}}(z) = - \frac{a_1 z}{\sqrt{z^2+z_0^2}} - a_2 z
,\end{equation}
with $a_1 = 1.42 \times 10^{-3}$ kpc/Myr$^{2}$,
$a_2 = 5.49 \times 10^{-4} \, \mathrm{Myr}^{-2}$ , and
$z_0 = 0.18$ kpc
\citep{kuijkenMassDistributionGalactic1989, joungTurbulentStructureStratified2006}.
This gravitational force adds up to the self-gravity of the gas.

Sink particles \citep{bleulerMoreRealisticSink2014} were used to follow the dense gas and model star formation. 
Sink creation was triggered when the gas number density passed a threshold of $10^3 \ \mathrm{cm^{-3}}$. The simulation with an initial column density higher than $50 \coldens$ used a higher threshold of $5\E{3} \ \mathrm{cm^{-3}}$. Since the SFR in these simulations is high, this allowed us to reduce the number of sinks without changing the SFR much \citep{collingImpactGalacticShear2018}.  Gas with a density higher than the threshold within a radius of four cells (16 pc) was then progressively accreted into the sink, with no more than 10~\% of the gas being accreted in one time step.
 
The gas cooling was as described in \cite{auditThermalCondensationTurbulent2005}, which takes
into account all the standard processes taking place in the atomic gas. It is essentially identical to the cooling described in 
\citet{koyamaMolecularCloudFormation2000}, for instance.

\subsection{Stellar feedback}

The simulations included models for the formation and expansion of HII regions, SN explosions, and the far-ultraviolet (FUV) feedback. They are the same as in \citetalias{brucyLargescaleTurbulentDriving2020}. The HII and SN feedbacks were attached to the sinks and were described in length in \cite{collingImpactGalacticShear2018} and references therein.
Each time a sink had accreted a mass of 120$\Msun$, a massive star particle with
a mass randomly determined from the Salpeter initial mass function \citep{salpeterLuminosityFunctionStellar1955} between 8 and 120 $\Msun$ was created and attached to the sink.
The lifetime $\tau_*$ of this star was computed using the model
\begin{equation}
    \tau_*\left(M\right) = \tau_0 \exp \left( - a  \left(\log \left(\frac{M}{M_0} \right)\right)^{b} \right)
,\end{equation}
with $\tau_0=3.265$ Myr, $M_0=148.16 \Msun$, $a=0.238$, and $b=2.205$ \citep{woosleyEvolutionExplosionMassive2002}.
When this massive star reached the end of its lifetime $\tau_*$, it exploded in a random location within a sphere with a radius $\tau_* \times$ 1 km $\cdot$ s$^{-1}$. 
The gas that was located inside a sphere of 12 pc radius around the location of the SN was heated up to inject a thermal energy of $10^{51} \mathrm{erg}$. Since the cooling radius was almost never resolved, the gas would cool down immediately and the SN would have no effect. To avoid this, the explosion also injected $4\times 10^{43}$ g $\cdot$ cm $\cdot$ s $^{-1}$ momentum into the same region.
When SNe explode in a low-density environment, a very high temperature and very high velocities can be generated, which translate into very short time steps due to the Courant condition.
A limitation of the temperature to $10^6~\mathrm{K}$ and of the velocity to $300~\kms$ generated by a SN explosion was therefore implemented.
We discuss the impact of this limitation on our results in Appendix \ref{sec:vsat}.

We also included self-consistent feedback from HII regions. Energy and momentum were injected according to the flux of ionising photons emitted by the massive star \citep{vaccaLymanContinuumFluxesStellar1996}. The evolution of the HII regions themselves was computed via the so-called M1 radiative transfer method \citep{rosdahlRAMSESRTRadiationHydrodynamics2013}.

The FUV heating was uniform. However, it was not kept constant at the solar neighbourhood value because young O-B stars contribute significantly to the FUV emission. As a first approximation, we considered the UV heating to be proportional to the SFR \citep{ostrikerRegulationStarFormation2010}. The mean FUV density relative to the solar neighbourhood value $G_0^{\prime}$ can then be written as
\begin{equation}
\label{eq:guv}
    G_0^\prime = \frac{\Sigma_{\mathrm{SFR}}}{\Sigma_{\mathrm{SFR,}\astrosun}}
               = \frac{\Sigma_{\mathrm{SFR}}}{2.5\times10^{-9}~\mathrm{M}_{\astrosun}\cdot\mathrm{pc}^{-2}\cdot\mathrm{yr}^{-1}}
.\end{equation}
In our model, $G_0^{\prime}$ has a minimum value of $1$ (as a background contribution) and follows equation \ref{eq:guv} when the SFR increases. 

\subsection{Injection of turbulence}
\label{subsec:turb_inj}

\cite{bournaudISMPropertiesHydrodynamic2010}, \cite{krumholzTurbulenceInterstellarMedium2016}, and
\cite{krumholzUnifiedModelGalactic2018} showed that for galaxies with a high column density or high SFR, large-scale gravitational instabilities are significant 
 sources of turbulent energy and might even 
 dominate stellar feedback.
 The ISM in a kiloparsec region of the galaxy is far from an isolated environment. 
It interacts with the rest of the galaxy.
Adding turbulent driving to simulations of restricted 
regions of a galaxy enables us to take a part of 
these complex interactions into account while keeping the computational cost 
of the simulation reasonable. 
The turbulence added in the kiloparsec box is clearly not fully self-consistent.
 We numerically investigated the effect of turbulent 
driving on star formation. We used a model for turbulent driving 
that we adapted from the generalisation of Ornstein-Uhlenbeck that was developed 
and used by several authors  
\citep{eswaranExaminationForcingDirect1988,schmidtNumericalDissipationBottleneck2006,schmidtNumericalSimulationsCompressively2009,federrathComparingStatisticsInterstellar2010}.

The turbulent forcing was implemented by adding an external force 
density $\bm{f}$ that accelerated the fluid on large scales. 
Our main hypothesis is that this force is generated by large-scale
 processes at the galactic scale within the galactic disk. 
As a consequence, the driving is bidimensional (the force has no vertical 
component and no vertical mode), and it was applied only at low altitude. 
The driving procedure is slightly different from the procedure used in 
\citetalias{brucyLargescaleTurbulentDriving2020}
and is therefore described below at length.

The fluid equations that we solved were
\begin{align}
\label{eq:cont}
\pd{\rho}{t} + \bm{\nabla}\cdot \left(\rho \bm{v} \right) &= 0, \\
\label{eq:euler}
\rho \left(\pd{\bm{v}}{t} + \left(\bm{v}\cdot\bm{\nabla}\right)\bm{v}\right) &= -\bm{\nabla} P + \dfrac{\left( \bm{\nabla} \times \bm{B} \right) \times \bm{B}}{4 \pi} + \rho \bm{g} + \rho \bm{f}, \\
\label{eq:energy}
\rho \left( \pd{e}{t} + \left(\bm{v} \cdot \bm{\nabla} \right) e \right) &= 
- P \left(\nabla\cdot\bm{v}\right) +  \rho \bm{f} \cdot \bm{v} - \rho \mathcal{L}, \\ 
\label{eq:induction}
\pd{\bm{B}}{t} - \bm{\nabla} \times \left(\bm{v} \times \bm{B}\right) &= 0.
\end{align}
In these equations, $\rho$ is the gas density, $\bm{v}$ is the gas speed, $P$ is the thermal pressure, $e$ is the internal energy, $\bm{B}$ is the magnetic field, $\bm{g}$ is the gravitational potential, and $\mathcal{L}$ is the energy loss function.

\begin{table*}[ht]
\caption{List of simulations. Each group of simulations corresponds to an experiment discussed in the article. The simulations of the group \textsc{noturb} were discussed in \citetalias{brucyLargescaleTurbulentDriving2020}. The parameter $n_0$ is the initial midplane density and sets the initial column density $\Sigma_{0, \mathrm{gas}}$ (see Eqs. \eqref{eq:n0} and \eqref{eq:Sigma0}). The strength of the driving is given by parameter $f_{\mathrm{rms}}$ (see Eq. \eqref{eq:injection}). A value of $f_{\mathrm{rms}} = 10^4$ in code units corresponds to an RMS acceleration due to the driving of $1.46 \ \mathrm{km}\cdot\mathrm{s}^{-1}\cdot\mathrm{Myr}^{-1}$. 
The parameter $\chi = 1 - \zeta$, introduced in Eqs. \eqref{eq:pzeta1} and \eqref{eq:projection}, is the compressibility fraction of the turbulent driving. 
Finally, $B_0$ is the initial value of the magnetic field, which was initialised as described in Eq. \eqref{eq:B}.  }\label{tbl:simu}
  \begin{center}
        \begin{tabular}{llccccc}
        \hline\noalign{\smallskip}
        Group & $n_0$ & $\Sigma_{0, \mathrm{gas}}$  & $f_{\text{rms}}$  & $\chi$ & $B_0$ \\
        &{\scriptsize $[\mathrm{cm}^{-3}]$ }& { \scriptsize $[\Msun\cdot\mathrm{pc}^{-2}]$} & {\scriptsize $[1.46 \E{-4}~\mathrm{km}\cdot\mathrm{s}^{-1}\cdot\mathrm{Myr}^{-1} ]$} & & $[\mu \mathrm{G}]$ \\
        \noalign{\smallskip}
        \hline
        \noalign{\smallskip}
        \hline
        \noalign{\smallskip}
                
        \multirow{2}{*}{\textsc{strength}}
        &  3  & 38.74 &  $10^4$ to $10^5$  & 0.25 & 3.8 \\
        &  6 & 77.4 &   $ 5\E{4}$ to $8\E{5}$  & 0.25 & 3.8  \\
        \hline
        \multirow{2}{*}{\textsc{comp}}
        &  3 & 38.7 & $8 \E{4}$ & 0 to 1 & 7.6 \\
        &  6 & 77.4 & $2 \E{5}$ & 0 to 1 & 7.6 \\
        \hline
        \multirow{2}{*}{\textsc{mag}}
        &  3 & 38.7 & $8 \E{4}$ & 0.25 & [0, 1.9, 3.8, 5.3, 7.6] \\
        &  6 & 77.4 & $2 \E{5}$ & 0.25 & [0, 1.9, 3.8, 7.6, 15, 23] \\

        \noalign{\smallskip}
        \hline 
        \noalign{\smallskip}
        \hline
        \noalign{\smallskip}
        \multirow{7}{*}{\textsc{noturb}} 
        & 1 & 12.9 & 0 & N/A & 3.8 \\
        & 1.5 & 19.4 & 0 & N/A  & 3.8 \\
        &  2 & 25.8 & 0 & N/A  & 3.8 \\
        &  3 & 38.7 & 0 & N/A  & 3.8 \\
        &  4 & 51.6 & 0 & N/A  & 3.8 \\
        &  6 & 77.4 & 0 & N/A  & 3.8 \\  
        &  12 & 155 & 0 & N/A  & 3.8 \\  
        \hline
        \multirow{4}{*}{\textsc{noturb\_hB}} 
    &  1 & 12.9 & 0 & N/A  & 7.6 \\  
        &  3 & 38.7 & 0 & N/A  & 7.6 \\  
        &  6 & 77.4 & 0 & N/A  & 7.6 \\  
        &  12 & 155 & 0 & N/A  & 7.6 \\  
        \hline
        \multirow{4}{*}{\textsc{turb}} 
        & 1.5 & 19.4 & $5 \E{3}$ & 0.25 & 3.8 \\
        &  3 & 38.7  & $6 \E{4}$& 0.25 & 3.8 \\
        &  6 & 77.4 & $4 \E{5}$ & 0.25 & 3.8 \\
         &  12 & 155 & $3 \E{6}$ & 0.25 & 3.8 \\
        \hline
        \multirow{4}{*}{\textsc{turb\_hB}}
        & 1.5 & 19.4 & $1 \E{3}$  & 0.25 & 7.6 \\
        &  3 & 38.7 & $3 \E{4}$ & 0.25 & 7.6 \\
        &  6 & 77.4 & $10^5$  & 0.25 & 7.6 \\
        &  12 & 155 & $10^6$  & 0.25 & 7.6 \\ 
        \hline
    \multirow{3}{*}{\textsc{turb\_Bvar}}
        &  3 & 38.7 & $5 \E{4}$  & 0.25 & 5.3 \\
        &  6 & 77.4 & $2 \E{5}$ & 0.25 & 7.6 \\
        &  12 & 155 & $10^6$ & 0.25 & 10.7 \\
        \hline
        \multirow{2}{*}{\textsc{turb3D}}
        &  3 & 38.74 & $6 \E{4}$ & 0.25 & 7.6  \\
        &  6 & 77.4 & $2 \E{5}$ & 0.25 & 7.6  \\
    &  12  & 155 & $8 \E{5}$ & 0.25 & 7.6  \\
    \noalign{\smallskip}
        \hline
        \noalign{\smallskip}
        \hline
        
        \end{tabular} 
  \end{center}
\end{table*}

\noindent The evolution of the Fourier modes of the turbulence driving the acceleration field $\bm{\hat{f}}(\bm{k}, t)$ follows a stochastic differential equation \citep{schmidtNumericalDissipationBottleneck2006,schmidtNumericalSimulationsCompressively2009},
\begin{equation}
\label{eq:ed_fourier}
    \mathrm{d}\bm{\hat{f}}(\bm{k}, t) = - \bm{\hat{f}}(\bm{k}, t)\frac{\mathrm{d}t}{T} + F_0(\bm{k})\bm{P_\zeta}\left(\bm{k}\right)\mathrm{d}\bm{W}_t.
\end{equation}
In this equation, $ \mathrm{d}t$ is the time step for integration, 
and $T$ is the autocorrelation timescale. 
In our simulations, we chose  $T = 5$~Myr and $ \mathrm{d}t/T = 1/100$. 
The value of $5$ Myr was chosen to be of the same order of magnitude as 
the crossing time of the over-densities created by the turbulent driving. 
$d\bm{W_t}$ is a small vector that was randomly chosen following the Wiener process, 
as described in \cite{schmidtNumericalSimulationsCompressively2009}.
The main modification to the Ornstein-Uhlenbeck process 
is the weighting function of the driving modes $F_0$ and the projection operator 
$\bm{P_\zeta}$.
We selected only large 2D modes, and more weight was given to modes 
with a wavelength of 500 pc (half the box size),
\begin{equation}
    \label{eq:F0}
    F_0(\bm{k}) = 
    \begin{cases} 
    1 - \left(\dfrac{\bm{k}}{2\pi} - 2\right)^2\text{ if } 1 < \dfrac{\vert k \vert}{2\pi} < 3 \text{ and } k_z = 0\\
    0 \text{ if not.}
    \end{cases}
.\end{equation}
In other words, the driving modes (divided by $2 \pi$) were 
\[ 
\left(\begin{array}{c} 2 \\ 2 \\ 0\end{array}\right),
\left(\begin{array}{c} 1 \\ 1 \\ 0\end{array}\right),
\left(\begin{array}{c} 2 \\ 1 \\ 0\end{array}\right),
\left(\begin{array}{c} 1 \\ 2 \\ 0\end{array}\right),
\left(\begin{array}{c} 2 \\ 0 \\ 0\end{array}\right), 
\left(\begin{array}{c} 0 \\ 2 \\ 0\end{array}\right),
\]
and the respective weights were approximately
\[ 
 0.31,\; 0.65,\; 0.65,\;  0.94,\; 0.94,\;  1,\; 1.
\]

\begin{figure*}[htbp!]
    \centering
    \includegraphics[width=0.9\textwidth]{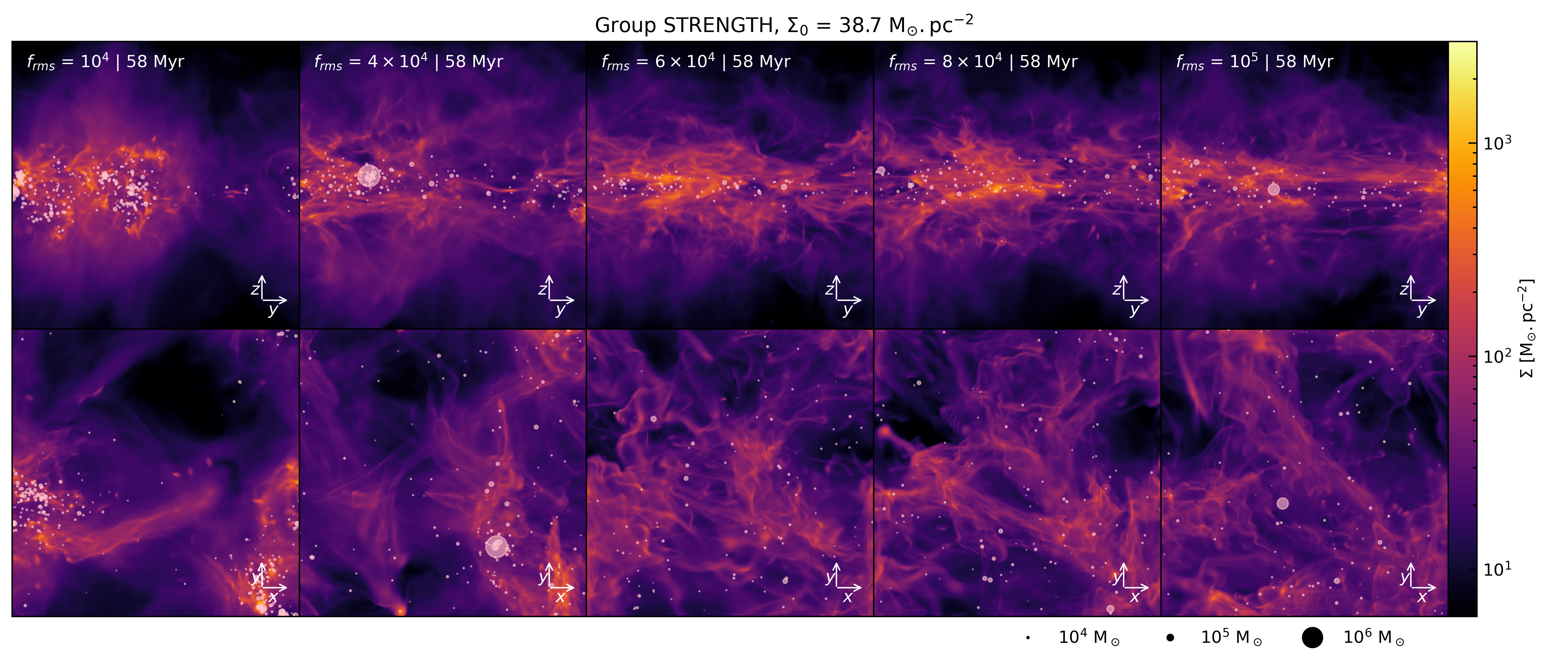}
    \includegraphics[width=0.9\textwidth]{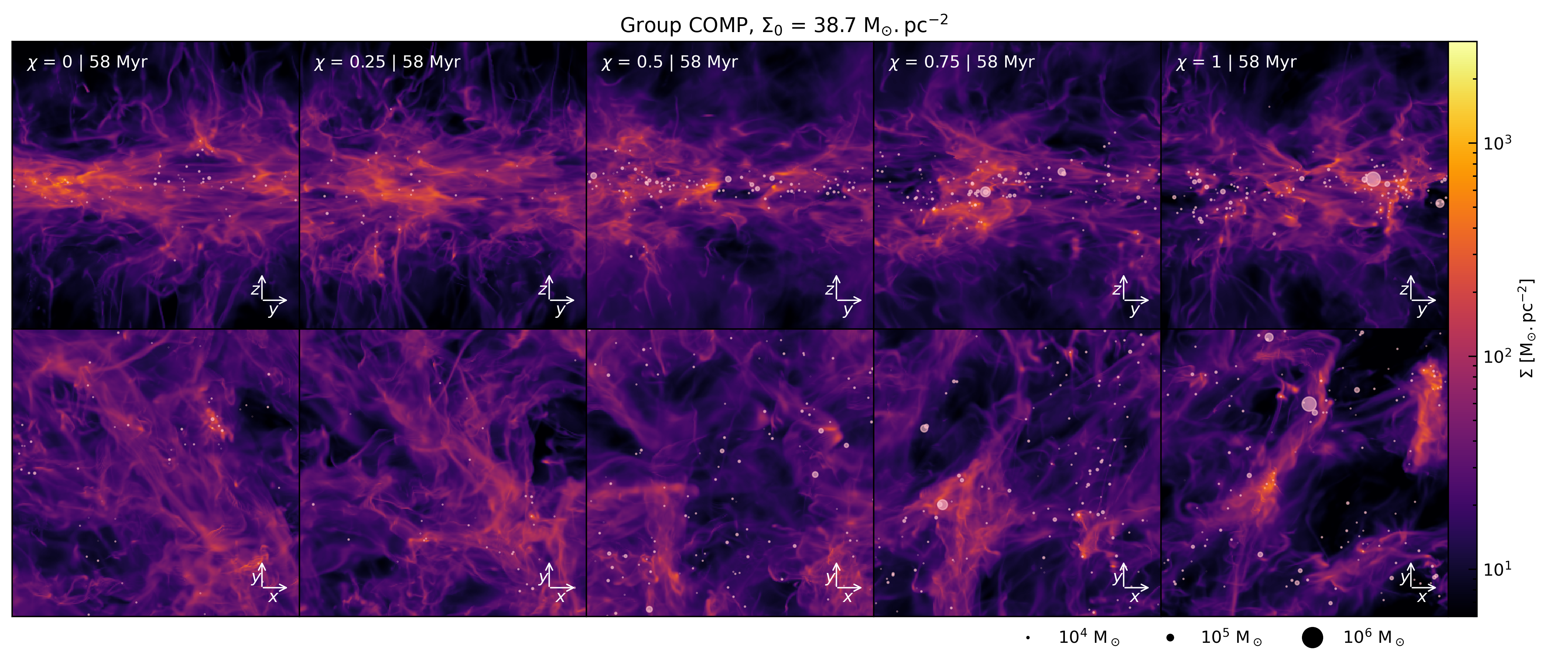}

    \caption{
     Column densities for the group \textsc{strength} (top) and \textsc{comp} (bottom) for a medium column density. For each panel, the top subpanel is viewed edge-on along the x-axis and the bottom subpanel is viewed face-on.
    White dots represent sink particles, with a radius proportional to their mass.  
    Top:  RMS acceleration of the turbulent driving is increased from left to right. 
    Bottom: Compressibilty of the turbulent driving is increased from left to right. $\chi = 0$ and $\chi = 1$ refer to purely solenoidal and compressive driving, respectively.}
    \label{fig:coldens_strcomp}
\end{figure*}

The projection operator $\bm{P_\zeta}$ is the weighted sum of 
the components of the Helmholtz decomposition of compressive versus solenoidal modes, 
projected in 2D,

\begin{equation}
\label{eq:pzeta1}
 \bm{P_\zeta}(\bm{k}) = 
    \begin{pmatrix}
    \zeta + (1 - 2 \zeta)  \left(\dfrac{k_x^2}{k^2} \right) & (1 - 2 \zeta) \left(\dfrac{k_x k_y}{k^2} \right)  & 0 \\
     (1 - 2 \zeta)  \left(\dfrac{k_x k_y}{k^2} \right) &   \zeta + (1 - 2 \zeta)  \left(\dfrac{k_y^2}{k^2} \right)  & 0 \\
    0 & 0 & 0 \\
    \end{pmatrix}
.\end{equation}
In other words,
\begin{equation}
 \label{eq:projection}
    \bm{P_\zeta}(\bm{k}) =  \zeta \bm{P}^{\perp}(\bm{k}) + (1 - \zeta) \bm{P}^{\parallel}(\bm{k}) 
,\end{equation}
with $\bm{P}^{\perp}$ and $\bm{P}^{\parallel}$ the projection operators perpendicular and parallel to $\bm{k}, $ respectively \citep{federrathComparingStatisticsInterstellar2010}.
The weight $\zeta$ is the solenoidal fraction of the driving.
In the following, we also refer to the compressive fraction, $\chi = 1 - \zeta$.

\begin{figure*}[!ht]
    \begin{center}
        \includegraphics[width=0.42\textwidth]{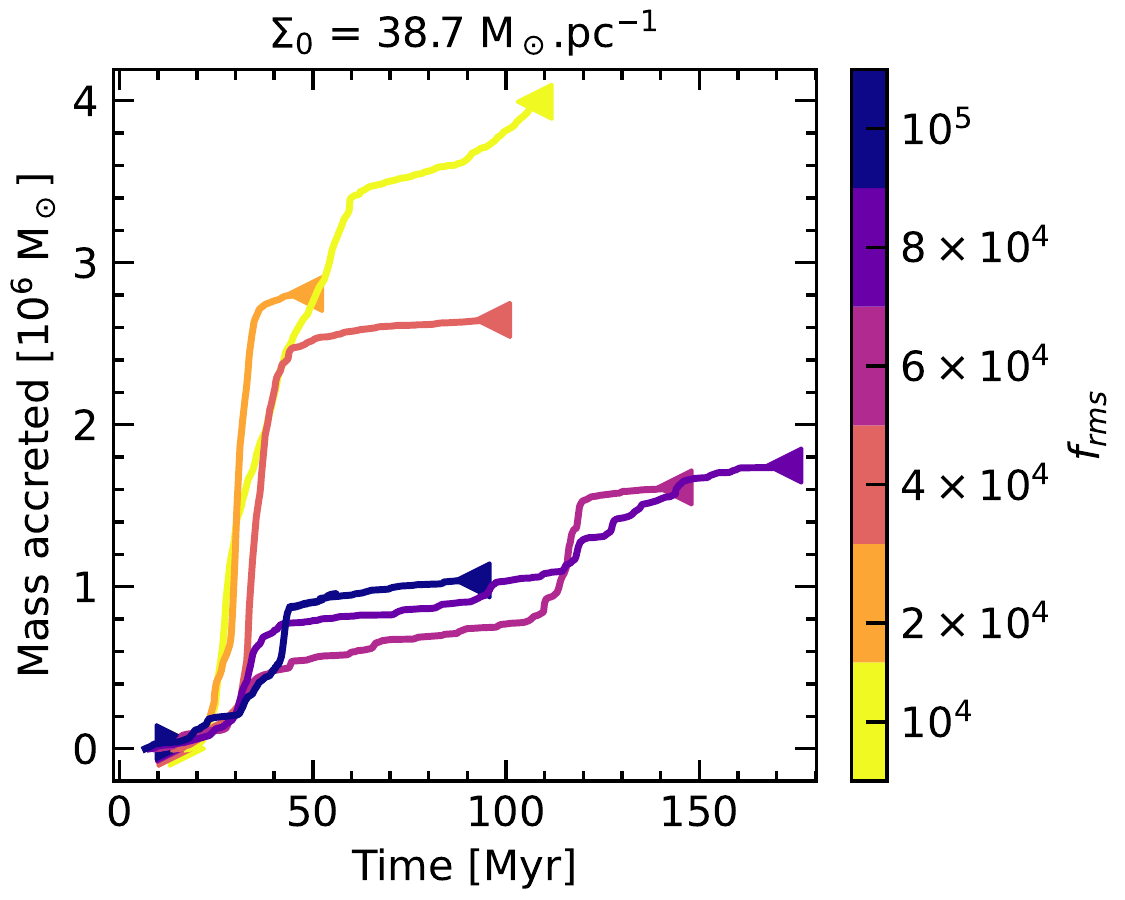}
        \vspace{0.05 \textheight}
        \includegraphics[width=0.42\textwidth]{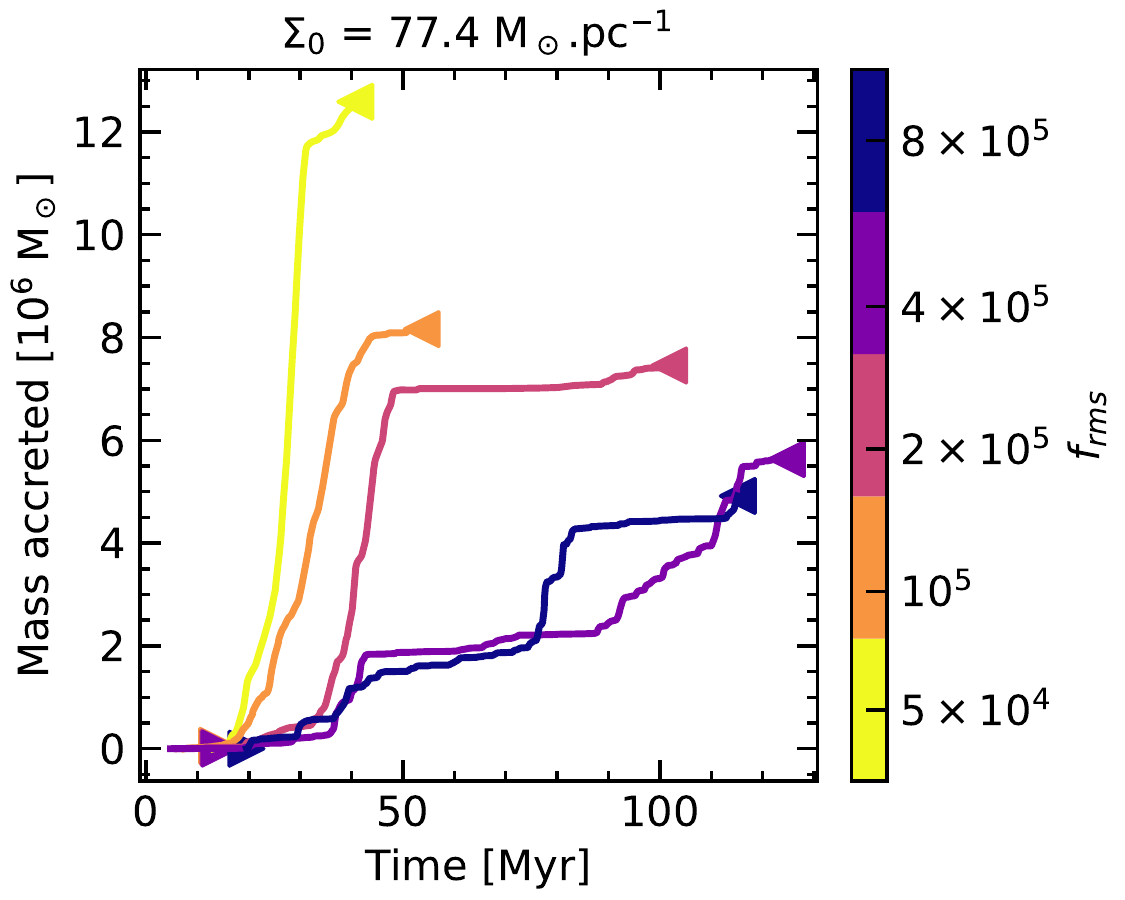}
        \includegraphics[width=0.83\textwidth]{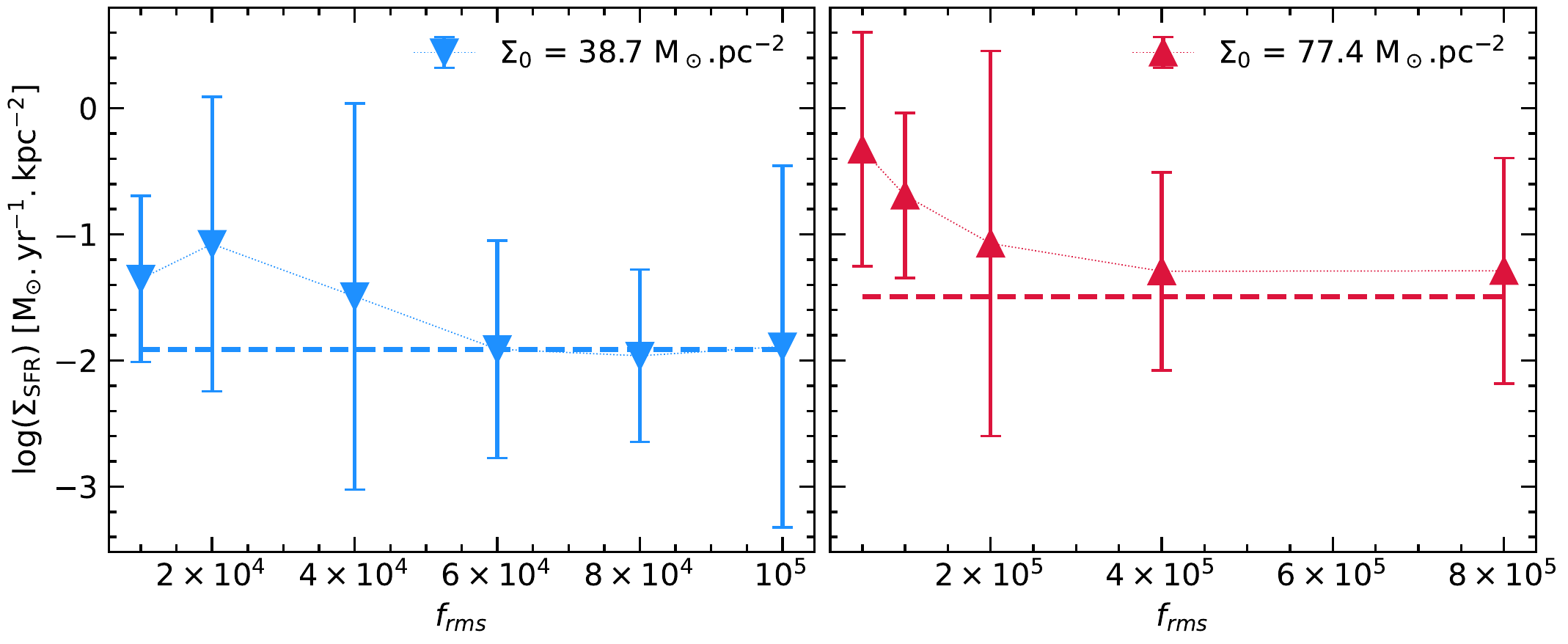}
    \end{center}

    \caption{Effect of the intensity of the turbulent driving on the SFR (group \textsc{strength}). $f_{\mathrm{rms}}$ is a measure of the driving strength. More precisely, $f_{\mathrm{rms}}$ is the RMS acceleration of the force density in code units. Top: Mass accreted in sinks. The SFR computation started when 3\% were consumed or lost ($t_\mathrm{3 \%}$, left triangle) and ended when 40\% of gas was consumed or lost ($t_\mathrm{40 \%}$, right triangle). Bottom: Surface density of the SFR as a function of the strength of the driving $f_\mathrm{rms}$.
    The plotted value is the accretion rate between $t_\mathrm{3 \%}$ and $t_\mathrm{40 \%}$ , and the error bars reflect the standard deviation of all the accretion rates we obtained by choosing starting and end points between these two values. The dashed line is the value of the SFR given by the SK relation \citep{kennicuttStarFormationMilky2012}.}\label{fig:turbvar}

\end{figure*}

\begin{figure*}[!ht]
    \begin{center}
        \includegraphics[width=0.82\textwidth]{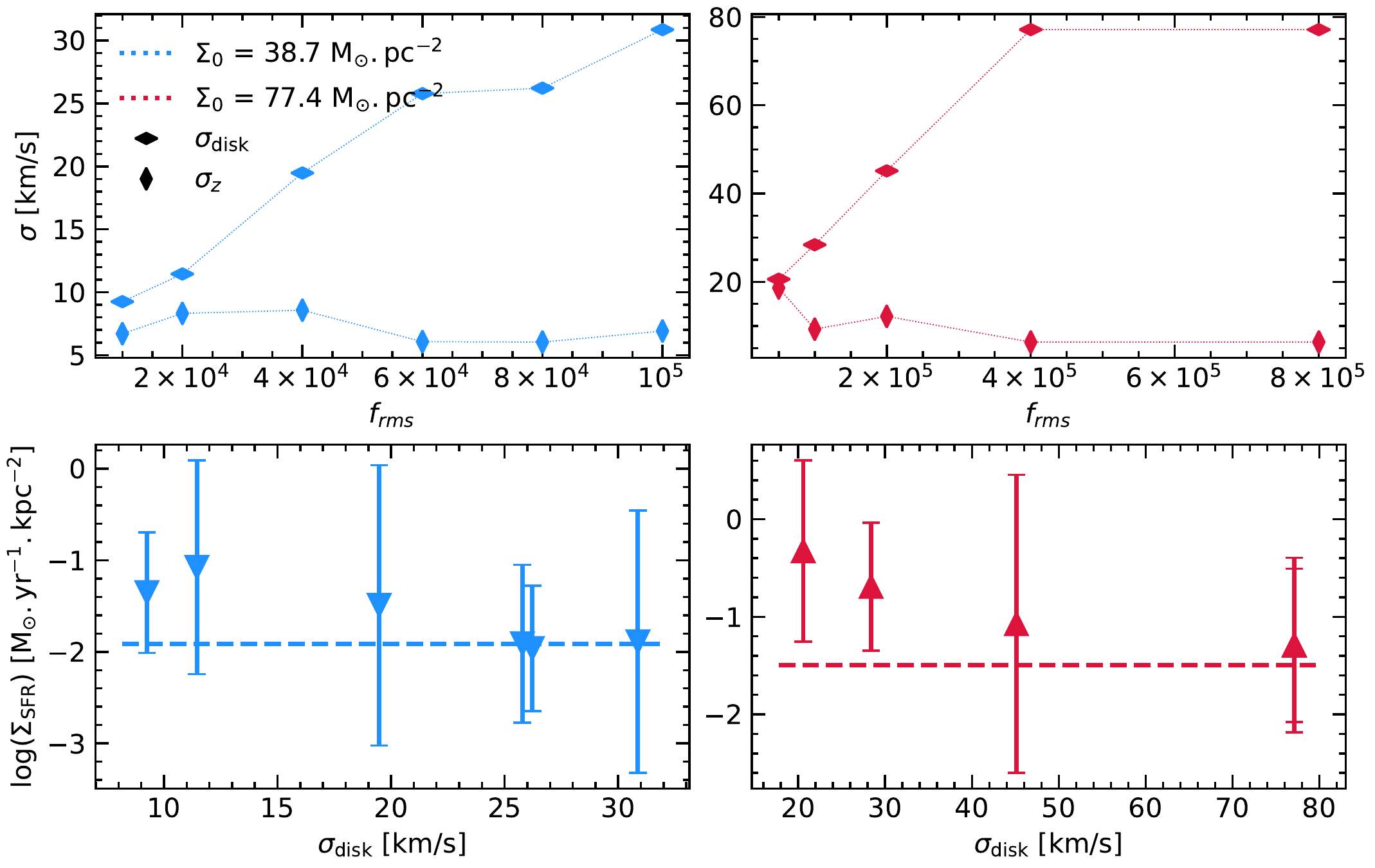}

    \end{center}

    \caption{Relation between the strength of the driving, the velocity dispersion, and the SFR. \textit{Top panel}: Mass-weighted velocity dispersion measured in the  \textsc{strength} simulations at 58 Myr at the kiloparsec scale. The velocity dispersion parallel to the disk plane $\sigma_\mathrm{disk}$ is defined by $\sigma_{\mathrm{disk}} = \sqrt{\sigma_x^2 + \sigma_y^2} / \sqrt{2}$. \textit{Bottom panel}: SFR as a function of the velocity dispersion parallel to the disk. The dashed line is the value of the SFR given by the SK relation.}\label{fig:turbveldisp}

\end{figure*}

The forcing field $\bm{f}(\bm{x}, t)$ was then computed from the Fourier transform,
\begin{equation}
\label{eq:injection}
\bm{f}(\bm{x}, t) = \mathrm{att}(z) g(\zeta) f_{\mathrm{rms}}  \int\bm{\hat{f}}(\bm{k}, t) e^{i\bm{k}\cdot x} \mathrm{d}t^3\bm{k}
.\end{equation}
The parameter $f_{\mathrm{rms}}$ controls the total power that is injected by the turbulent force into the simulation. 
The $g(\zeta)$ factor is an empirical correction so that the resulting time-averaged 
root mean square (RMS) of the power of the Fourier modes is equal to 
$f_{\mathrm{rms}}$, regardless of the solenoidal fraction $\zeta$ 
(see the discussion in Appendix \ref{sec:calibration}).
The attenuation function $\mathrm{att}$ ensures that the driving occurs only on the disk,
\begin{equation}
    \label{eq:att}
     \mathrm{att}(z) = \exp\left(\dfrac{-z^2}{2 z_t^2}\right)
,\end{equation}
where $z_t = 75$ pc.

There are two main differences with the driving method used in \citetalias{brucyLargescaleTurbulentDriving2020}. 
First, the modes aligned with the $x$ - and $y$ -axis were not used in  \citetalias{brucyLargescaleTurbulentDriving2020}, leading to global motions than can be seen through the diagonal features in the lower right panel of Fig.~1 of that paper.
Second, the attenuation function \eqref{eq:att} was not used in \citetalias{brucyLargescaleTurbulentDriving2020}.
It is introduced here so that the driving occurs only within the scale height of the disk.

\subsection{Computation of the SFR and the column density}
\label{subsec:computation_sfr}

The total mass of gas accreted by the sinks was considered to form stars and is denoted $\mathrm{M}_\star(t)$. 
As described in section \ref{subsec:ismfeed2:num_models}, only a fraction of the dense gas is accreted onto the stars so that the efficiency of the conversion of dense gas into stars is strictly lower than 1. 
The gas in the box is continually consumed by sinks or expelled from the box through the vertical boundaries. 
As a consequence, the gas available for star formation decreases with time. 
Since we used the simulations to test the star formation relation that links the SFR with the column density of gas,  
and to avoid too huge variations of the gas reservoir, we restricted the analysis to a short enough time span.
In detail, we started the analysis at $t_\mathrm{3 \%}$ when 3\% of the gas was consumed or lost, and we stopped it when this depletion fraction reached 40\% ($t_\mathrm{40 \%}$). The surface density of SFR was then computed via
\begin{equation}
    \label{eq:sfr_num}
    \Sigma_{\mathrm{SFR}}(t_{3 \%}, t_{40 \%}) = 
    \dfrac{\mathrm{M}_\star(t_{40 \%}) - \mathrm{M}_\star(t_{3 \%})}
    {L^2 \left(t_{40 \%} - t_{3 \%} \right)}
.\end{equation}
The uncertainties on this value were computed by taking the standard deviation of all values of the SFR between $ t_{3 \%}$ and $t_{40 \%}$, that is, $\Sigma_{\mathrm{SFR}}(t_i, t_j)$ for $ t_{3 \%} \leq t_i < t_j \leq t_{40 \%}$.

The mean column density in the considered time span varies between  $0.97 \Sigma_0$ and $0.6 \Sigma_0$, where $\Sigma_0$ is the initial column density.
For each simulation, the associated column density is then $\Sigma = 0.8 \Sigma_0 \pm 0.2 \Sigma_0$.
The determination of $\Sigma_{\mathrm{SFR}}$ and of $\Sigma$ is different from the determinations used in \citetalias{brucyLargescaleTurbulentDriving2020}, where the SFR was averaged over a fixed period of time and only the initial column density was considered. This means that the depletion of the gas during the time span covered by the simulation was not taken into account.

\subsection{List of simulations}

We ran several simulations that we list in Table \ref{tbl:simu}. We separated them into groups, each of which corresponds to an experiment. There are two families of experiments. 

The first family of experiments (\textsc{strength}, \textsc{comp,} and \textsc{mag}) aimed to quantify the effect of one given parameter on the SFR, and the properties of the ISM in general.
For these experiments, we set all parameters except for the tested parameter to fiducial values.
The tested parameters were the strength of the turbulent driving, its compressibility, and the initial magnetic field.

The goal of the second family (\textsc{noturb}, \textsc{noturb\_hB}, \textsc{turb}, \textsc{turb\_hB}, \textsc{turb\_Bvar}, and \textsc{turb3D}) was to test whether it is possible to derive an SFR that is compatible with the SK relation in simulation boxes of 1 kpc$^3$ even in a dense environment under various conditions.
Each of the groups contained three to seven simulations with increasing column density.
We computed the SFR for each of the simulations and compared it to what is expected from the SK relation.
The main parameters we varied are the strength of the turbulent driving and the intensity of the initial magnetic field. For the latter, 
several values were considered for each column density. 
For the simulation with turbulent driving, we chose the driving strength by trial and error to find an acceptable SFR. 
The goals here were to show that it is possible to obtain an SFR that was similar to the observations when turbulent driving was added, but also to analyse the resulting velocity dispersion and gas distribution.

\section{Effects of turbulence}
\label{sec:turbexplo}

In the following experiments, we consider dense regions of galaxies with an initial column density of $38$ and $77 \Msun  \cdot \mathrm{pc}^{-2}$. They are referred to as medium and high column density in what follows. 
We kept all parameters constant except for one parameter. 

We started by investigating the properties of turbulence, 
in particular, the strength of the driving and the 
fraction of compressible modes. Whereas we expect that 
higher compressibility leads to a higher SFR, the impact of 
higher Mach numbers is less straightforward. 
One of the difficulties that arises when trying to understand the effect of the turbulent driving on the SFR is that turbulence can both trigger and quench star formation \citep{maclowControlStarFormation2004}. 
In classical models for star formation \citep{padoanStarFormationRate2011,hennebelleAnalyticalStarFormation2011,federrathStarFormationRate2012}, the SFR is principally set by the width of the density probabilty distribution function (PDF), the density threshold above which the gas collapses under its own gravity, and the time needed for the collapse. 
When the Mach number is increased, the width of the PDF as well as the density threshold also increase.

\subsection{Influence of the strength of the turbulent driving}
\label{subsec:turb_strength}

\begin{figure}[!ht]
    \begin{center}
        \includegraphics[width=0.39 \textwidth]{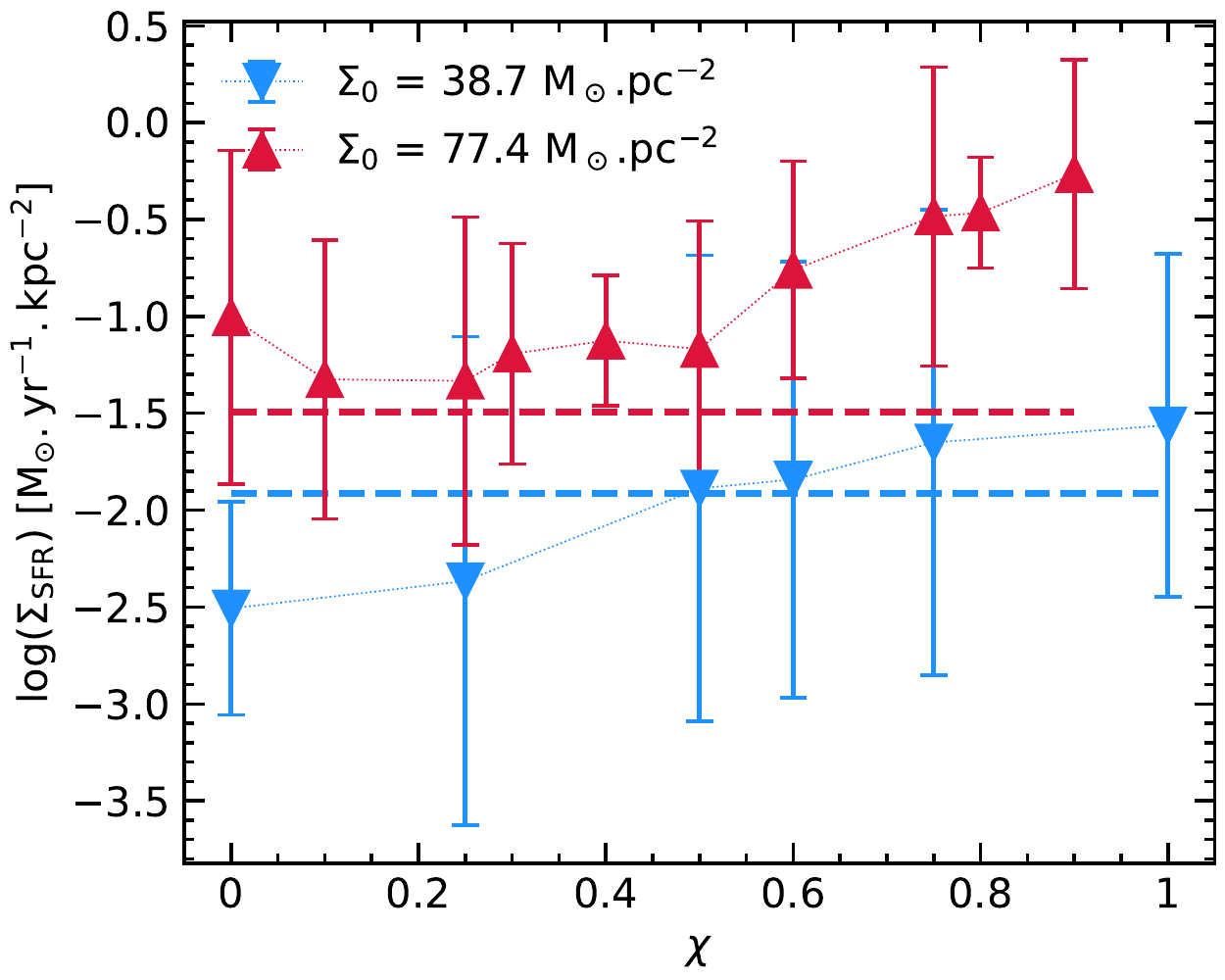}
    \end{center}

    \caption{Effect of the compressibility of the turbulent driving on the SFR (group \textsc{comp}). Surface density of the SFR as a function of the compressibility $\chi$ of the driving. The dashed line is the value of the SFR given by the SK relation.}\label{fig:compvar}

\end{figure}

\begin{figure}[htb!]
    \begin{center}

        \includegraphics[width=0.39 \textwidth]{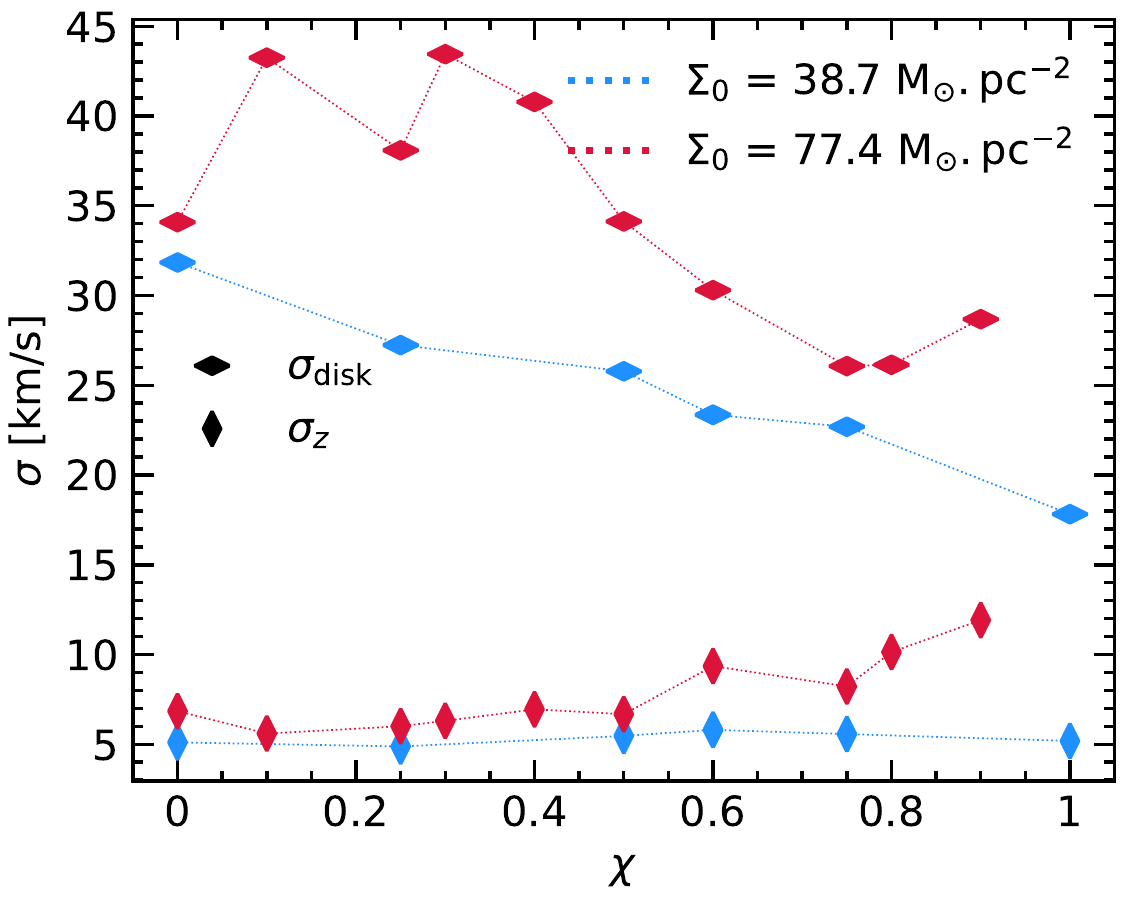}

    \end{center}
    \caption{Mass-weighted velocity dispersion measured in the \textsc{comp} simulations at 58 Myr at the kiloparsec scale.  }\label{fig:compveldisp}
\end{figure}

\begin{figure}[htb!]
    \begin{center}
        \includegraphics[width=0.42\textwidth]{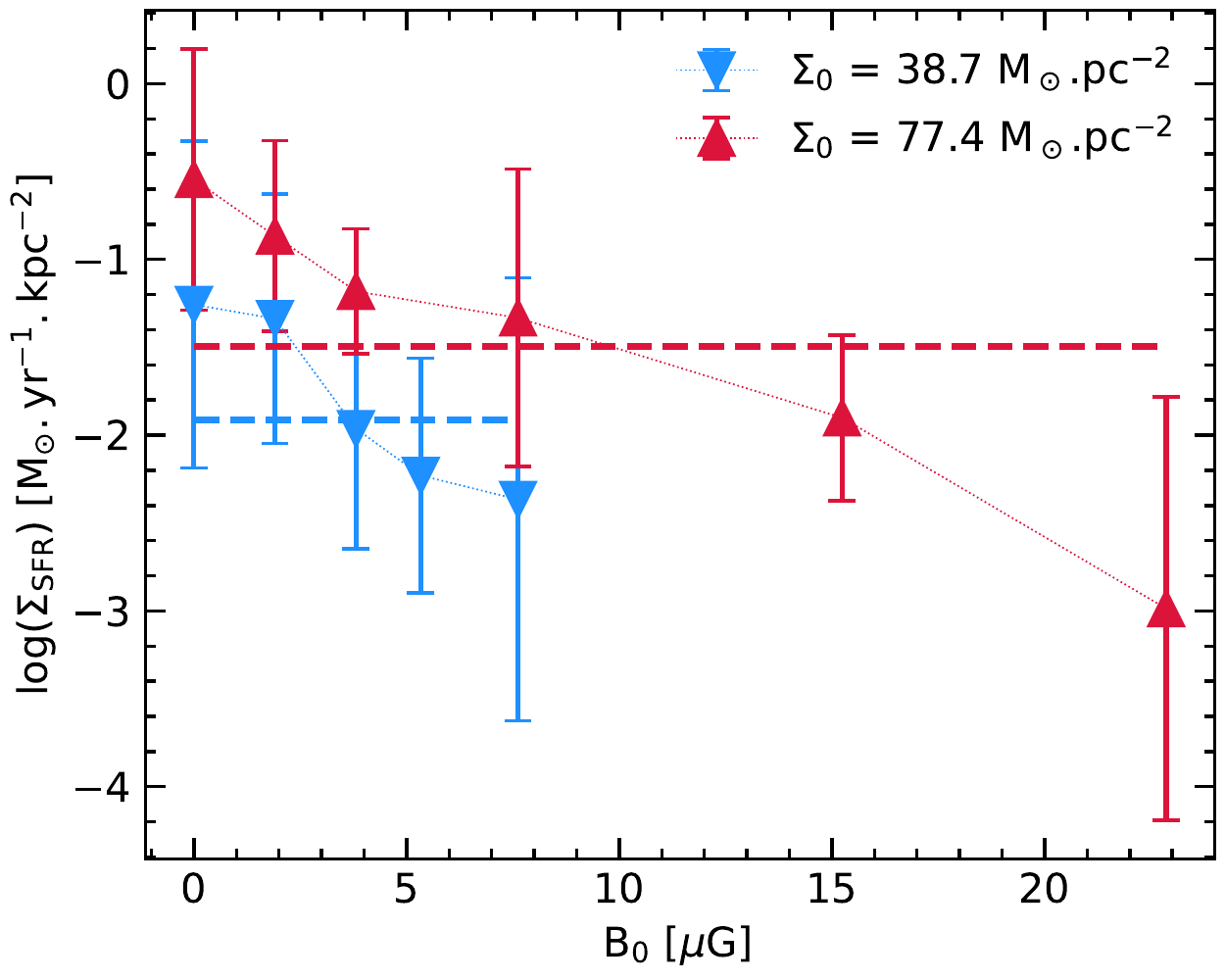}
    \end{center}

    \caption{Effect of the intensity of the initial magnetic field on the SFR (group \textsc{mag}). $B_0$ is the initial midplane intensity of the magnetic field. The legend is the same as in Fig.~\ref{fig:turbvar}. The dashed line is the value of the SFR given by the SK relation.}\label{fig:mag_sfr}
\end{figure}

\begin{figure*}[htbp!]
    \centering
    \includegraphics[width=0.90\textwidth]{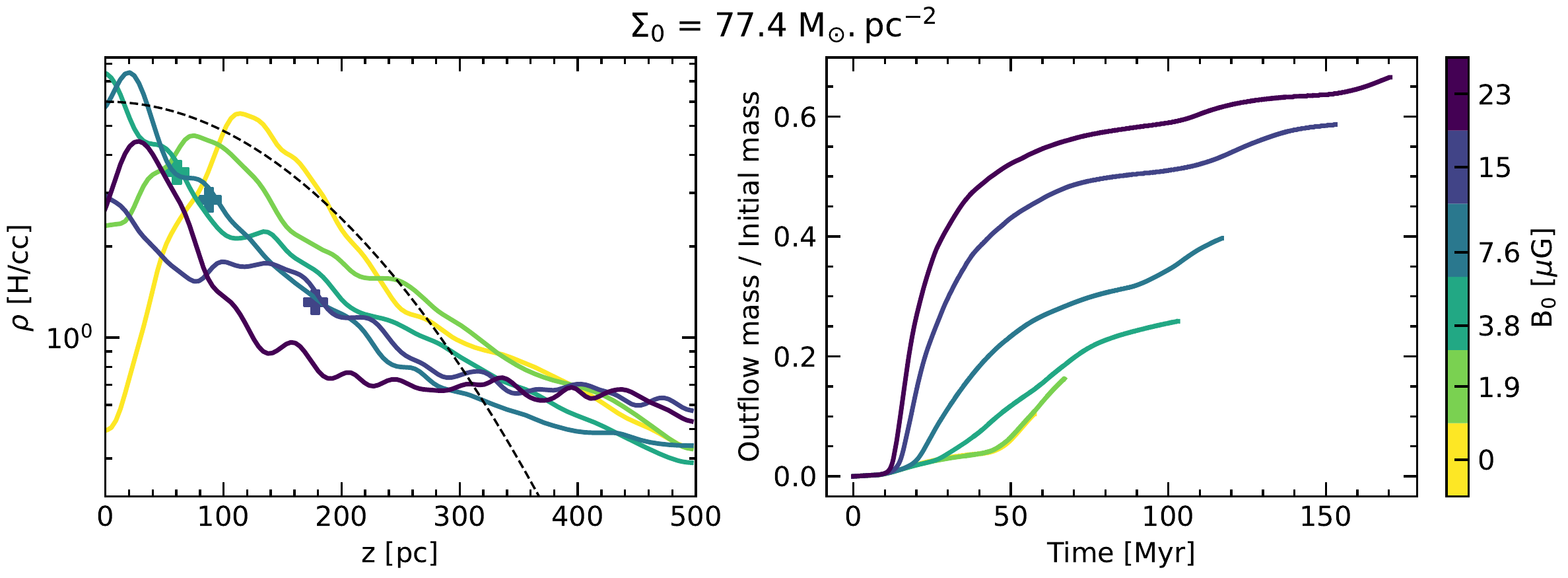}
     \includegraphics[width=0.91\textwidth]{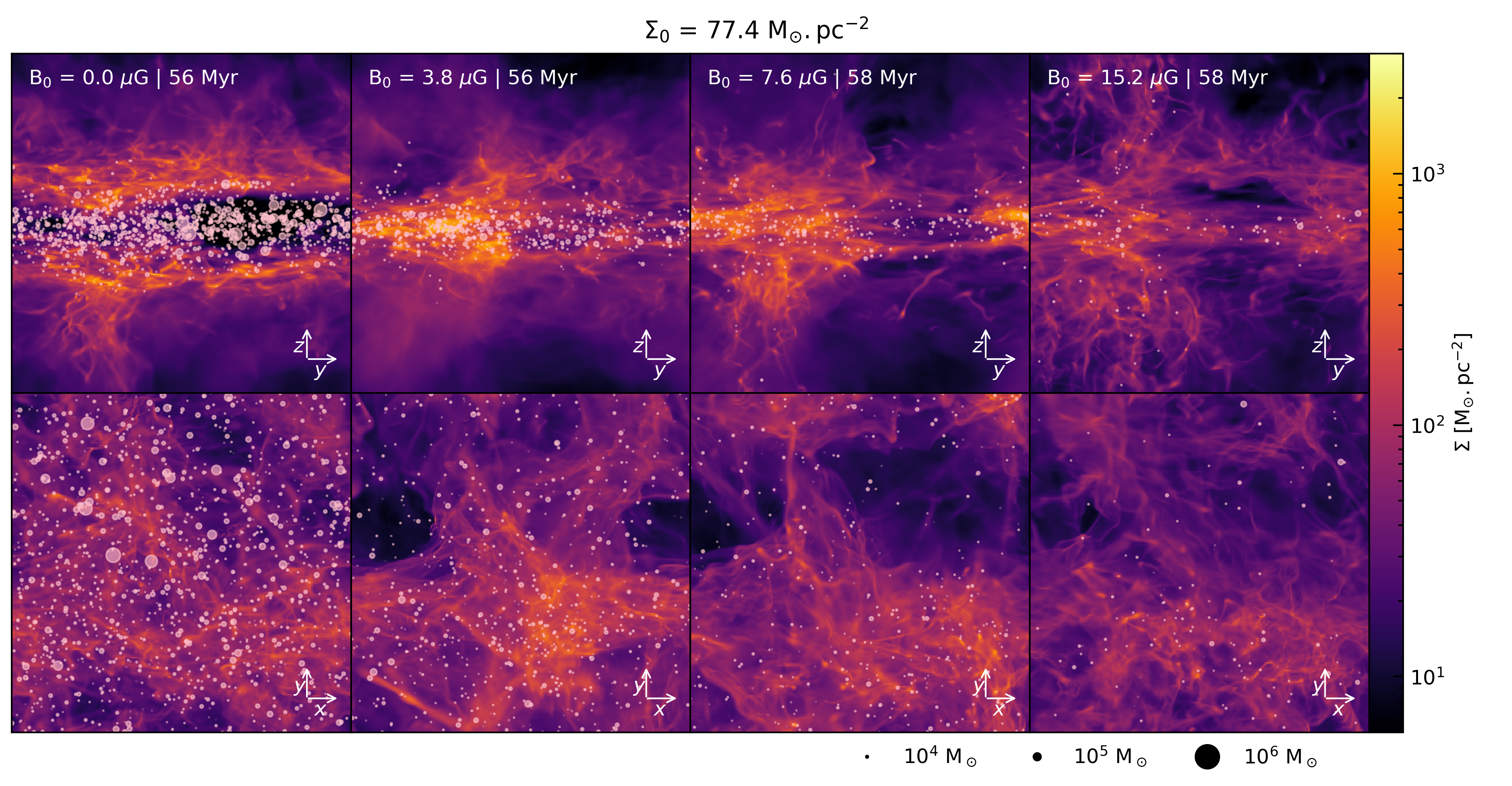}

    \caption{Effect of the magnetic field on the structure of the disk and the outflows (group \textsc{mag})´ 
   for a high  column density. \emph{Top left panel:} Averaged density profile (slightly smoothed for readability) at $t \approx 58~\mathrm{Myr}$. For runs in which the midplane density was close to the maximum density, the position at which the density is half of the midplane density is marked with a plus. The dashed black line corresponds to the initial profile. The colours of the lines refer to the same colour bar as in the plot of the top right panel. \emph{Top right panel}:  Mass loss via outflows through the open up and down boundaries. \emph{Lower panel:} Edge-on (top) and face-on (bottom) column densities. 
   }
    \label{fig:mag_diskheight}
\end{figure*}

\begin{figure}[ht!]
    \begin{center}
        \includegraphics[width=0.37\textwidth]{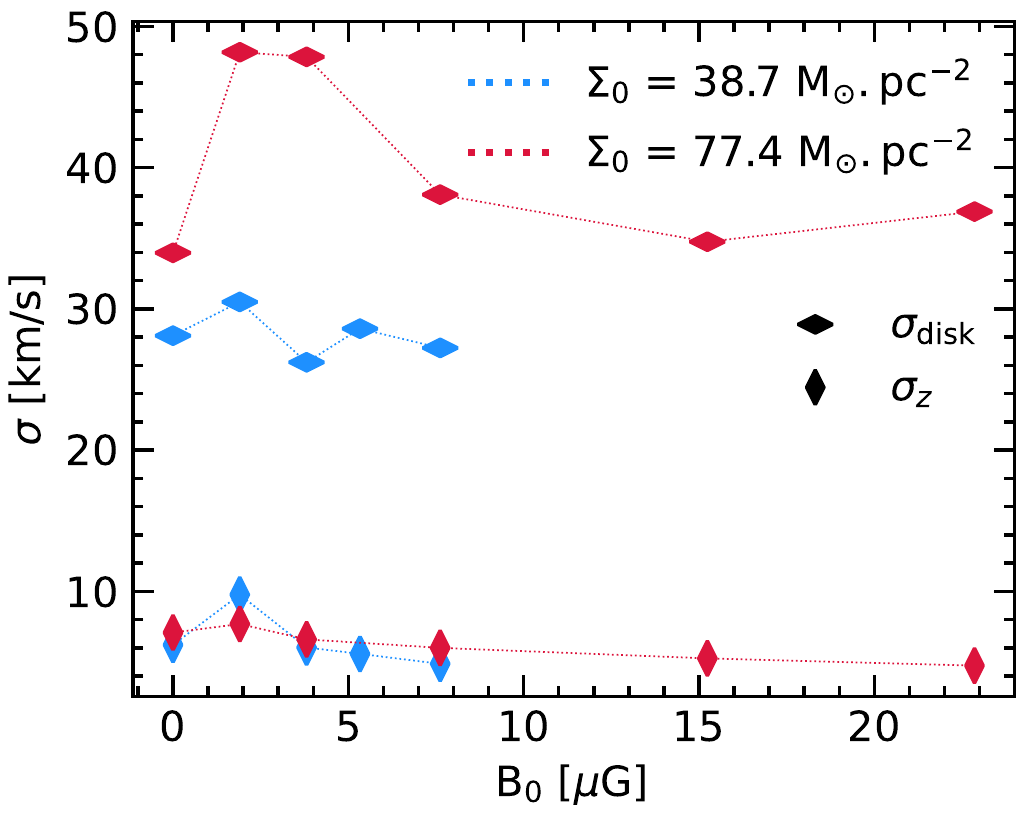}
    \end{center}
    \caption{Velocity dispersion measured in the \textsc{MAG} simulations at 58 Myr.}\label{fig:mag_veldisp}
\end{figure}

We used our numerical setup to probe the effect of the strength of turbulence (simulation set \textsc{strength} in Table~\ref{tbl:simu}) by
varying the RMS acceleration of the turbulence between $1\E{4}$ to $1\E{5}$ in code units\footnote{the code units for the RMS acceleration of the turbulence convert into 1.46 $\E{-4}~\mathrm{km}\cdot\mathrm{s}^{-1}\cdot\mathrm{Myr}^{-1}$.} for medium column densities and from  $5\E{4}$ to $8\E{5}$ for a high column density.

These values were selected because they lead
to SFR values that are not too far to the value inferred from the SK relation.
Figure~\ref{fig:coldens_strcomp} portrays column density images
for the medium column density simulation with several values of the
turbulent forcing strength, 
projected along the x-axis (top panel) and z-axis (bottom panel).
When the turbulent forcing is weak, the gas and the recently formed stars are organised in a few dense regions. That is 
to say, the gas is organised as few giant molecular complexes. 
As turbulent driving is increased, the density field is less strongly concentrated and the stars tend to be distributed more uniformly. 
We also note that their number is clearly reduced.
Figure~\ref{fig:turbvar} shows that with stronger turbulent driving, accretion onto the sinks is indeed slower, and the SFR decreases.

 Fig.~\ref{fig:coldens_strcomp} shows that for the highest forcing values, the anisotropy
that results from the stratification is even more pronounced. 
This is because not only the forcing alone applies in the 
xy-plane, but also because the star formation and therefore the 
stellar feedback are reduced. 

Increasing the strength of the driving naturally leads to an increase in the velocity dispersion (Fig.~\ref{fig:turbveldisp}).
Because the driving is 2D, projected on the plane of the disk, only the velocity dispersion along the disk plane, defined as $\sigma_{disk} = \sqrt{\sigma_x^2 + \sigma_y^2}/\sqrt{2}$, is increased. We would expect the vertical speed dispersion, which is assumed to be mainly due to the stellar feedback, to be reduced with stronger driving, but no such  trend is clearly observed.
The velocity dispersion scales linearly with the driving strength, but saturates for very strong driving.
Interestingly, the velocity field is generally 
significantly non-isotropic, with a ratio of 
the velocity dispersion parallel to the disk $\sigma_{disk}$ 
and the vertical velocity dispersion $\sigma_z$ that can reach a factor of 5.

The outcome of this increased velocity dispersion is a reduction in the SFR (Figs.~\ref{fig:turbvar} and \ref{fig:turbveldisp}). 
An increase in the velocity dispersion by a factor 
of a few leads to a decrease in the SFR by nearly a factor 10, which 
is a very significant drop. This illustrates the fact that 
 the externally driven turbulence can substantially modify the ISM evolution in this regime.

\subsection{Influence of the compressive fraction on the turbulence}
\label{subsec:compressibility}

The dual effect of turbulence, which both triggers and quenches star formation, is particularly clear when the compressive fraction of the driving is varied.
The velocity field can be divided through the Helmotz decomposition into compressible modes $\bm{v_c}$ and solenoidal modes $\bm{v_s}$. 
Compressible modes verify $\nabla \times  \bm{v_c} = 0$ and trigger compression and dilation of the density field.
By contrast, solenoidal modes verify $\nabla \cdot  \bm{v_s} = 0$ and thus do not affect the density field.
When the compressive fraction $\chi$ of the turbulent driving is changed (see Eq. \eqref{eq:projection}), the amount of energy in the different velocities modes is changed accordingly. 
However, it is important to note that the correspondence between the compressive fraction of the turbulent driving $\chi$ and the resulting ratio of the compressive and solenoidal modes in the velocity field is not perfect. Rotation motions induced by solenoidal driving eventually generate compression as well \citep{vazquez-semadeniInfluenceCoolinginducedCompressibility1996}, especially as the turbulence cascades to lower scales \citep{federrathUniversalitySupersonicTurbulence2013}.
With solenoidal driving, a smaller fraction of the kinetic energy is globally in compressive modes that can produce density fluctuations.
This is well illustrated in the study of the PDF in purely hydrodynamic simulations \citep{federrathDensityProbabilityDistribution2008}. The density PDF is broader and explores higher densities when the compressive fraction $\chi$  is closer to one (compressive driving) than when it is closer to zero (solenoidal driving).
To quantify the effect of the compressibility of the turbulent driving on the SFR, we ran the \textsc{comp} experiment. We again selected galaxies with an initial column density of $38$ and $77 \Msun  \cdot \mathrm{pc}^{-2}$. 
 With every other parameter fixed, we modified the compressive fraction $\chi$ from $0$ to $1$.

\begin{figure*}[htbp!]
    \includegraphics[width=\textwidth]{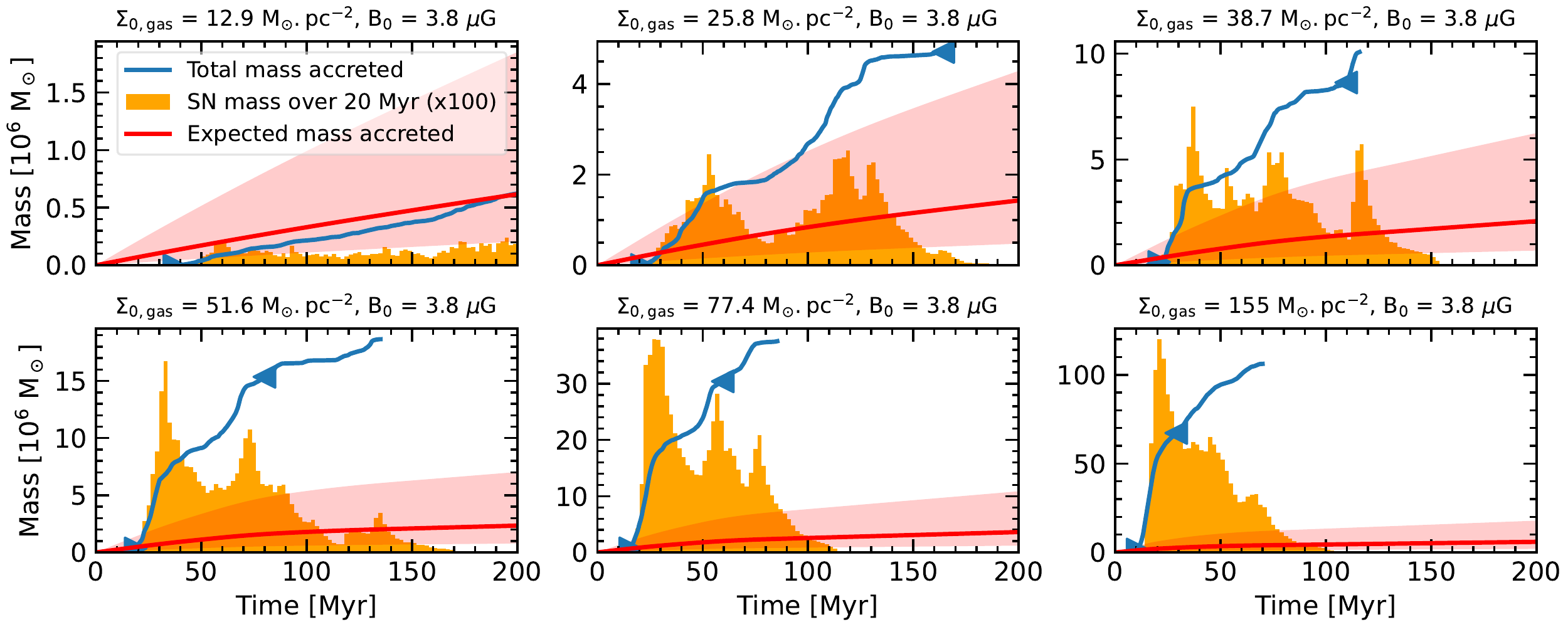}

    \caption{Stellar mass for the group \textsc{noturb}. The red line is the stellar mass creation when the SFR in the kiloparsec box is equal to $ 2.5 \E{-3}  (\Sigma / 10  \coldens)^{1.4}  \Msun\cdot\mathrm{yr}^{-1}$ \citep{kennicuttStarFormationMilky2012}. The red zone corresponds to an SFR that is three times higher and lower and aims to represent the spread of the SK relation.
    The orange histogram is the mass  (multiplied by 100 for readability) of stars that blow up into SNe in a time span of 20 Myr. 
    The SFR computation for Fig.~\ref{fig:SK} starts when 3\% of the gas is consumed or lost ($t_\mathrm{3 \%}$, left triangle) and ends when 40\% of the gas is consumed or lost ($t_\mathrm{40 \%}$, right triangle).
    }\label{fig:sink_mass_noturb}
\end{figure*}

The column density maps of snapshots at 58 Myr of these simulations are shown in the lower panel of Fig.~\ref{fig:coldens_strcomp}. 
From left to right, the compressibility is increased 
from 0 to 1. In the edge-on view (top), 
 the purely solenoidal case produces horizontal filaments, 
while more clumpy structures are produced when the compressibility increases.
The face-on view shows that dense structures tend to be smaller 
when the compressive fraction is smaller.
These larger and denser structures are actively forming stars. 
As a result, and this is depicted in Fig.~\ref{fig:compvar}, increasing the compressibility increases the SFR by a factor of about 10. 
However, even a purely compressive run produces fewer stars than a run
without turbulent driving (compared with Fig.~\ref{fig:sink_mass_noturb}).
Therefore, the effect of turbulent driving at this strength is always to quench star formation, regardless of the compressibility. 

Interestingly, even though the driving
was calibrated to produce an identical 
velocity dispersion regardless of the compressibility (see Appendix \ref{sec:calibration}),
 a dependence of the velocity dispersion on the compressibility (see Fig.~\ref{fig:compveldisp}) is nevertheless observed. 
  The variations in the velocity dispersion remain 
  well below the variations induced by the variations
  of the driving strength, however.

Overall, these results clearly indicate that the nature and thus the sources of the turbulence are important to identify,
not only regarding the driving intensity of the turbulence,
but also regarding the compressibility of the ISM turbulence. Both play a significant role in setting the SFR.

\section{Influence of the magnetic field}
\label{sec:mag}

The magnetic field plays an important role on the gas dynamics
and in the structure of the ISM \citep{hennebelleRoleMagneticField2019}. 
In the following, we study the impact of stronger initial magnetic fields on the SFR in our simulations when all other parameters remain the same.
We considered the simulations from the group \textsc{mag} (cf Table~\ref{tbl:simu}).
We varied the initial midplane magnetic field $B_0$ from 0 $\mu \mathrm{G}$ 
to 15 $\mu \mathrm{G}$.
As a reminder, the initial magnetic field is uniform in the galactic plane and follows a Gaussian distribution in the vertical direction (see Eq.~(\ref{eq:B})). The initial column density for these runs was also $33.7$ (medium) or $77.4 \Msun\cdot\mathrm{pc}^{-2}$ (high).

Figure~\ref{fig:mag_sfr} shows that an increased magnetic field has a quasi-linear effect on the SFR. For the two cases 
(medium and high column density), as $B_0$  increased
from 0 to 7 $\mu$G, the SFR has dropped by a factor of nearly 10. 
Since high magnetic intensities lead to a substantial reduction in SFR, 
this implies that a strong magnetic field could reduce the turbulent 
energy that would otherwise be required to sustain a low SFR. 
This point is further investigated in section \ref{subsec:bturb}.

\begin{figure}[htbp!]
    \includegraphics[width=0.5\textwidth]{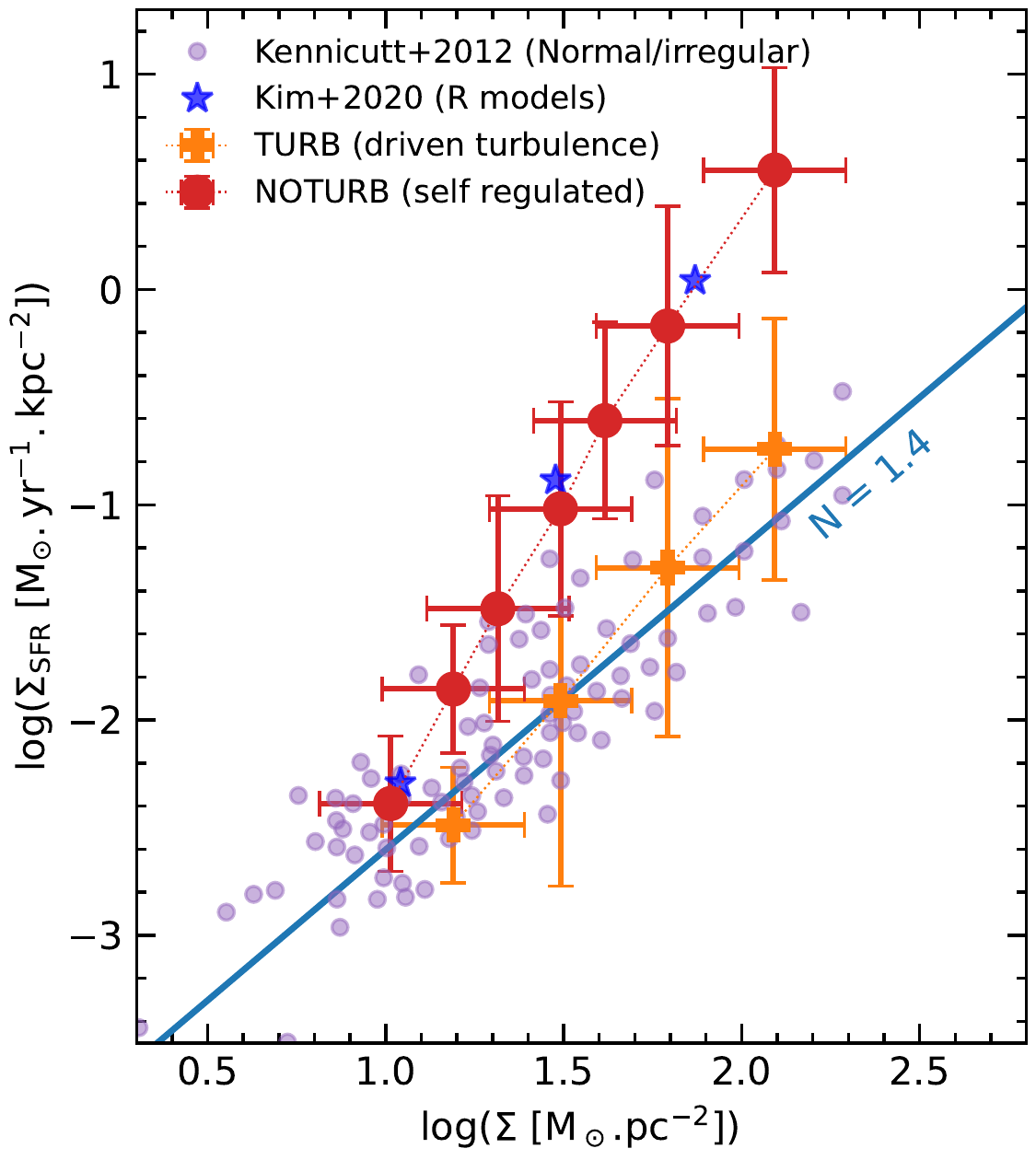}
    \caption{Star formation relation for the groups \textsc{noturb} and \textsc{turb}. The SFR and the associated uncertainties are computed as explained in section \ref{subsec:computation_sfr}.
    We added data from observations of normal and irregular galaxies, and the blue line represents the SK relation \citep{kennicuttStarFormationMilky2012}. We also compare our results with the results from the R models of \cite{kimFrameworkMultiphaseGalactic2020}.
    }
    \label{fig:SK}
\end{figure}

The effect of the magnetic field on the SFR is a consequence of magnetic support, which locally resists 
the impact of self-gravity, but also of the global magnetic support that leads to a large disk scale height \citep{iffrigStructureDistributionTurbulence2017} and therefore to reduced densities.
To further verify this point, we investigate the effect of the magnetic intensity 
on the density profile in Fig.~\ref{fig:mag_diskheight}.
For low magnetic fields, the SFR is so high that the gas in the midplane of the disk is accreted
in stars or is swept up by the SNe, and this depletes the 
gas near the equatorial plane, as we show in the column density map and in the upper left panel, which features the averaged density profile.

The scale height, denoted with plus marks, increases when we increase the initial magnetic from $3$ to $15 \mu$G. 
However, the density profile is not smooth, and the effect of the magnetic field is better understood by comparing the profile at $t \approx 58~\mathrm{Myr}$ (in color) with the initial profile (dashed black line).
The gas is lifted, and a significant amount of gas is lost through outflows (upper right panel). 
A stronger magnetic field triggers stronger outflows. 
This effect is also seen in simulations reported by \cite{girichidisSILCCProjectImpact2018}.
Magnetically driven outflows can be distinguished from the SN outflows that are observed in the low magnetic field cases because they occur earlier in the simulation.
It is important to note that increased outflows also contribute to the diminution of the SFR. Less gas remains available for the formation of gas. 

The bottom panel of Fig.~\ref{fig:mag_diskheight} shows that the value of the initial magnetic field 
does not seem to affect the distribution of the gas in the horizontal direction, 
where turbulent motion generated by the driving dominates the dynamics of the gas. 
For the same reason, it has very little effect on the velocity dispersion, as we show in Fig.~\ref{fig:mag_veldisp}. 
The regulation effect of the magnetic field is thus mainly due to how it changes the vertical structure.

Because the driving is not computed self-consistently, hypothetical stabilising effects of the magnetic field analogous
to what was reported by \cite{heitschMagnetizedNonlinearThinShell2007} and \cite{zamora-avilesMagneticSuppressionTurbulence2018} at smaller scales would not be captured by our model. 
However, the turbulent driving aims to model turbulence that is generated by instabilities at the galactic scale.
Even if the magnetic field intensity is locally high, the averaged magnetic field at the galactic scale is probably not strong enough to have a significant influence on the development of the instability.

\section{Reproducing the Schmidt-Kennicutt relation}
\label{sec:sk}

\subsection{Self-regulation by stellar feedback?}
\label{subsec:noturb}

\begin{figure}[htp]
    \includegraphics[width=0.45\textwidth]{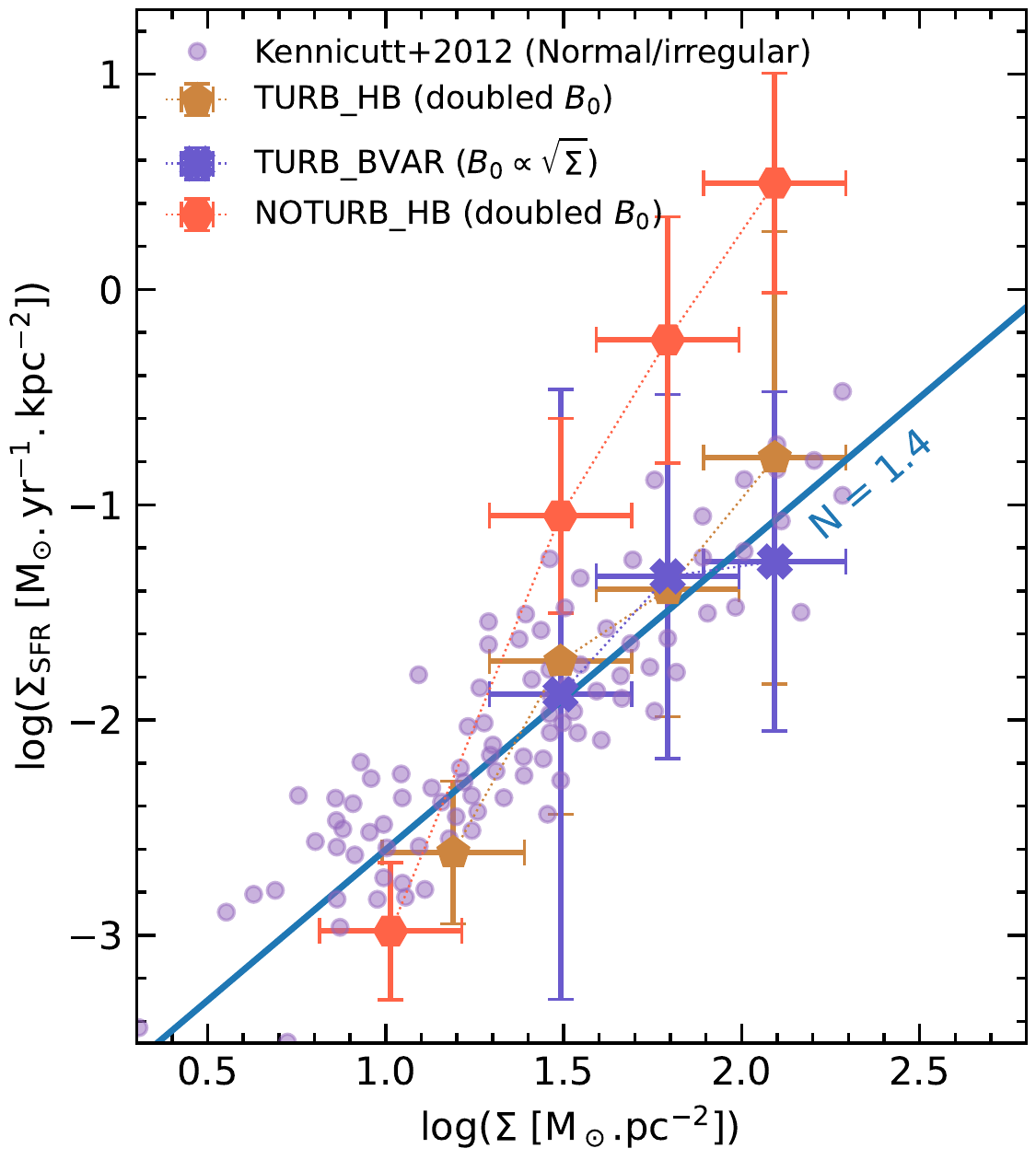}
    \caption{Star formation relation for the simulations with a high magnetic field.}
    \label{fig:SK_hB}
\end{figure}

In \citetalias{brucyLargescaleTurbulentDriving2020} we found that stellar feedback alone cannot regulate the SFR to match the observed values, especially when the column density is high. 
In this section, we focus on the simulation of the \textsc{noturb} group, without injection of turbulence from the large scale. 

Self-regulation works well for galaxies with column densities similar to the mean value of the Milky Way ($\approx 10 \Msun\cdot \mathrm{pc}^{-2}$). In our simulation with a starting column density of approximately $12 \Msun\cdot \mathrm{pc}^{-2}$ ($n_0 = 1$), star formation is moderate and well in line with the observations. However, for higher column densities, 
Fig.~\ref{fig:sink_mass_noturb} reveals that 
the SFR, that is, the slope of the blue solid line, is far above the observationally determined values represented by the slope of the thick red line.
Moreover, the star formation versus column density relation that can be derived is much steeper (see Fig.~\ref{fig:SK}), with typically ${\dot \Sigma} \propto \Sigma^{2.5}$.
Interestingly, a similar result was found by \cite{kimFrameworkMultiphaseGalactic2020}.
However, this comparision should be taken carefully because the simulations from  \cite{kimFrameworkMultiphaseGalactic2020} have a different box size and shape, a different integration time, a different feedback implementation, and a higher external potential for high column density.

Figure~\ref{fig:sink_mass_noturb} shows that high SFR generates a peak in SN mass. 
After this peak, star formation is reduced for a while. 
However, the accumulation of gas at the large scale is not destroyed, 
and the very dense gas reservoir is replenished over a period of a few tens of million years. 
As a consequence, star formation can continue at a fast rate.
This adds up to the fact that was already highlighted in previous studies 
\citep{brucyLargescaleTurbulentDriving2020,nusserRegulationStarFormation2022},
that the energy injection via stellar feedback is lower than what would be needed 
to maintain the SFR at a rate compatible with the SK relation, 
as well as the energy injected via gravo-turbulence to maintain the Toomre parameter $Q \sim 1$.
Therefore, this constitutes strong evidence that stellar feedback cannot regulate star formation sufficiently 
in high column density environments.

\begin{figure}[ht]
    \includegraphics[width=0.45\textwidth]{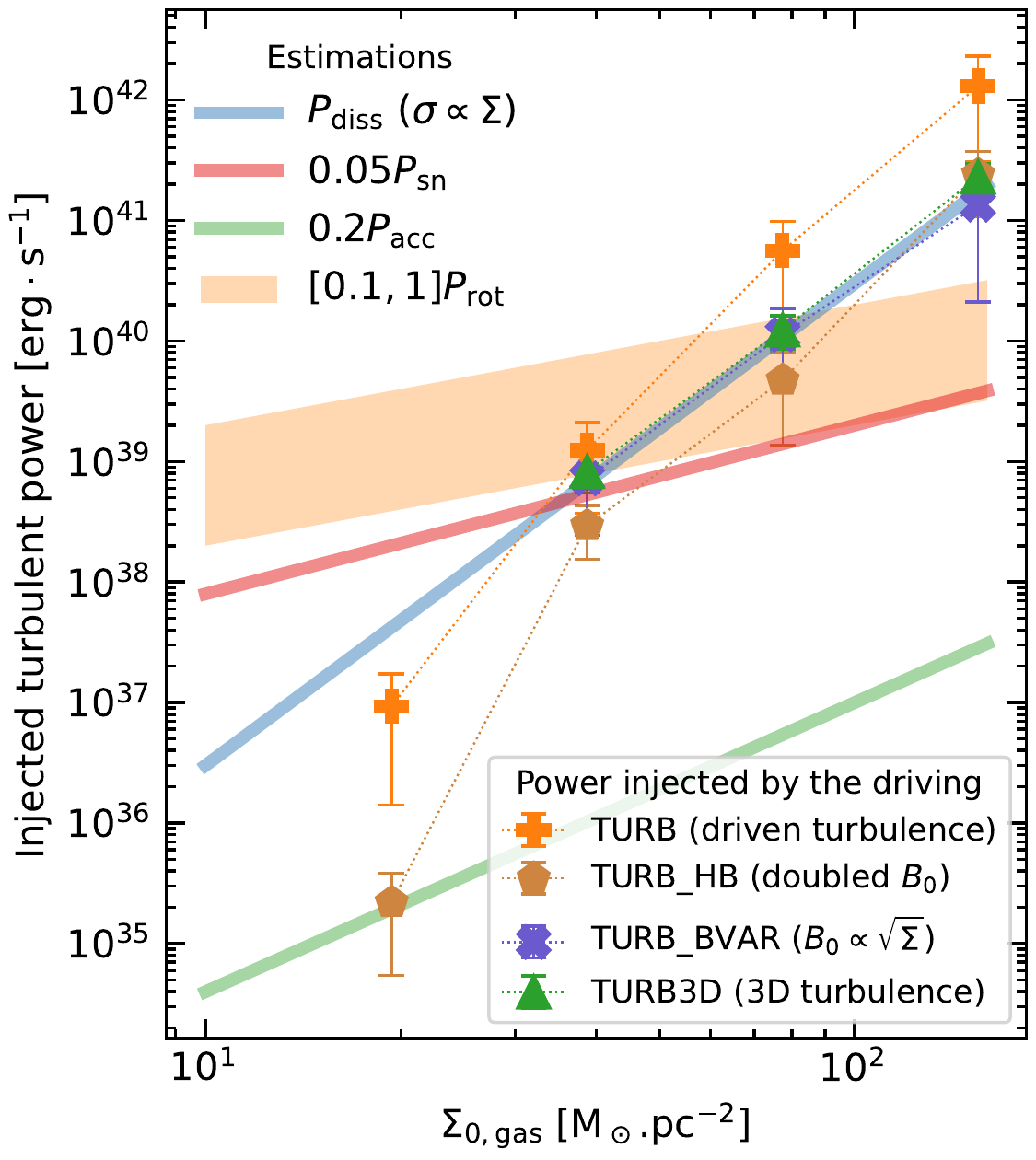}
    \caption{Markers represent the time-averaged power injected by the turbulent driving force in order to reproduce the SK relation. The power is computed in the whole simulation box and averaged between 4 and 60 Myr. In these simulations, the driving strength was calibrated so that the SFR was close to the SK relation (see section \ref{subsec:turb} and \ref{subsec:bturb}, as well as Figs. \ref{fig:SK} and \ref{fig:SK_hB}). The mean value is the total energy injected during the period divided by the time span, while the error bar is the standard deviation from the power computed at each time step. The solid lines refer to the estimates of the possible source of turbulent driving, and the notation is the same as in section \ref{sec:energy_diss}. 
    The efficiency coefficients ($\epsilon_\mathrm{sn} \sim 0.05, \epsilon_\mathrm{acc} \sim   0.2, \epsilon_\mathrm{rot} \in [0.1, 1]$) are reasonable estimates, as discussed in section \ref{sec:energy_diss} as well. 
    To draw the blue line, we assumed that the velocity dispersion was proportional to the column density.}
    \label{fig:inj_power}
\end{figure}

\subsection{Impact  of turbulent driving on the Schmidt-Kennicutt relation}
\label{subsec:turb}

\citetalias{brucyLargescaleTurbulentDriving2020} has shown that it was possible 
to obtain a star formation relation that was compatible with the SK relation 
when the turbulent driving was scaled with the initial column density.
In this work, we reproduced this result with a slightly different method 
for turbulent injection that we presented in section \ref{subsec:turb_inj} 
and with the improved SFR computation presented in section \ref{subsec:computation_sfr}. 
For the group of simulations \textsc{turb}, 
we kept the same magnetic field for all the simulations, 
and we studied the increasing column densities. 
The compressive fraction was set at 0.25. This choice can be discussed: it is lower than the natural mix, which would be 0.5 \citep{federrathDensityProbabilityDistribution2008}, but the velocity field in the galactic scale simulation of \cite{jinEffectiveTurbulenceDriving2017} indicates that the solenoidal mode may dominate at large scales.

For each column density, we ran several simulations with different driving strengths (as in section \ref{subsec:turb_strength}) and selected the simulation in which the SFR was more similar to the SK relation. Figure~\ref{fig:SK}
shows the result, where the differences between the 
SFR from simulations with and without driving are rather clear.

This proves that the turbulent driving from the large scale is a valid candidate for closing the gap between the observations and the simulations with stellar feedback alone.
The question clearly is whether the added driving is realistic in terms of injected energy and resulting speed dispersion, and also regarding the consequence on the structure of the ISM.

One possible diagnostic is to compute the energy injected within the simulation, as has been done in \citetalias{brucyLargescaleTurbulentDriving2020}. Another diagnostic is 
to consider the velocity dispersion. Both are 
discussed below.

\subsection{Impact of the magnetic field on the Schmidt-Kennicutt relation}
\label{subsec:bturb}

\begin{figure*}[htbp]
\begin{center}
    \includegraphics[width=0.37\textwidth]{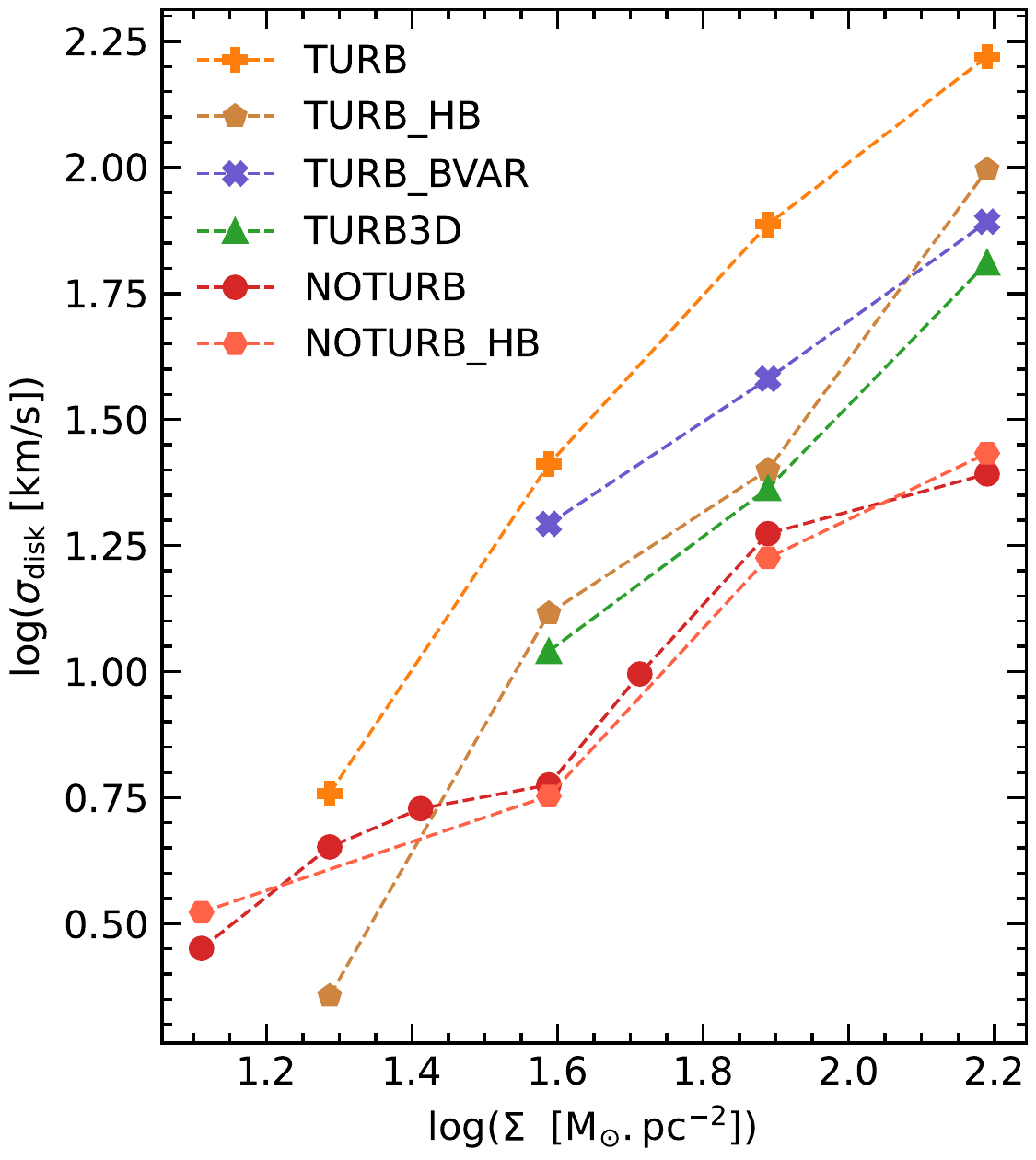}
    \hspace{3em}
    \includegraphics[width=0.37\textwidth]{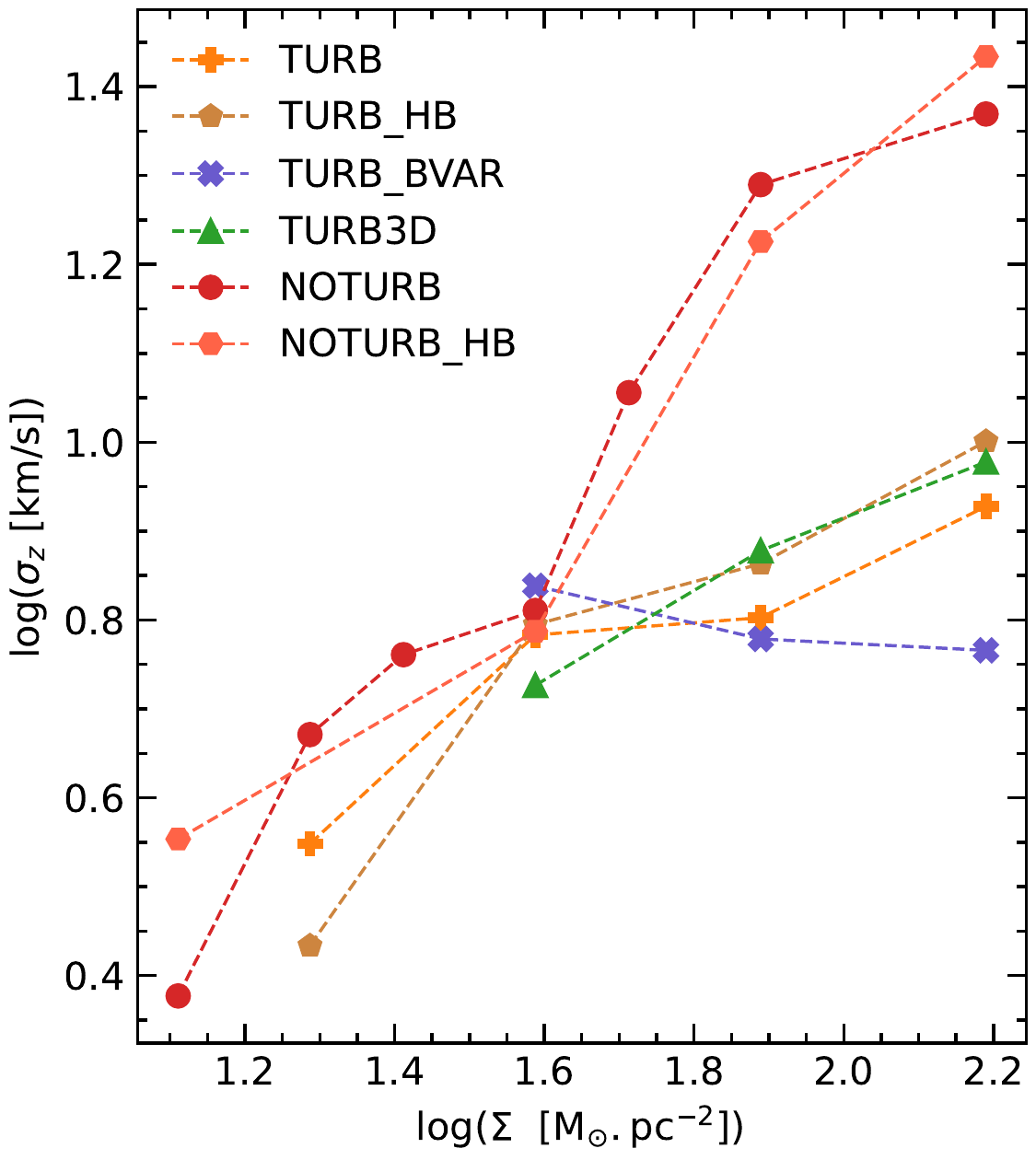}
\end{center}
    \caption{Mass-weighted velocity dispersion in the full computational volume of 1 kpc$^3$ computed at $t = 58$ Myr. \textit{Left panel}: Velocity parallel to the galactic plane. \textit{Right panel}:
    Vertical velocity.}
    \label{fig:veldisp2D}
\end{figure*}

The actual value of the magnetic field in galaxies
\citep{beckMagneticFieldsSpiral2015,hanObservingInterstellarIntergalactic2017} remains controversial 
because it is very difficult to measure. However, 
based on energy equipartition, 
it is reasonable to think that gas-rich galaxies are subject to stronger magnetic fields. As shown in section \ref{sec:mag}, a stronger magnetic field results in a lower SFR.

It is thus interesting to investigate how the magnetic field,
in conjunction with turbulence, may
contribute to shaping the SK relation. To investigate this, we ran three series of simulations called \textsc{noturb\_hB},  \textsc{turb\_hB}, and \textsc{turb\_Bvar}.
The description of all simulations can be found in 
Table~\ref{tbl:simu}.
We recall that 
 \textsc{noturb\_hB} and  \textsc{turb\_hB}  are the same as   \textsc{noturb} 
and \textsc{turb,}  respectively, but with doubled magnetic intensities. 
For the group \textsc{turb\_Bvar},  the initial magnetic field, parametrised by its midplane value $B_0$, scales as the square root of the column density. 
In this way, the ratio of the kinetic and magnetic energy at the beginning of the simulation remains the same. 

Figure~\ref{fig:SK_hB} displays the SFR for the runs 
. In spite of the initial field, which is twice
as intense, the trend remains similar to the 
SFR of the \textsc{noturb} runs. In particular, the 
SFR dependence on the column density is clearly too steep.

To investigate the impact that magnetic fields 
have on the SK relation in conjunction with turbulence, 
we proceeded as with the \textsc{turb} group. For 
each value of the initial column density and 
magnetisation,
we made an educated guess of what would be 
a good strength of the turbulent RMS acceleration to obtain an SFR similar to the SK (equal 
to within a factor of 2).
We then launched the simulation, computed the SFR, and corrected the RMS acceleration if necessary.
The simulations we finally selected are described in Table \ref{tbl:simu} . They produce a mean SFR that is more similar to the SK value.
The goal is to determine for an initial given magnetic field intensity and column density
how much turbulent energy is needed to reproduce SK. 

First, Fig.~\ref{fig:SK_hB} shows that with our choices of RMS acceleration, 
we reproduce the SK relation for all groups, as expected. 
The SFR is globally slightly lower for the \textsc{turb\_Bvar} group, 
meaning that we slightly overestimated the turbulent energy required
for this group.

Second, Fig.~\ref{fig:inj_power} shows how much  power 
is injected into the simulation by the driven turbulence for all groups
as a function of the column density.
The injected power was computed at each step by computing the difference between the kinetic energy before and after applying the turbulent driving and dividing by the time step.
This reveals that the required power is significantly 
reduced in the presence of substantial
magnetic fields. 
Compared to the series of runs \textsc{turb}, the power with which 
it is necessary to drive the turbulence is 
more than a factor of ten lower when the magnetic 
field is doubled, and the slope is roughly the same (group \textsc{turb\_hB}).
When the magnetic field scales as the square root of the column density (group \textsc{turb\_Bvar}),
the power-law slope of the power needed to reproduce the SK relation versus the column density 
is lowered from almost 5 to 4. 
This means that magnetic fields may significantly contribute 
to the origin of the SK relation.
The values of the magnetic field we considered are likely reasonable, 
but it is entirely possible that stronger intensities have to be considered,
which would further reduce the required turbulent driving.

Figure~\ref{fig:inj_power}  also shows the various estimates for the energy injection made in section \ref{sec:energy_diss}. 
In gas-rich galaxies, rotation could in principle provide enough energy up 
to $\Sigma \simeq 100$~M$_\odot$~pc$^{-2}$. For higher values, however, this does not seem to be 
the case, meaning that a source of energy that could explain the SK relation for column densities higher than 
$100$ M$_\odot$ pc$^{-2}$ is currently lacking.

\subsection{Velocity dispersion}
\label{subsec:veldisp}

The velocity dispersion is also a clear and 
simple diagnostic to assess the realism of simulations. 
Figure~\ref{fig:veldisp2D} displays the mass-weighted velocity 
dispersion computed at the scale of the full kiloparsec box as a function of column density for the 
various runs we performed. 

Several trends are worth noticing. First, in the absence of 
turbulent driving (runs \textsc{noturb} and 
\textsc{noturb\_hB}), the velocity dispersion
remains broadly isotropic while in the presence 
of driving, particularly at high column densities, 
the velocity dispersion in the disk plane is 
several times higher than the velocity along the 
z-axis. Interestingly, in the presence of driving, 
the velocity dispersion in the disk plane is higher 
than without driving, whereas the reverse is true for the 
z-velocity. This is because in the latter case, self-gravity 
is strong and triggers isotropic collapse in various places. 
In the presence of strong magnetic fields
(runs \textsc{turb\_hB} and  \textsc{turb\_Bvar}), 
the velocity dispersion necessary to reproduce the SK relations
is lower by a factor of about~2. Since the 
energy dissipation rate is expected to be proportional to $\sigma^3$, 
this agrees well with the factor of about ten induced 
for the driving power presented in Fig.~\ref{fig:inj_power}.

It is important to stress that the estimated injected power of Fig.~\ref{fig:inj_power}
and the velocity dispersion of Fig.~\ref{fig:veldisp2D} are not predictions,
but an estimate of the strength of the turbulence needed to quench star formation 
down to the SK relation.
The way in which the turbulence is injected and the parameters of the model, such as the strength 
of the magnetic field, may change these requirement estimates (see sections
\ref{subsec:turb3D} and \ref{subsec:bturb}, respectively).

Comparisons with observations are difficult because postprocessing steps 
beyond the scope of the work are needed to ensure that the same quantities are compared
in terms of scale and tracked gas.
Data from the ShiZELS galaxies \citep{swinbankPropertiesStarformingInterstellar2012}, where properties also measured at scales close to 1 kpc in the H$\alpha$ band, feature regions with a surface density of SFR $\Sigma_\mathrm{sfr}$ around 0.1~$\ssfr$  and measured velocity dispersion that can reach up to 200~$\kms$.
Similar results were found by \cite{lawKiloparsecscaleKinematicsHighredshift2009} and in the WiggleZ survey \citep{wisnioskiWiggleZDarkEnergy2011}.
The values from the \textsc{turb} group, with a initial magnetic field at $3.8~\mu\mathrm{G}$, are quite extreme, 
indicating that for these column densities, the magnetic field is probably stronger.

\begin{figure}[htbp]
\centering
    \includegraphics[width=0.39\textwidth]{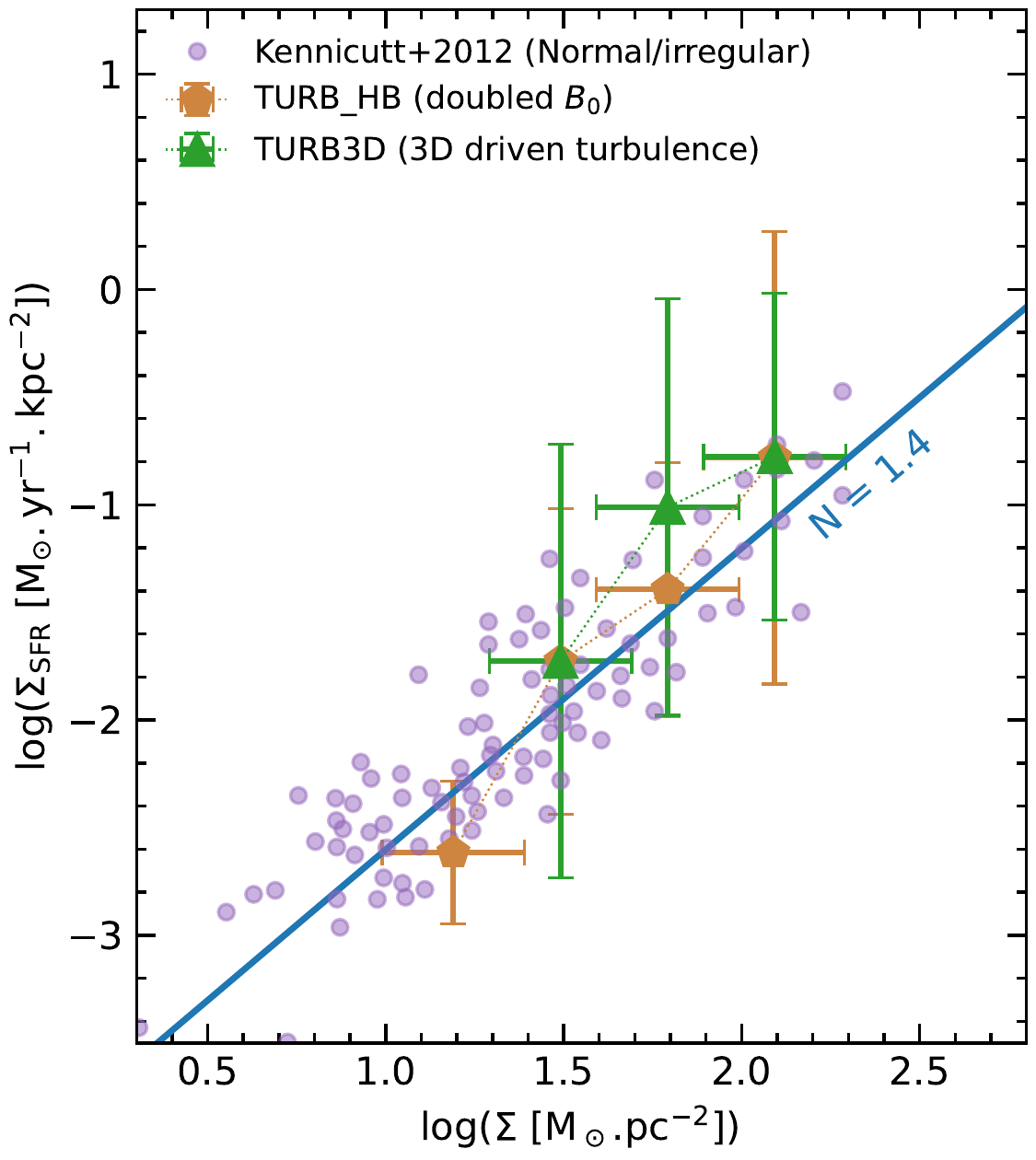}
    \caption{SFR surface density  for the \textsc{turb3D} group compared to the \textsc{turb\_hB}. The legend is the same as in Fig.~\ref{fig:SK}.}
    \label{fig:t3d}
\end{figure}

\subsection{Three-dimensional driving}
\label{subsec:turb3D}

The high anisotropy of the velocity field found in the driven simulations 
is a consequence of the 2D turbulent driving.
The driving force is parallel to the disk. 
Furthermore, it does not depend on the vertical coordinate because the wave vectors of the Fourier decomposition of the force are also aligned with the disk.
Both of these assumptions are quite strong and can be relaxed. 

Even though large-scale motions are expected to be aligned with the galactic plane, turbulent eddies can spread in the vertical direction if their size is below the scale height of the disk.
In this last part, we try to modify our driving model to take this into account. 
The driving force follows the same equations as in section \ref{subsec:turb_inj} with the following modifications.

We changed the power spectrum $F_0$ to add vertical modes with a wavelength of about 32 pc (the disk scale height is initially 150 pc), 
\begin{equation}
    \label{eq:F0_3D}
    F_0(\bm{k}) = 
    \begin{cases} 
    \left(\dfrac{k}{2 \pi}\right)^{-5/3} \text{ if } 2 \leq \dfrac{k_x}{2\pi}, \dfrac{k_y}{2\pi} \leq 3 \text{ and } 30 \leq \dfrac{k_z}{2\pi}   \leq 32, \\
    \\
    0 \text{ if not.}
    \end{cases}
\end{equation}

The projection operator $\bm{P_\zeta}$ was the normal 3D projection operator defined in \cite{schmidtNumericalSimulationsCompressively2009}. However, we applied a factor of $0.3$ to the vertical motion of the resulting force to avoid motions that were too strong in the vertical direction that may destroy the disk.

We were able to find the value of the turbulent
 driving for the 2D driving that leads to SFR compatible with the SK relation (Fig~\ref{fig:t3d}, left panel). 
Compared with the most similar set of simulation, \textsc{turb\_hB}, the energy needed to achieve SK is slightly increased, as we show in Fig.~\ref{fig:inj_power}. However,  Fig.~\ref{fig:veldisp2D} shows that the velocity dispersion in the disk is lower, as is the anisotropy. 
The vertical velocity dispersion does not increase significantly, probably because of the strong attenuation of the vertical component of the force. 
With this experiment, a less anisotropic velocity dispersion can be produced by releasing the assumption on 2D turbulence. 
When more accurate constraints on the turbulent driving are available, the presence and strength of a vertical component are an important item to monitor.

\section{Caveats}
\label{sec:caveats}

\subsection{Shear and Coriolis effect} 

\label{subsec:cav_shear}

The fiducial model includes neither shear nor the Coriolis effect because the driven turbulence module and the shearing box used in aprevious study ( \cite{collingImpactGalacticShear2018}) require incompatible boundary conditions. 
Our 2D and mainly solenoidal turbulent driving reproduces some features of the shear and the Coriolis effect, but in a less self-consistent way. 
In its current state, it does not allow us to distinguish between the effect of large-scale unordered turbulence and ordered rotation.

The effect of shear and the Coriolis effect were studied by \cite{collingImpactGalacticShear2018} and \cite{kimFrameworkMultiphaseGalactic2020}. Both studies showed that a shear value comparable to it did not reduce the SFR enough to match the SK relation and just slightly improved the discrepancy observed in group \textsc{turb} in Fig.~\ref{fig:SK}. 
 \cite{collingImpactGalacticShear2018} showed that for an initial column density of $19.4~\coldens$, the shear associated with an angular speed of $28~ \kms\cdot\mathrm{kpc}^{-1}$ , which is typical of the solar neighbourhood, only reduced the SFR by a factor $1.2$. 
This is almost negligible given the high uncertainties of the type of computations. 
On the other hand, a higher angular speed of $56~\kms\cdot\mathrm{kpc}^{-1}$, which may be found in denser environments at a galactocentric radius around $4~\mathrm{kpc}$, leads to a reduction of the SFR of a factor $3.6$. 
The turbulent driving we chose for the same initial column density of $19.4~\coldens$ reduced the SFR by a factor $4.5$ and led to an SFR below the SK relation. 
This means that a region with this moderate column density undergoing a strong shear would not need additional driving.
The question that arises is whether the shear may reduce the much higher energy needed to reproduce the SK law (see Fig.~\ref{fig:inj_power}) for the cases with a high column density. 
While we cannot answer this question for the shear with our current setup, it is possible to answer this for the Coriolis effect alone, as we did in Appendix \ref{sec:coriolis}. 
The conclusion is that for a high column density of $77 \coldens$, the Coriolis effect has a limited impact on the SFR and cannot replace or complete the large-scale turbulent driving to match the SK relation.

\subsection{Nature of the turbulence}
Following this line of thought, an important point to keep in mind is that the turbulence injected in groups \textsc{turb},  \textsc{turb\_Bvar},  \textsc{turb\_hB,} and  \textsc{turb3D} is calibrated so that the resulting SFR stays close to the observed value at a given column density and is not generated self-consistently. 
We showed in section \ref{sec:turbexplo} that the exact nature of the turbulence matters greatly for the SFR.
The choice of the Ornstein-Uhlenbeck process makes it hard to model a very coherent driving.
These drivings may produce velocity fields that are even more solenoidal that the field yielded by a purely solenoidal Ornstein-Uhlenbeck driving \citep[e.g. with the magneto-rotational instability in][Fig.~2]{gongImpactMagnetorotationalInstability2020}.
To proceed with more precise studies of gas-rich galaxies, a better modelling of the driving source is required.

\subsection{Sources of feedback}
Several sources of stellar feedback are not included in the model, such as stellar winds and cosmic rays. 
Both have been studied in other works (\cite{gattoSILCCProjectIII2017} and \cite{rathjenSILCCVIMultiphase2021}, respectively). Stellar feedback effects do not add up linearly for the SFR: including  at least one kind of early feedback (photo-ionisation, stellar winds, or infrared irradiation) may change the result dramatically, but combining them does not change the SFR by more than a factor of three. 

\subsection{Integration time}
Another possible caveat is that unlike other similar studies (e.g. \cite{kimThreephaseInterstellarMedium2017}), the time span over which the simulations were run was only 150 - 200 Myr. 
This ensured that the total mass of gas within the box did not vary by more than 40\% because of outflows and star formation.
Previous studies have often solved the issue of outflowing gas by using very elongated boxes that kept a large fraction of the outflowing gas within the computational domain, maybe artificially.

\subsection{Numerical parameters}
Finally, as always with numerical simulations, the influence of the choice of the numerical parameters may be a source of concern. This includes the influence of the sink creation criterion, which was discussed in \cite{collingImpactGalacticShear2018}. 
They showed in their figure 11b that the SFR is unchanged even when they modified the threshold for sink accretion by a factor 16, but the addition of a test to verify the boundedness of the gas was able to increase the SFR by almost a factor of two.
We discuss the details of the SN recipe in Appendix \ref{sec:vsat}. 
The limitation we applied for numerical reasons to the SN subgrid model altered the properties of the low-density gas, but had a limited impact on the dense gas and the SFR.
We present the convergence study with resolution in Appendix \ref{sec:convergence} and show that an increase in the resolution only modifies the SFR by about 10\%.

\section{Discussion and conclusions}
\label{sec:conclusion}

We extended the study of \citetalias{brucyLargescaleTurbulentDriving2020} of what sets the SFR at the kiloparsec scale. Using a very similar numerical setup of a kiloparsec cube region of a galaxy, we explored the parameter space, and more particularly, the effect of the compressibility of large-scale turbulence and the strength of the magnetic field. 
 In \citetalias{brucyLargescaleTurbulentDriving2020} we showed that stellar feedback alone was not sufficient to quench star formation to match the observed SFRs, and that large-scale turbulent driving may be the main effect that is missing, explaining the gap.
To refine the study, both the mechanism of turbulence injection (section \ref{subsec:turb_inj}) and the computation of the SFR   (section \ref{subsec:computation_sfr}) were improved to allow a better comparison with the observed SFR.
Compared to the reworked method for turbulence injection with the method used in \citetalias{brucyLargescaleTurbulentDriving2020}, quenching via large-scale turbulence is more efficient in low column density regions and slightly less efficient in high column density regions.

The main results of this study are the following:
\begin{enumerate}
    \item In high column density environments, stellar feedback is unable to destroy a large accumulation of dense gas and quench star formation efficiently enough to yield a rate that is compatible with observations (section \ref{subsec:noturb}).
    \item If it is strong enough, 2D large-scale turbulent driving can destroy these giant clouds and can efficiently quench star formation down to levels that are compatible with observations (section \ref{subsec:turb}).
    \item Increasing the 2D turbulent driving force linearly increases the velocity dispersion parallel to the disk up to a maximum limit at which it saturates. Below this saturation limit, stronger turbulent driving translates into a lower SFR (section \ref{subsec:turb_strength}). In the case of a low magnetic field, the velocity dispersion associated with the turbulent-driving strength that is required to match the SK relation is high compared with observations (section \ref{subsec:veldisp}).
    \item The compressibility of the turbulent driving force matters as more compressive driving schemes are ten times less efficient than solenoidal ones in quenching star formation (section~\ref{subsec:compressibility}).
    \item An increased magnetic field can also have a dramatic effect on the SFR, where an increase of roughly $10~\mu G$ reduces the SFR by a factor of ten (section \ref{sec:mag}). It cannot explain the observed SFR with stellar feedback alone, but it reduces the amount of energy needed from large-scale driving (section \ref{subsec:bturb}) and the associated velocity dispersion (section \ref{subsec:veldisp}).
    
\end{enumerate}

With this work, we provided an overview of how the large-scale 
driving and the mean value of the magnetic field can influence the 
SFR.
The next step towards comprehending what 
regulates the SFR at the kiloparsec scale in general and the 
role of the turbulent driving in particular is to obtain more precise constraints on the 
energy that can be tapped from the large scale, as well as the nature of the driving.

Several analytical works have investigated the possible source of this turbulence, gravitational instabilities 
\citep{nusserRegulationStarFormation2022}, mass transfer within the 
galactic disk \citep{krumholzUnifiedModelGalactic2018, meidtModelOnsetSelfgravitation2020}, 
or accretion onto the disk from the circumgalactic medium.
Observations from local \citep[e.g.][]{sunMolecularCloudPopulations2022} 
or high-redshift galaxies, as well as isolated and cosmological simulations, are other 
ways to better know how energy is transferred from large to small scales.

\section*{Acknowledgements}

The authors thank the referee, Enrique Vazquez-Semadeni, for his useful comments which helped to significantly improve the article. 
The authors acknowledge Interstellar Institute's program "With Two Eyes" 
and the Paris-Saclay University's Institut Pascal for hosting discussions 
that nourished the development of the ideas behind this work.
NB thanks Jérémy Fensch, Chang Goo Kim  and Munan Gong for interesting and fruitful discussions.
This work was granted access to HPC resources of CINES and
CCRT under the allocation x2020047023 made by GENCI (Grand
Equipement National de Calcul Intensif) 
and the special allocation 2021SA10spe00010. 
NB, PH and TC acknowledges financial support from the European Research Council (ERC) via the ERC Synergy Grant "ECOGAL: Understanding our Galactic ecosystem -- From the disk of the Milky Way to the formation sites of stars and planets" (grant 855130).

\subsubsection*{Software}

We made use of the following software and analysis tools:
GNU/Linux, \textsc{Ramses} \citep{teyssierCosmologicalHydrodynamicsAdaptive2002}, \textsc{Python}, \textsc{Matplotlib} \citep{hunterMatplotlib2DGraphics2007}, \textsc{Numpy} \citep{vanderwaltNumPyArrayStructure2011}, 
\textsc{Pymses} \citep{guilletPyMSESPythonModules2013}, \textsc{Astrophysix},
\textsc{Astropy} \citep{astropycollaborationAstropyCommunityPython2013, astropycollaborationAstropyProjectBuilding2018}. 
Many thanks to the authors for making them publicly available.
Special thanks to all the contributors of \textsc{ramses}, and in particular to Andrew McLeod for the introduction of the turbulence driving module that has been heavily used in this work.

\subsubsection*{Data availability}

The data underlying this article are available in the Galactica Database at \url{http://www.galactica-simulations.eu}, and can be accessed with the unique identifier \textsc{\href{http://www.galactica-simulations.eu/db/STAR_FORM/ISMFEED}{ISMFEED}}.
Additional data and the source code used to run simulations and perform analysis will be shared on reasonable request to the corresponding author.

\bibliographystyle{aa}
\bibliography{phD} 

\appendix

\section{Calibration of the turbulent driving}
\label{sec:calibration}
In section \ref{subsec:compressibility} we studied the effect of changing the solenoidal fraction $\zeta$ of the driving on the SFR. 
In order to have a fair comparison, it is important to ensure that the strength of the driving is the same, regardless of $\zeta$. 
A way to  check this is to consider the RMS power of the Fourier modes,
\begin{equation}
    \label{eq:rms}
    \mathrm{RMS(t)} = \sqrt{\int\left\vert\bm{\hat{f}}(\bm{k},t)\right\vert^2  \mathrm{d}^3\bm{k}}
.\end{equation}

\begin{figure}[ht]
    \includegraphics[width=0.48\textwidth]{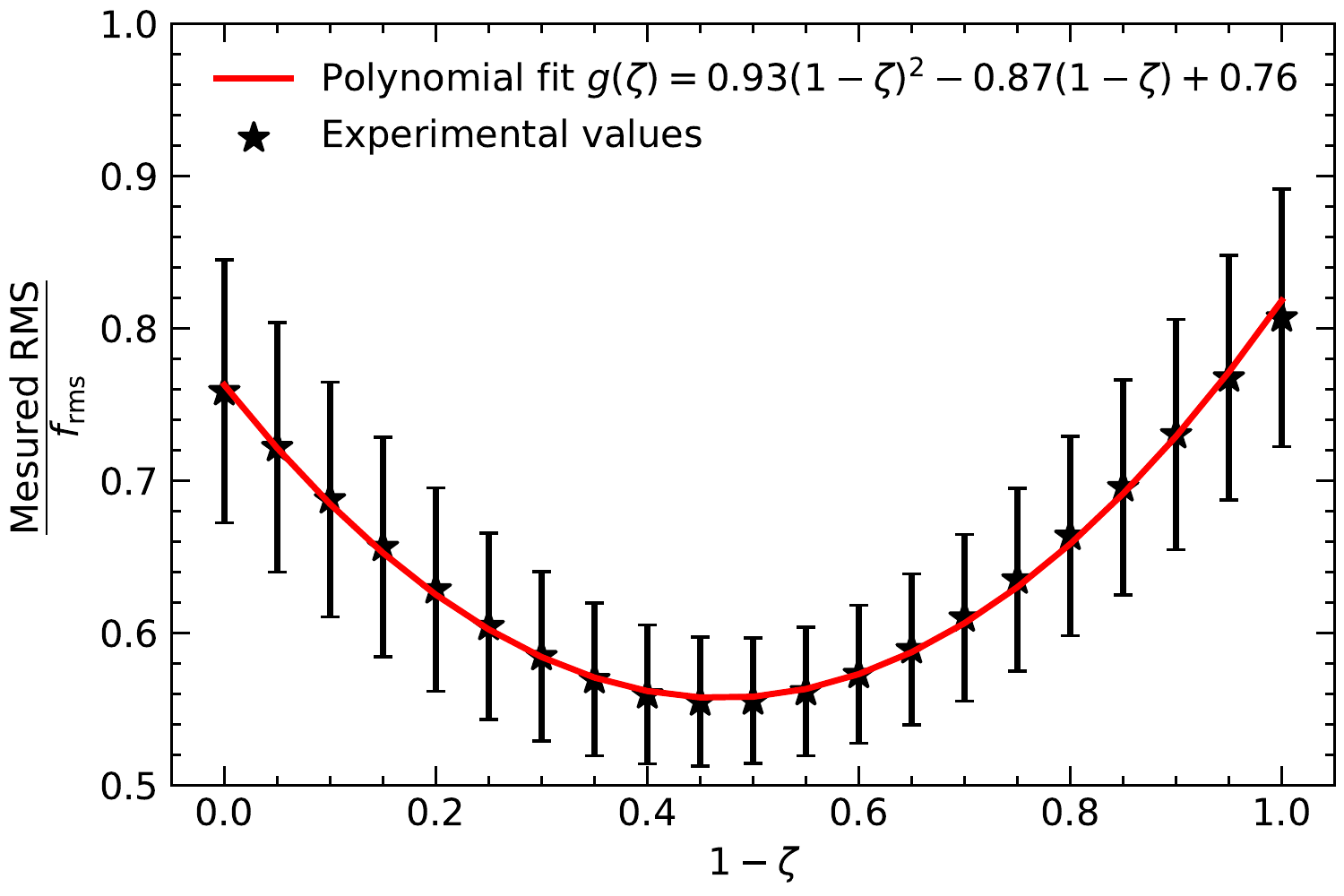}
    \caption{Ratio of the time-averaged RMS  (as in Eq. \eqref{eq:rms}) over the $f_\mathrm{rms}$ parameter for unnormalised simulations. The fit of the curve gives the normalisation factor $g_\zeta$.}
    \label{fig:calibration}
\end{figure}

The time-averaged RMS depends on the solenoidal fraction (see Fig.~\ref{fig:calibration}), and the role of the normalization factor $g_\zeta$ of Eq. \eqref{eq:injection} is to correct for this.
Because we used  2D driving instead of 3D driving, we cannot use the normalisation factor given by \cite{schmidtNumericalSimulationsCompressively2009} in their Equation (9). 
Instead, we empirically computed the relevant factor by running the turbulent generation code without any normalisation and compared the measured RMS in the Fourier space with the input $f_\mathrm{rms}$ (see Fig.~\ref{fig:calibration}). A fit of the obtained curve then gives the $g_\zeta$ factor we used in the code,

\begin{equation}
    g_\zeta = 0.93(1 - \zeta)^2 - 0.87 (1- \zeta) + 0.76
.\end{equation}

\section{Limitation of the supernova feedback}
\label{sec:vsat}

\begin{figure}[ht]
\includegraphics[width=0.49\textwidth]{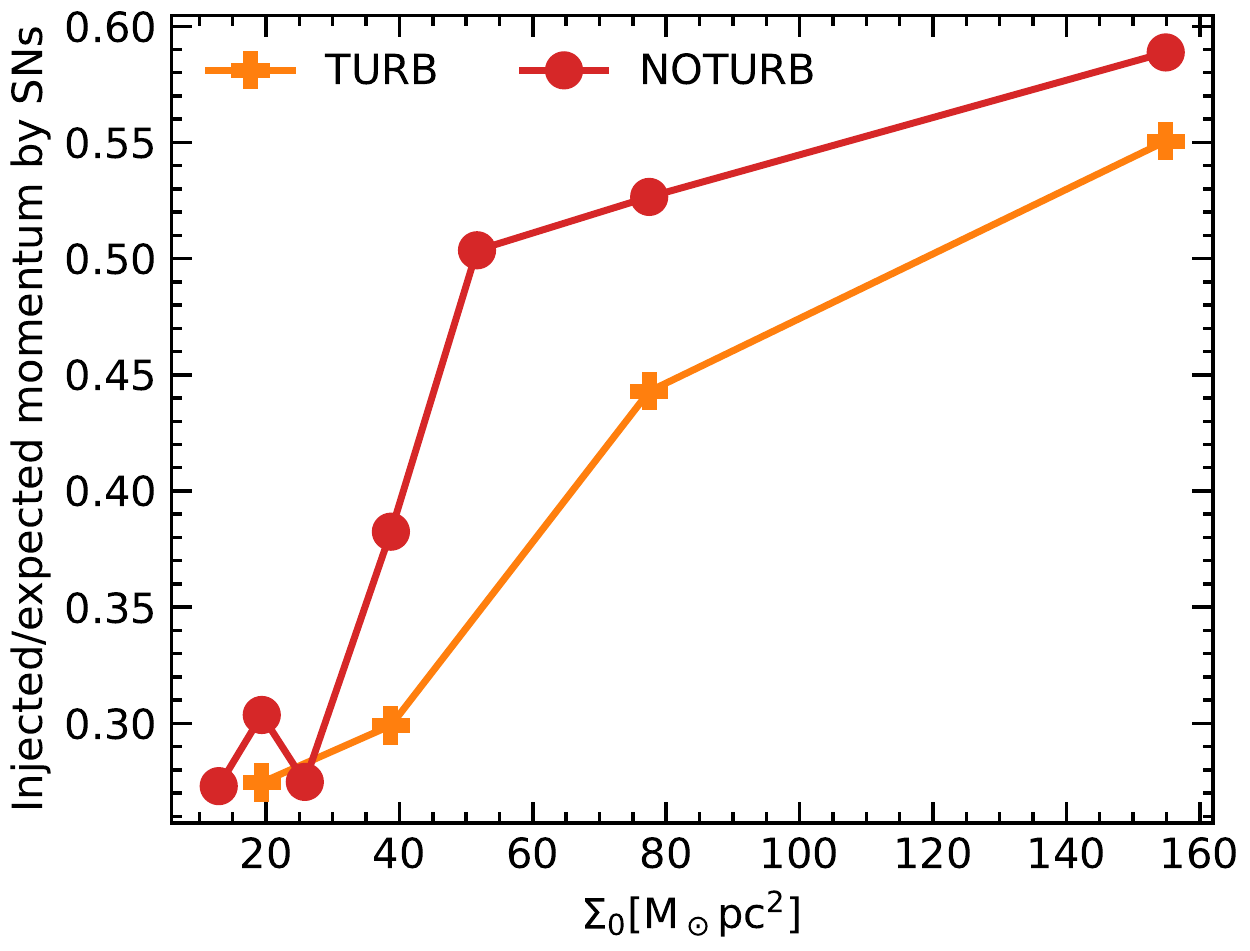}
\caption{Averaged ratio of the injected momentum over the expected reference momentum of $4\times 10^{43}$ g $\cdot$ cm $\cdot$ s $^{-1}$ for all the SNe that blew up during the simulation.
For simulations with higher initial column densities, the injected momentum approaches the reference value.}
\label{fig:sn_inj_mom}
\end{figure}

\begin{figure}[ht]
\includegraphics[width=0.49\textwidth]{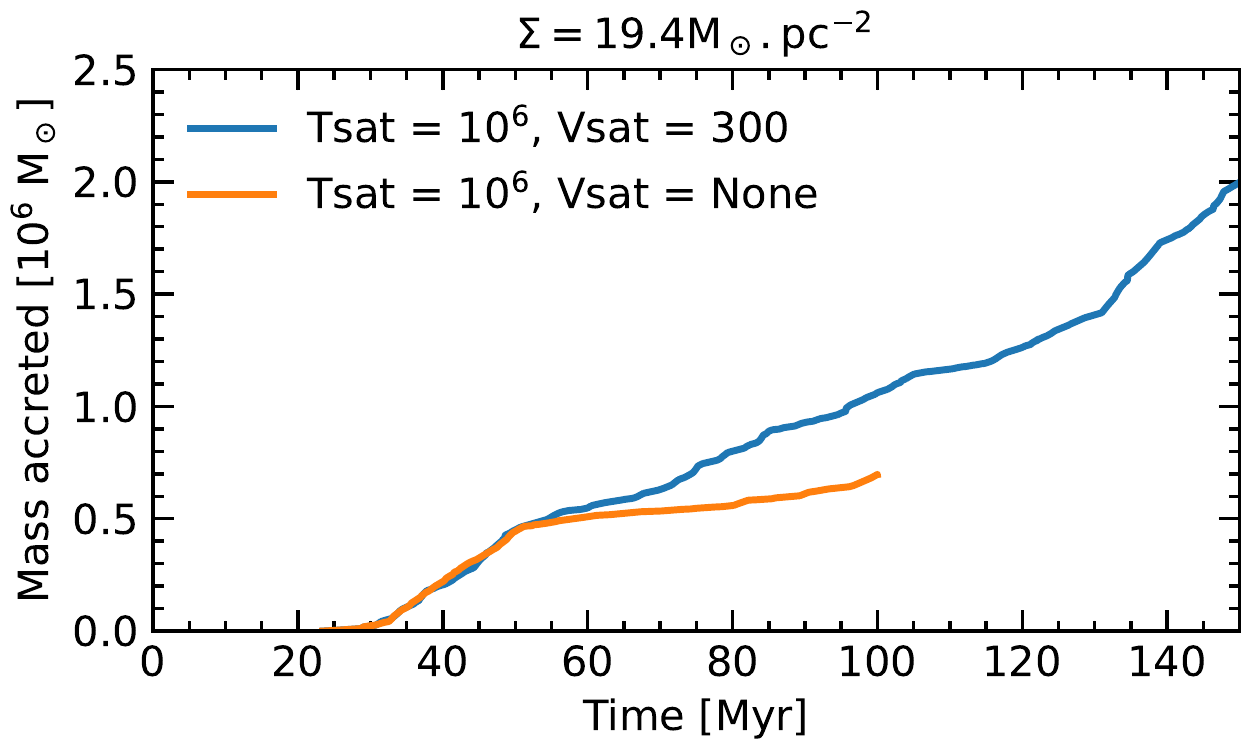}
\includegraphics[width=0.49 \textwidth]{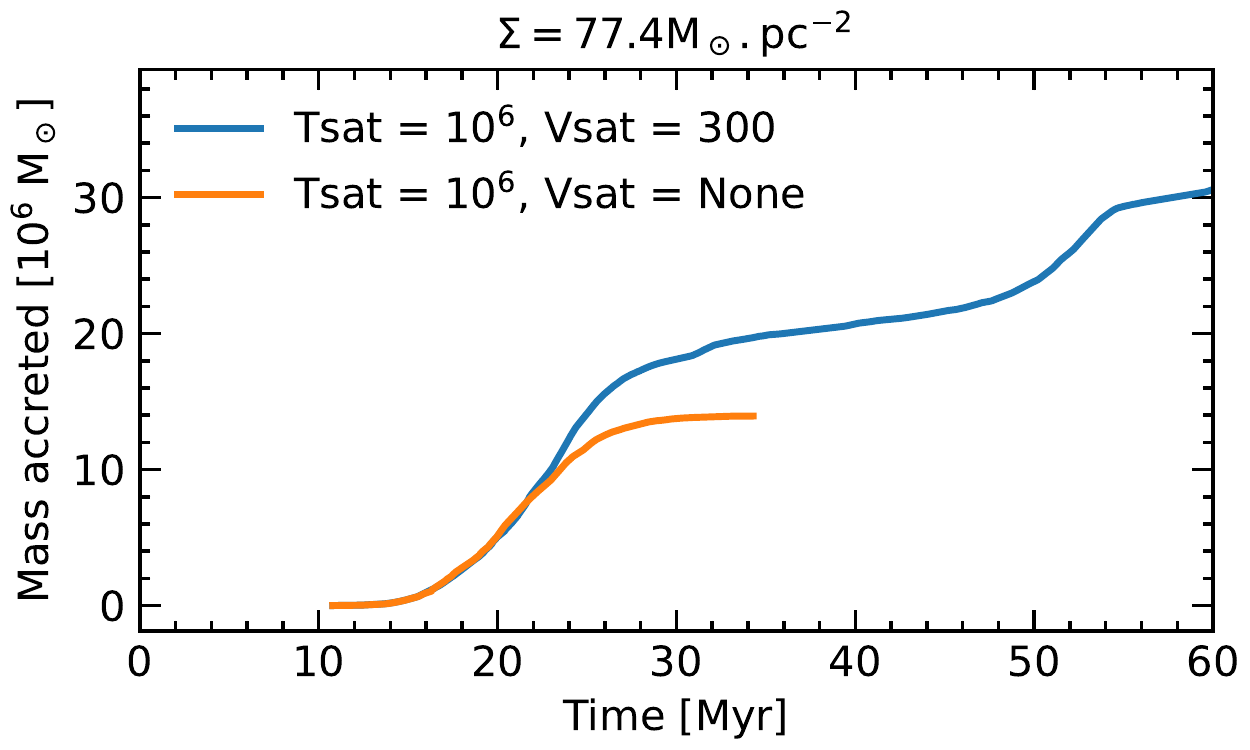}
\caption{Comparison of the SFR with (blue line) and without (orange line) limiting the velocity of the SN ejecta for $n_0 = 1.5$ (top) and $n_0 = 6$ (bottom).}
\label{fig:compsn_sfr}
\end{figure}

\begin{figure*}[ht]
\begin{center}
\hfill
\includegraphics[height=0.34 \textheight]{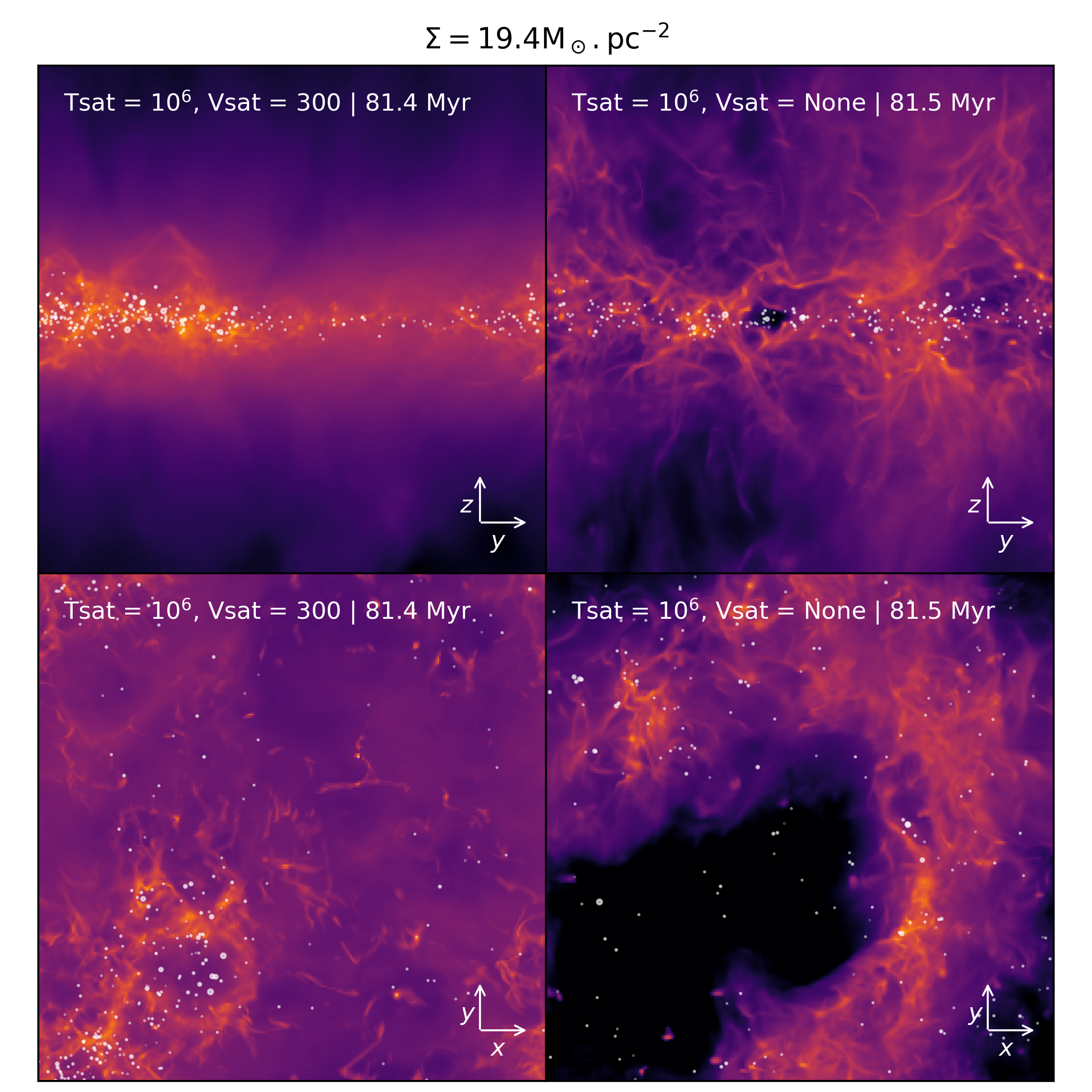}
\hfill
\includegraphics[height=0.34 \textheight]{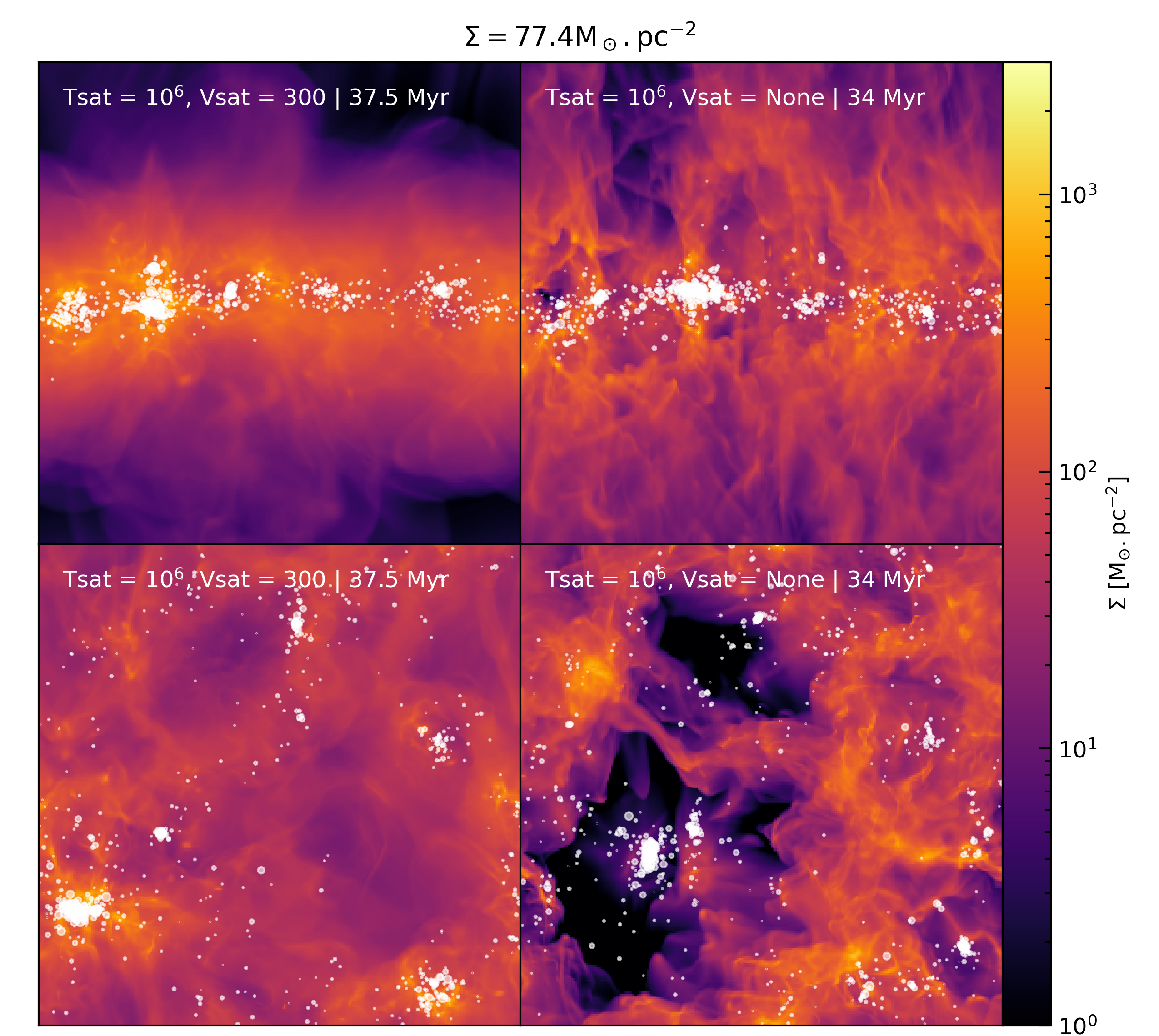}
\hfill
\end{center}
\caption{Comparison of the edge-on (top) and face-on (bottom) column density map with and without (left and right of each panel, respectively) limiting the velocity of the SN ejecta for $n_0 = 1.5$ (left panel) and $n_0 = 6$ (right panel).}
\label{fig:compsn_coldens}
\end{figure*}

\begin{figure*}[ht]
\begin{center}
\includegraphics[width=0.97 \textwidth]{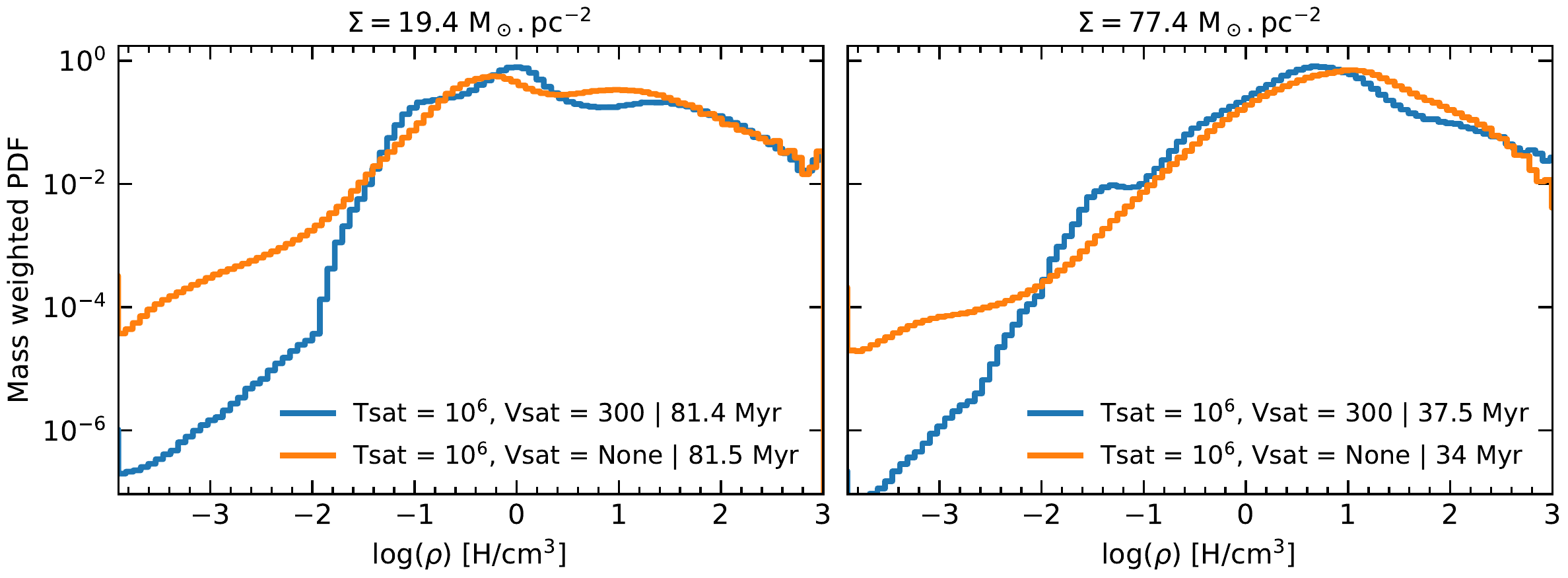}
\end{center}

\label{fig:compsn_pdf}
\caption{Mass-weighted probability distribution function of the gas density for the same snapshots as in Fig.~\ref{fig:compsn_coldens}.}

\end{figure*}

When SNe explode in a low-density environment, they can generate very high velocities and temperatures.
This leads to a numerical issue because it implies very short time-steps, to a point that the simulation can hardly progress. 
In order to avoid this in our simulation, the energy from SNe was mainly injected as a momentum, and both the gas temperature and the gas velocity cannot exceed $\mathrm{T}_\mathrm{sat} = 10^6~\mathrm{K}$ and $\mathrm{V}_\mathrm{sat} = 300~\kms$.
As a consequence, we did not model the very hot phase of the gas, which, despite representing only a small fraction of the total mass of the ISM, fills the majority of the volume.
However, the hot gas cools down very slowly and thus cannot collapse to form stars at the timescale we studied.
Furthermore, with our 4 pc resolution, the cooling radius is not resolved for most of the SNe, especially for the most massive simulations.

Limiting the velocity means that overall, the momentum injected is lower than prescribed by the SN model, as we show in Fig.~\ref{fig:sn_inj_mom}. The figure depicts the ratio of the total momentum effectively injected by SNe over the total expected momentum, which is the number of SNe multiplied by the reference value of $4\times 10^{43}$~g~$\cdot$~cm~$\cdot$~$\mathrm{s}^{-1}$. 
On average, $30$ to $60 \%$ of the reference momentum is injected, with a higher ratio in a dense environment.
This may seem very low, but it is worth noting that SNe exploding in a low-density environment have a quasi-negligible impact on dense gas, as was demonstrated by \cite{iffrigMutualInfluenceSupernovae2015}.

In order to verify that the limitation of the velocity of the gas does not change the outcome of the SFR significantly, we ran two simulations analogous to \textsc{noturb} with $n_0 = 1.5$ and $6$ without the limitation. 
When the SNe ignited, the time step of the simulation without the limited velocity was ten times lower than the fiducial time step. 
Therefore, the simulations without a limited velocity evolved for a shorter time.
Since our main focus is the computation SFR, we compared the amount of gas accreted by the sink particles in both cases in Fig.~\ref{fig:compsn_sfr}.
It shows that for the period we considered, the limitation of the velocity due to SN feedback leads to an overestimation of the SFR by a factor $1.6$. 
This is comparable to the uncertainty due to other factors, such as specific choices of the sink creation recipe, implementation of other stellar feedback, or resolution.
We were unable to run the simulation longer, but similar simulations run by \citet{kimThreephaseInterstellarMedium2017} and \citet{ostrikerPressureRegulatedFeedbackModulatedStar2022} with an elongated box in the vertical direction indicates that the SFR is bound to oscillate, with bursts of similar or slightly lower amplitude as the first event of star formation depicted in Fig.~\ref{fig:compsn_sfr}.

The column density map (Fig.~\ref{fig:compsn_coldens}) reveals that our fiducial simulation fails to capture an important feature of the gas structure: superbubbles of hot gas. 
However, the dense gas does not seem to be strongly affected by the change in the SN feedback recipe: it is displaced, but its internal structure and its quantity are roughly the same. 
The latter point can be verified through the mass-weighted probability distribution function of the gas density shown in Fig.~\ref{fig:compsn_pdf}. 
This figure shows that the amount of very low density gas increases when the SN speed is not limited, but the amount of dense gas stay the same.

\section{Convergence with the resolution}
\label{sec:convergence}

All the simulations we discussed were run with a uniform grid resolution of about $4$ pc, with $256^3$ cells. 
This allowed the wide parameter study presented here, with 63 simulations discussed out of a total of more than 200.
It is important, however, to determine the influence of the resolution and whether convergence is reached.
To this end, we selected two simulations (group \textsc{turb} with $n_0=6$ and group \textsc{noturb} with $n_0=6$) and reran them with a doubled resolution ($512^3$ cells of size $\sim$ 2 pc) and half resolution  ($128^3$ cells of size $\sim$ 8 pc).
The sink creation threshold $n_\mathrm{sink}$ was adapted so that
$n_\mathrm{sink} \propto \dfrac{1}{\vert \Delta x \vert^2}$ , where $\Delta x$ is the size of a cell
\citep{kimThreephaseInterstellarMedium2017}.

\begin{figure}[hb]
\begin{center}
\includegraphics[width=0.41 \textwidth]{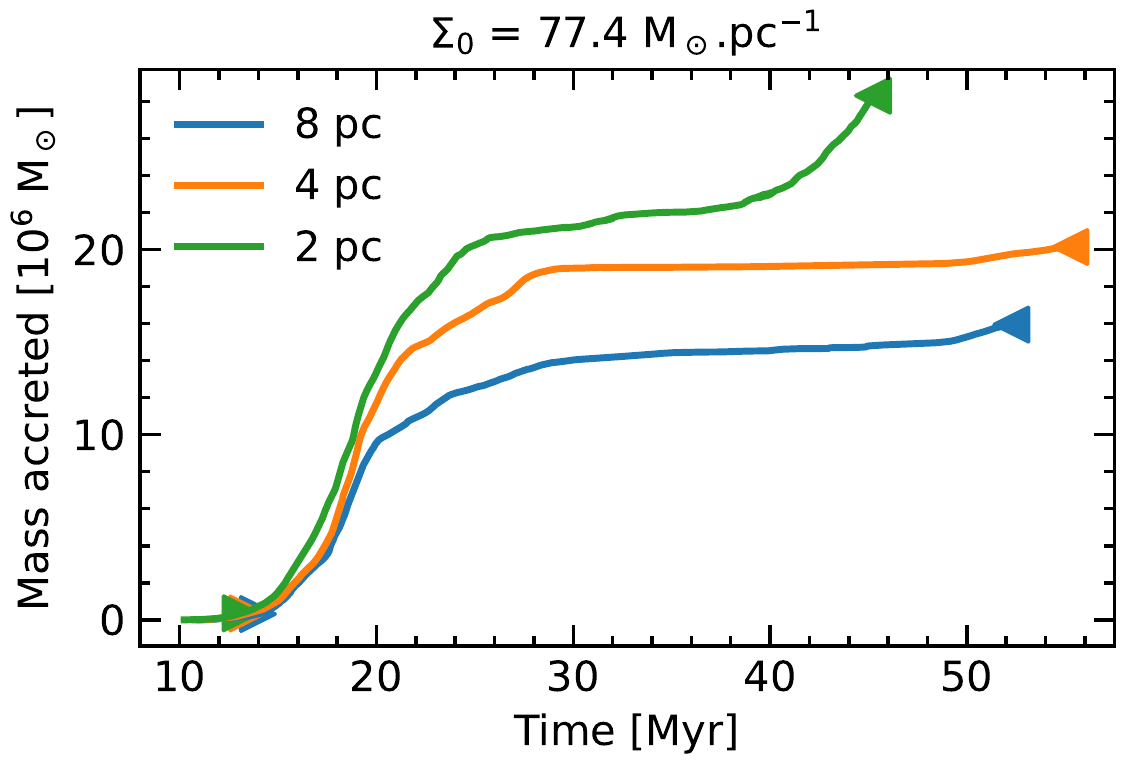}
\includegraphics[width=0.41  \textwidth]{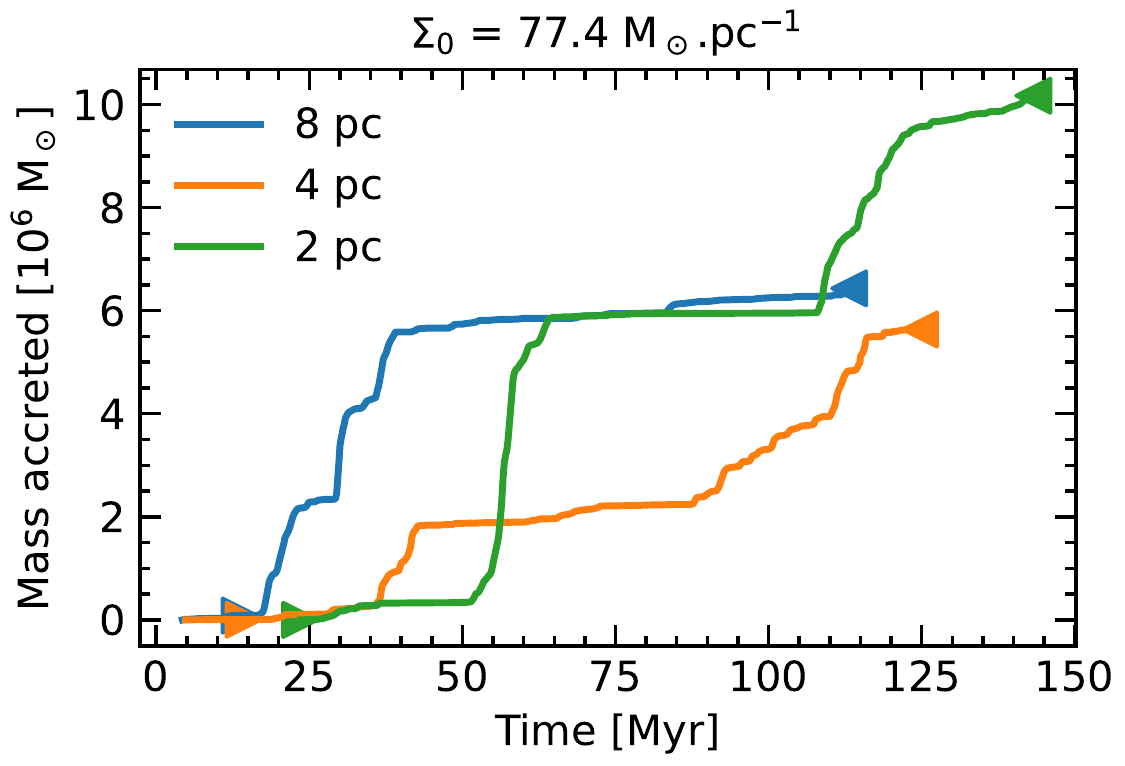}
\end{center}
\caption{Comparison of the total mass accreted by sinks as a function of time for different resolutions for a simulation with turbulence driving (top) and a simulation without it (bottom).}
\label{fig:res_sink_mass}
\end{figure}

\begin{figure}
\begin{center}
\includegraphics[width=0.45 \textwidth]{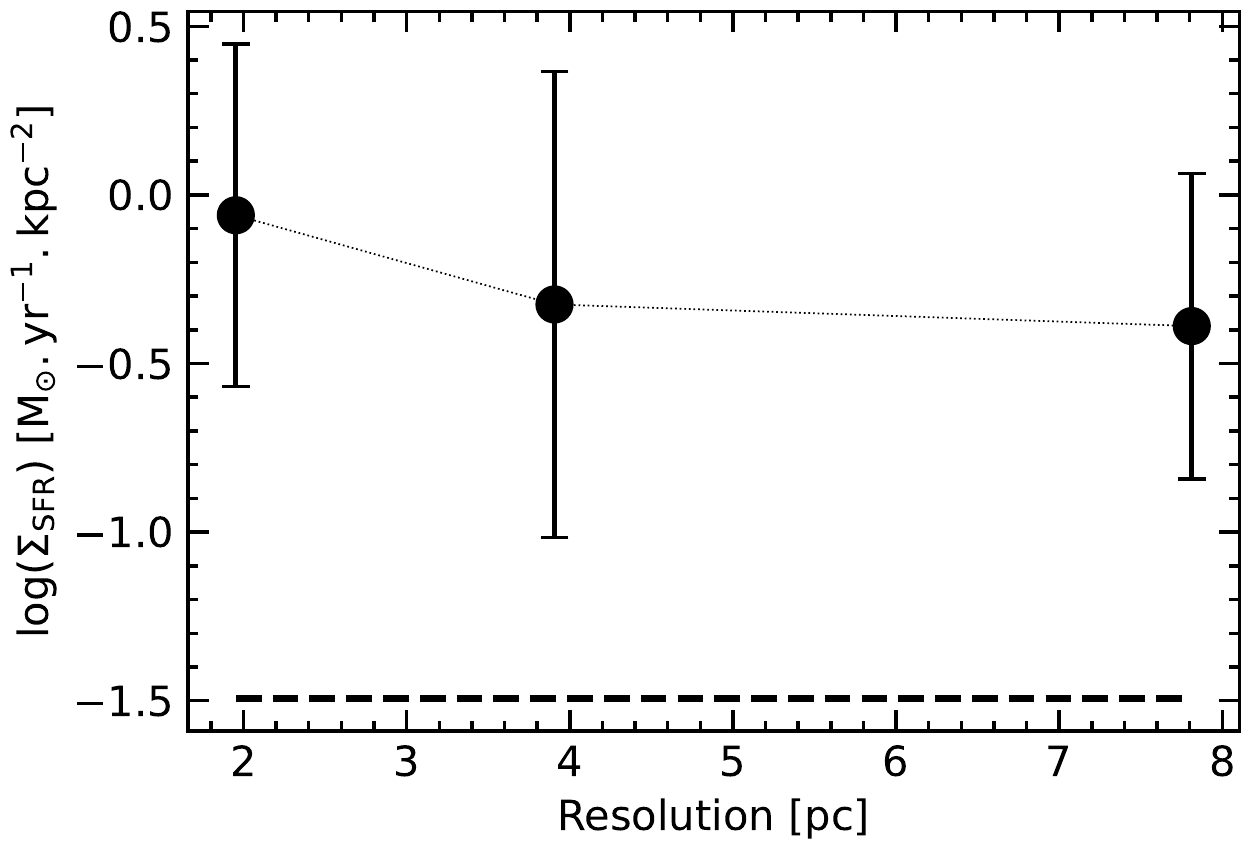}
\includegraphics[width=0.45  \textwidth]{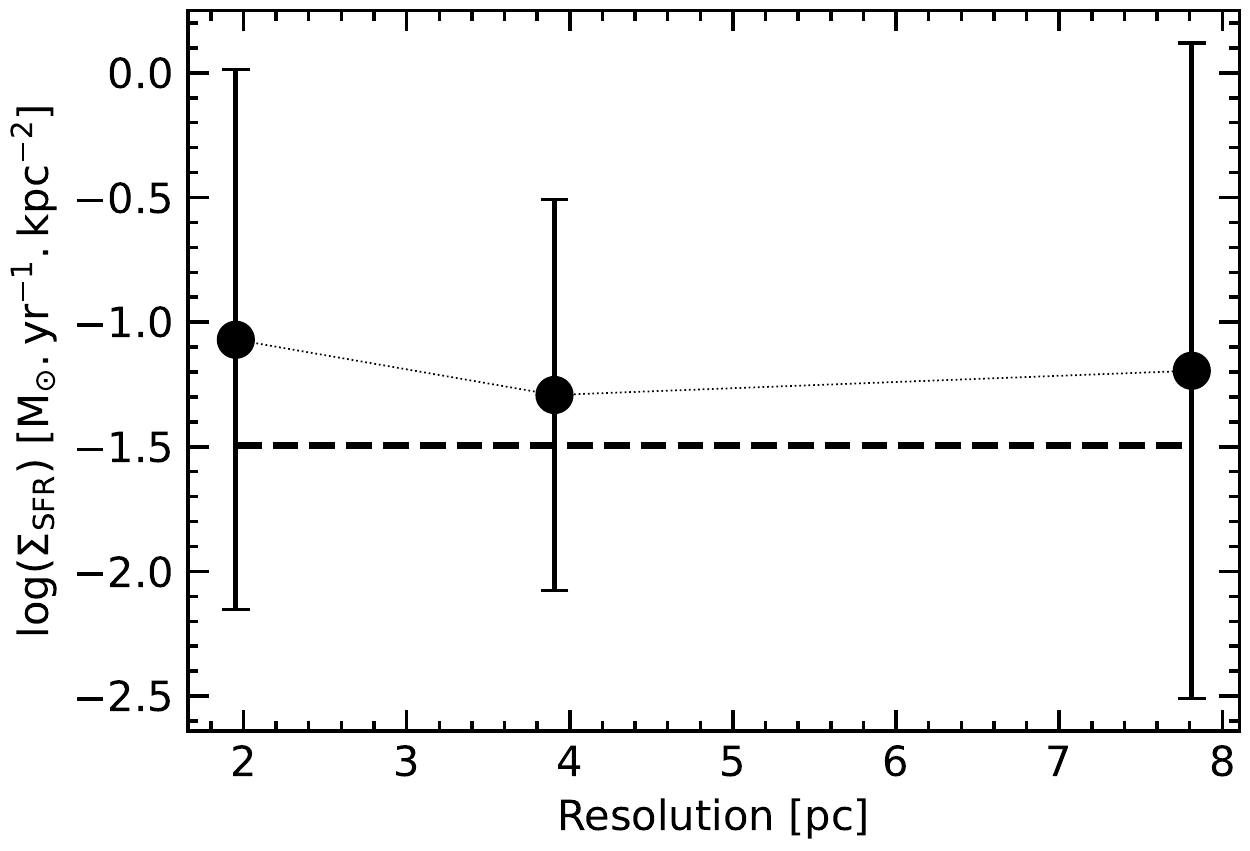}
\end{center}
\caption{Comparison of the computed SFR for different resolutions for a simulation with turbulence driving (top) and a simulation without it (bottom).  The dashed line is the value of the SFR given by the SK relation. }
\label{fig:res_sfr}
\end{figure}

Figure \ref{fig:res_sink_mass} shows the mass accreted by sinks in these simulations. 
In the case without turbulent driving (top), increasing the resolution results in a increased SFR.
The case with driving is less clear, especially because star formation is more stochastic in this configuration. 
In both cases, the variations in the amount of mass accreted by the sinks and the resulting SFR (Fig.~\ref{fig:res_sfr}) due to the resolution are about 10\%. 
This is small compared with the variations measured when changing the physical parameters (sections \ref{sec:turbexplo} and \ref{sec:mag}). 
It is within the same range as the other source of uncertainties, such as the choice of other numerical parameters (sink creation threshold and seed for the turbulence) or temporal variations.
This convergence result is in line with resolution studies that were carried out for a similar setup at a lower column density (\cite[Fig.~9]{collingImpactGalacticShear2018} and \cite[Fig.~14]{kimThreephaseInterstellarMedium2017}).

\begin{figure}
\begin{center}
\includegraphics[width=0.45 \textwidth]{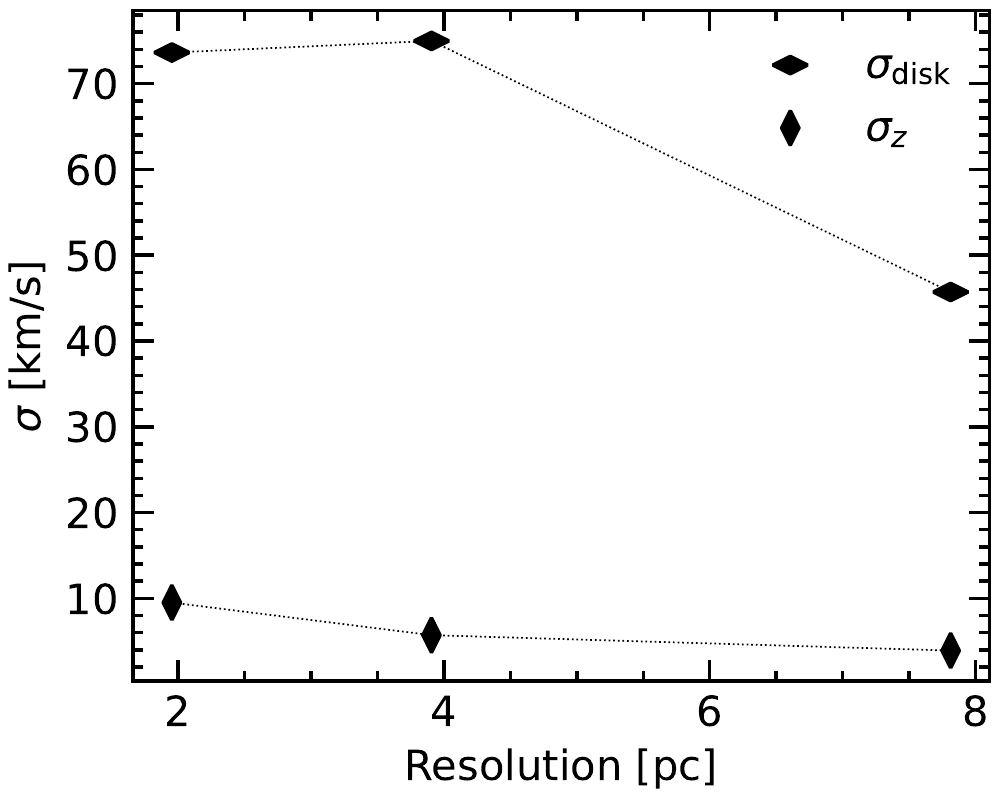}
\end{center}
\caption{Comparison of the mass-weighted velocity dispersion for different resolutions for a simulation with turbulence driving at $t = 72$ Myr.}
\label{fig:res_veldisp}
\end{figure}

Figure \ref{fig:res_veldisp} shows that similarly, the variation in the mass-weighted velocity dispersion are small between our fiducial resolution of $4$ pc and a doubled resolution of $2$ pc.

\newpage

~

\newpage

\section{Impact of the Coriolis effect on the SFR}
\label{sec:coriolis}

As stated in section \ref{sec:caveats}, our model does not include shear, mainly for numerical reasons. 
However, it is rather easy to implement the Coriolis effect in our simulations and assess how much it impacts the SFR.
To make this experiment, we added a source term $\rho \bm{f_\mathrm{cor}}$ in the momentum equation \eqref{eq:euler}.
We have
\begin{equation}
\bm{f_\mathrm{cor}} = - 2 \Omega \bm{e_z} \times  \bm{v}
,\end{equation}
where $\bm{e_z}$ is a unit vector in the vertical direction, and $\Omega$ is the angular speed, a constant and uniform parameter in our setup.
In practice, we updated the velocity field after each hydro and gravity step using an implicit Crank Nicholson scheme,
\begin{align}
    v_\mathrm{x,\mathrm{new}} &= \dfrac{v_\mathrm{x,\mathrm{old}} \left( 1 - \left(\Omega dt \right)^2 \right) + 2 v_\mathrm{y,\mathrm{old}}\Omega dt}{\left(\Omega dt \right)^2}, \\
    v_\mathrm{y,\mathrm{new}} &= \dfrac{v_\mathrm{y,\mathrm{old}} \left( 1 - \left(\Omega dt \right)^2 \right) - 2 v_\mathrm{x,\mathrm{old}}\Omega dt}{\left(\Omega dt \right)^2}, \\
    v_\mathrm{z,\mathrm{new}} &= v_\mathrm{z,\mathrm{old}}.
\end{align}
The implementation was similar to that used in \cite{collingImpactGalacticShear2018} and was dutifully tested.

We chose an initial column density of $\Sigma_0 = 77.4~\coldens$ and an initial magnetic intensity $B_0 = 3.8~\mu\mathrm{G}$.
We considered values of the angular speed of $27.5$ and $110~\kms\cdot\mathrm{kpc}^{-1}$.
When we assume a constant rotation velocity of $220~\kms$, these values would correspond to the angular speed at $8$ and $2~\mathrm{kpc}$, respectively. 
As a first step, we ran the simulations without additional large-scale turbulent driving\footnote{Self-consistent SN driving is still present.} to determine whether the Coriolis effect constituted an alternative mechanism to the large-scale driving.

\begin{figure}[h]
\begin{center}
\includegraphics[width=0.45 \textwidth]{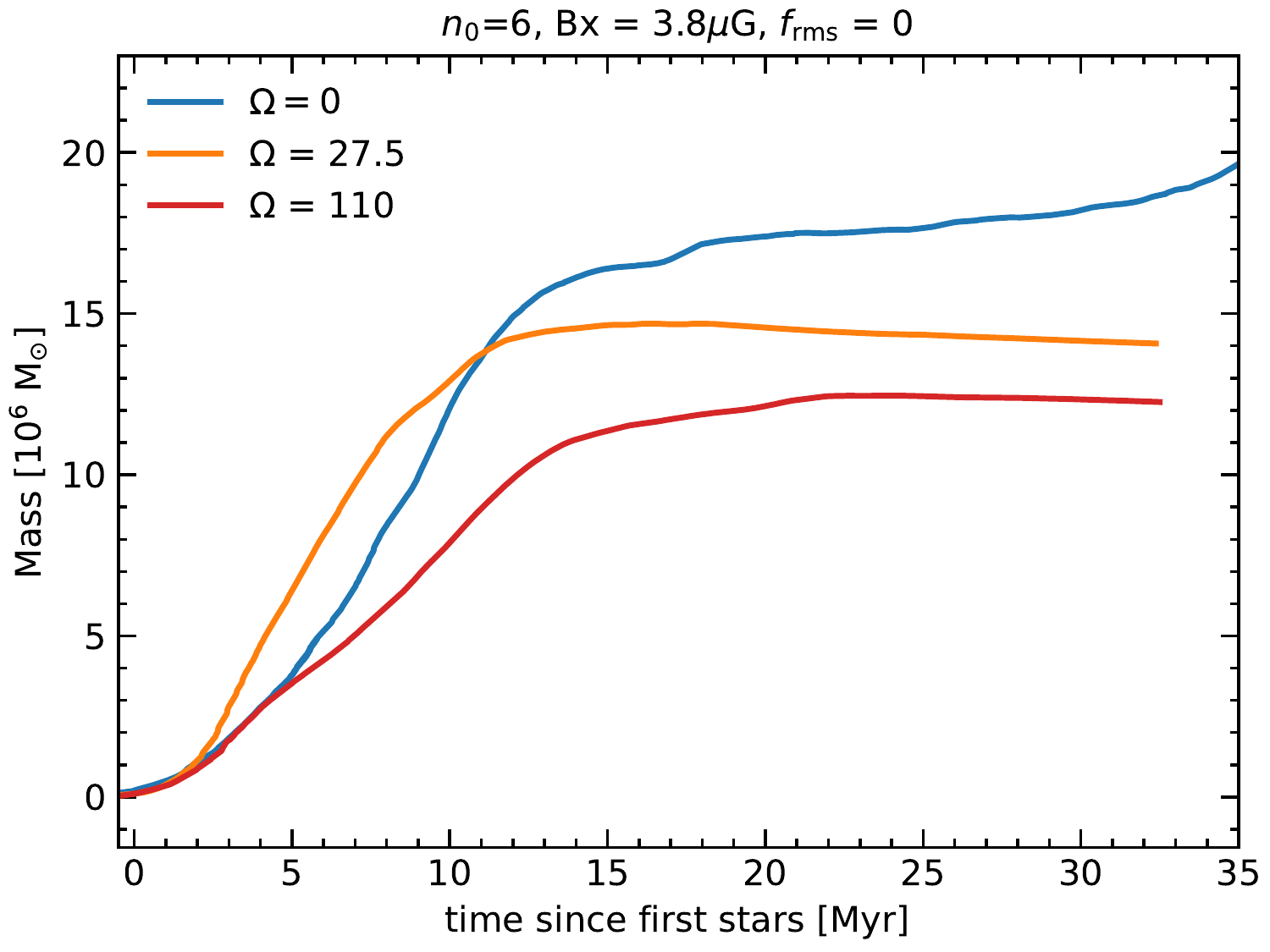}

\includegraphics[width=0.45 \textwidth]{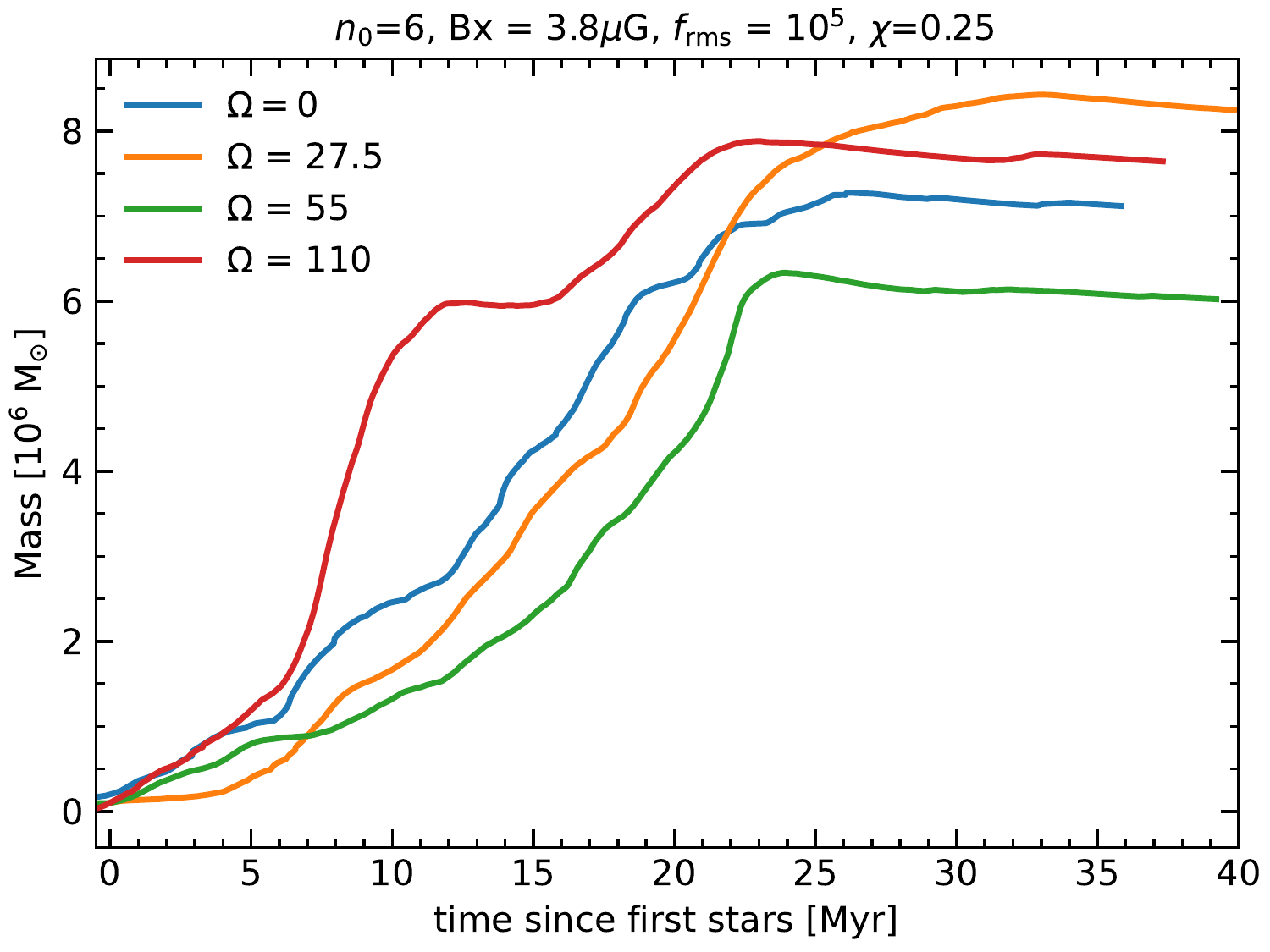}
\end{center}
\caption{Comparison of the accreted mass in sink particles for different angular speeds $\Omega$ (in $\kms\cdot\mathrm{kpc}^{-1}$) for the simulations without turbulent driving (top) and with turbulent driving $f_\mathrm{rms}$ of $10^5$ in code units (bottom). The time is rescaled so that 0 Myr corresponds to the time at which $10^5 \Msun$ were accreted into sinks.}
\label{fig:coriolis_sink_mass}
\end{figure}

Fig.~\ref{fig:coriolis_sink_mass} (top) shows that the Coriolis effect slightly stabilises the disk, with a reduction of the SFR of nearly a factor 1.5 with $\Omega = 110~\kms\cdot\mathrm{kpc}^{-1}$ compared to the simulation without the Coriolis effect. 
However, the resulting SFR is still much higher than the observed values. 
This means that even with a very strong Coriolis effect, an additional source of driving is still needed.

\begin{figure*}[hb]
\begin{center}
\includegraphics[width=0.7\textwidth]{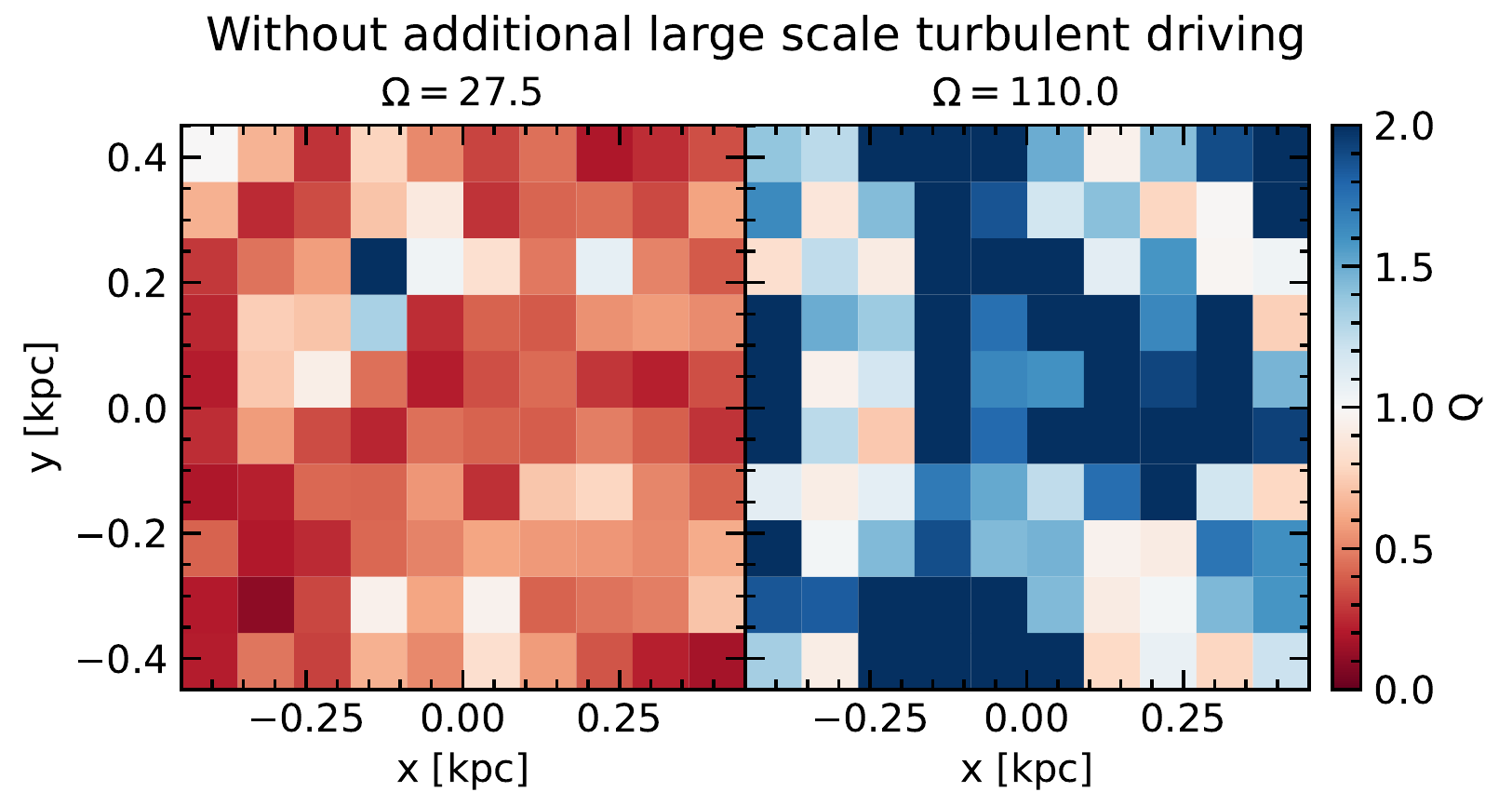}
\vspace{0.03 \textheight}

\includegraphics[width=\textwidth]{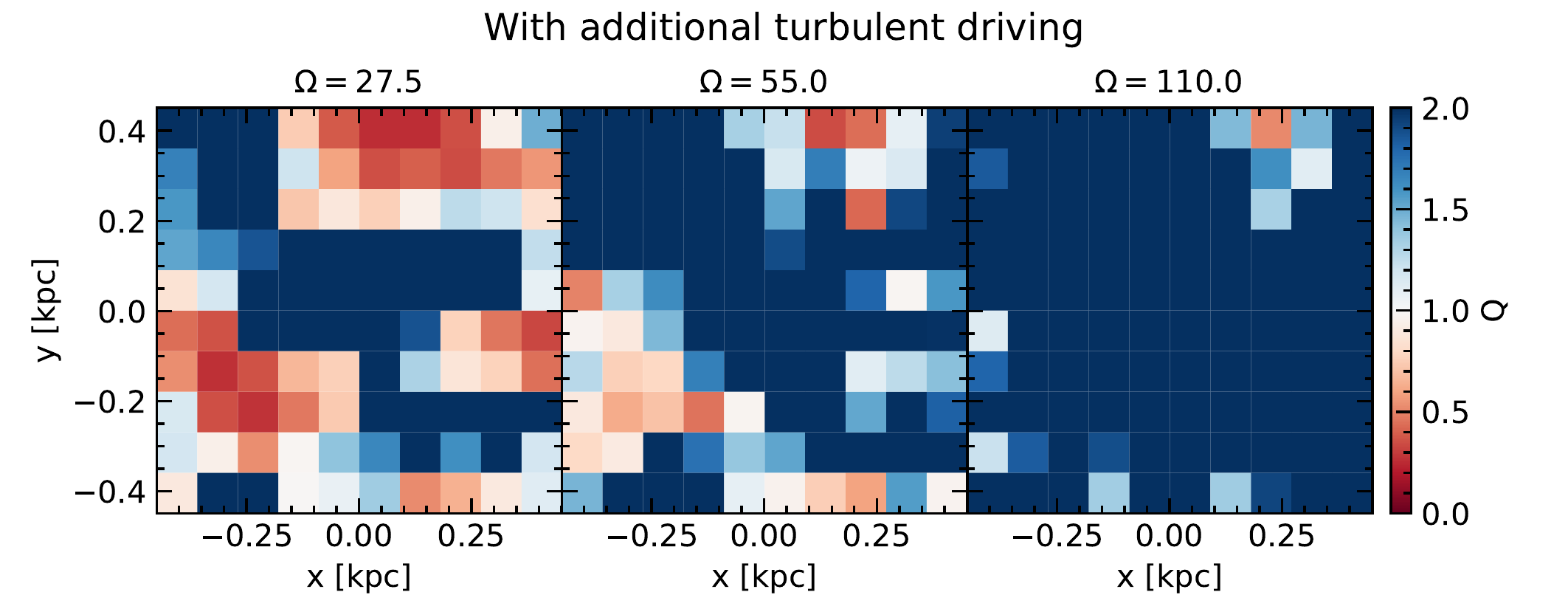}
\end{center}
\caption{Value of the Toomre parameter $Q$ (Eq. \eqref{eq:toomre_sigma}) computed in a square of $100~\mathrm{pc} \times 100~\mathrm{pc}$ for the simulations without turbulent driving (top) and with turbulent driving (bottom) at $t = 20$ Myr and for different angular speeds $\Omega$ (in $\kms\cdot\mathrm{kpc}^{-1}$).}
\label{fig:coriolis_Q}
\end{figure*}

This result is slightly surprising. Because the angular speed $\Omega$ is constant and uniform in our simulation domain, the implied Toomre parameter is given by
\begin{equation}
    Q = \dfrac{2 \Omega \sigma }{\pi G \Sigma}
    \label{eq:toomre_sigma}
.\end{equation}
If it is computed at the box scale ($1$ kpc), the velocity dispersion $\sigma = \sqrt{\sigma_x^2 + \sigma_y^2 + \sigma_z^2} / \sqrt{3}$ is about$10~\kms$ at $20~\mathrm{Myr}$ for the two runs.
For the chosen values of the angular speed $\Omega = 27.5$ and $110~\kms\cdot\mathrm{kpc}^{-1}$, we obtain $Q = 0.5$ and $2.1$, respectively.
We would expect the disk to be stabilised against gravitational instabilities for the simulation with $\Omega = 110~\kms\cdot\mathrm{kpc}^{-1}$, which has a value of the Toomre parameter $Q$ above one, the stability limit. 
However, because of the density fluctuations and because the velocity dispersion is lower when it is computed at a smaller scale, it is possible to reach lower values of Q locally.
Fig.~\ref{fig:coriolis_Q} shows the value of Q when it is computed in squares of $100~\mathrm{pc} \times 100~\mathrm{pc}$ at $t = 20   \mathrm{Myr}$. 
For all the tested values of the angular speed $\Omega$, at least one region has a Toomre parameter well below $1$.
It is also important to have in mind that the value of the velocity dispersion $\sigma$ used in Eq. \eqref{eq:toomre_sigma} also encompasses compressive motions, which tend to enhance star formation.

Because of the slight stabilisation effect observed, we may also wonder whether the additional turbulent energy required to reproduce the SK relation discussed in section \ref{sec:sk} is lowered when the Coriolis force is added.
To test this, we ran the same simulations, but with a turbulent driving $f_\mathrm{rms}$ of $10^5$ (code units) with a compressibility $\chi = 0.25$. 
We compared them with the run with the same parameters in the group \textsc{strength}, which has no Coriolis effect and yields an SFR that is ten times higher that the SFR predicted by the SK relation.
Fig.~\ref{fig:coriolis_sink_mass} (bottom) shows that the Coriolis effect has almost no impact on the SFR when it is combined with the large-scale turbulent driving. 
Despite a slightly different star formation history, the slope of the curve is indeed roughly similar, regardless of the angular speed, without any systematic trend.
As a consequence, we find that the additional turbulent energy required to reproduce the SK relation is not affected by the presence or absence of the Coriolis effect.
A more complete study of the exact role of the Coriolis force and why in this case it does not seem to play a major role in the stability of the disk is beyond the scope of this article, but can be the topic of further research work.

\end{document}